\input harvmac.tex
\input epsf


\lref\BauSin{L.~ Baulieu and I.~ Singer, ``Topological Yang-Mills Symmetry",
Nucl. Phys. B. Proc. Suppl.  {\bf 5B} (1988) 12.}
\lref\BauSinii{L.~ Baulieu and I.~ Singer, ``The Topological Sigma Model",
Commun. Math. Phys. {\bf 125} (1989) 227-237.}
\lref\bbrt{D.~ Birmingham, M.~ Blau, M.~ Rakowski, and
G.~ Thompson,  ``Topological Field Theories,'' Phys. Rep. {\bf 209} (1991)
129.}
\lref\Bis{J.-M.~ Bismut, ``Localization Formulas, Superconnections,
and The Index Theorem For Families,'' Comm. Math. Phys. {\bf 103} (1986) 127.}
\lref\blau{M. Blau, ``The Mathai-Quillen Formalism and Topological
Field Theory", Notes of Lectures given at the Karpacz Winter School on
`Infinite Dimensional Geometry in Physics', Karpacz, Poland, Feb 17-29, 1992;
J. Geom. and Phys. {\bf 11} (1991) 129.}
\lref\BlThlgt{M.~ Blau and G.~ Thompson, ``Lectures on 2d Gauge
Theories: Topological Aspects and Path Integral Techniques", Presented at the
Summer School in Hogh Energy Physics and Cosmology, Trieste, Italy, 14 Jun - 30
July 1993, hep-th/9310144.}
\lref\BlThqym{M.~ Blau and G.~ Thompson, ``Quantum Yang-Mills Theory On
Arbitrary Surfaces",  Int. J. Mod. Phys. {\bf A7} (1992) 3781.}
\lref\blauthom{M.~ Blau and G.~ Thompson,
``N=2 Topological Gauge Theory, the Euler Characteristic
of Moduli Spaces, and the Casson Invariant", Commun. Math. Phys. {\bf 152}
(1993) 41, hep-th/9112012.}
\lref\CMRLH{S.~ Cordes, G.~ Moore, and S.~ Ramgoolam;
``Lectures on 2d Yang-Mills Theory, equivariant Cohomology and Topological
Field Theories", http://xxx.lanl.gov/lh94.
Published in two parts.  Part I: Proceedings String Theory, Gauge Theory
and Quantum Gravity, Trieste 1994: 184-244.
Part II: Proceedings of the 1994 Les Houches summer school, session 62,
Fluctuating Geometries in Statistical Mechanics and Field Theory. }
\lref\DiVV{R.~ Dijkgraaf, H.~ Verlinde and E.~ Verlinde, ``Notes on Topological
String Theory and 2-d  Quantum Gravity", lectures given at {\it Spring School
on Strings and Quantum Gravity}, Trieste, Italy, Apr 24 - May 2, 1990.}
\lref\kalkman{J.~ Kalkman, ``BRST Model for Equivariant Cohomology
and Representatives for the Equivariant Thom Class", Commun. Math.
Phys. {\bf 153} (1993) 447;
``BRST Model Applied to Symplectic Geometry", PRINT-93-0637 (Utrecht),
hep-th/9308132.}
\lref\Man{Yu. I. Manin, {\it Quantized Fields and Complex
Geometry}, Springer Verlag.}
\lref\MatQui{V.~ Mathai and D.~ Quillen,
``Superconnections, Thom Classes, and Equivariant Differential Forms",
Topology {\bf 25} (1986) 85.}
\lref\Van{P. van Baal, ``An Introduction to Topological
Yang-Mills Theory,'' Acta Physica Polonica, {\bf B21}(1990) 73}
\lref\dsb{E.~ Witten, ``Dynamical Breaking of Supersymmetry",
Nucl. Phys. {\bf B188} (1981) 513.}
\lref\trminus{E.~ Witten, ``Constraints on Supersymmetry  Breaking",
Nucl. Phys. {\bf B202} (1982) 253.}
\lref\ssymrs{E.~ Witten, ``Supersymmetry and Morse Theory",
J. Diff. Geom. {\bf 17} (1982) 661.}
\lref\donaldson{E.~ Witten, ``Topological Quantum Field Theory",
Commun. Math. Phys. {\bf 117} (1988) 353.}
\lref\WitHS{E. Witten, ``The Search for Higher Symmetry in String Theory",
Phil. Trans. Royal Soc. London {\bf B320} (1989) 349-357.}
\lref\Witg{ E.~ Witten, ``Topological Gravity,'' Phys. Lett. {\bf 206B} (1988)
601.}
\lref\jonespoly{E.~ Witten, ``Quantum Field Theory and  the Jones Polynomial",
Commun.  Math. Phys. {\bf 121} (1989) 351.}
\lref\Witdgit{E.~ Witten, ``Two-dimensional Gravity and Intersection Theory on
Moduli Space", in {\it Cambridge 1990, Proceedings, Surveys in Differential
Geometry}, 243-310.}
\lref\Wiag{E.~ Witten, ``Algebraic Geometry Associated with Matrix Models
of Two-Dimensional Gravity", IASSNS-HEP-91/74.}
\lref\Winm{E.~ Witten, ``The N-Matrix Model and gauged WZW models",
Nucl. Phys. {\bf B371} (1992) 191.}
\lref\Witdgt{ E.~ Witten, ``On Quantum gauge theories in two dimensions,''
Commun. Math. Phys. {\bf  141}  (1991) 153.}
\lref\Witdgtr{E.~ Witten, ``Two Dimensional Gauge Theories Revisited",
J. Geom. Phys. {\bf G9} (1992) 303; hep-th/9204083.}
\lref\Witp{E.~ Witten, ``On the Structure of the Topological Phase
of Two Dimensional Gravity", Nucl. Phys. {\bf B340} (1990) 281}
\lref\Witr{E.~ Witten, ``Introduction to Cohomological Field Theories",
Lectures at Workshop on Topological Methods in Physics, Trieste, Italy,
Jun 11-25, 1990, Int. J. Mod. Phys. {\bf A6} (1991) 2775.}
\lref\wttnmirror{E.~ Witten, ``Mirror Manifolds and Topological Field Theory",
hep-th/9112056, in {\it Essays on Mirror Manifolds} International
Press 1992.}
\lref\wittcs{E.~ Witten, ``3D Chern-Simons as Topological  Open String"; \-
hep-th/9207094.}
\lref\wittphases{E.~ Witten, ``Phases of N=2 Theories in Two Dimensions",
Nucl. Phys. {\bf B403} (1993) 159; hep-th/9301042.}
\lref\wittsusygt{E.~ Witten, ``Supersymmetric Gauge  Theory on a
Four-Manifold", hep-th/9403193.}
\lref\monfour{E. Witten, ``Monopoles and four-manifolds,''
hep-th/9411102}


\def\unlockat{\catcode`\@=11}
\def\lockat{\catcode`\@=12}

\unlockat

\global\newcount\secno \global\secno=0
\def\newsec#1{\global\advance\secno by1\message{(\the\secno. #1)}
\global\subsecno=0\global\subsubsecno=0\eqnres@t\noindent
{\bf {\the\secno.} #1}
\writetoca{{\secsym} {#1}}\par\nobreak\medskip\nobreak}
\global\newcount\subsecno \global\subsecno=0
\def\subsec#1{\global\advance\subsecno by1\message{(\secsym\the\subsecno. #1)}
\ifnum\lastpenalty>9000\else\bigbreak\fi\global\subsubsecno=0
\noindent{\it\secsym\the\subsecno. #1}\writetoca{\string\quad
{\secsym\the\subsecno.} {#1}}\par\nobreak\medskip\nobreak}
\global\newcount\subsubsecno \global\subsubsecno=0
\def\subsubsec#1{\global\advance\subsubsecno
by1\message{(\secsym\the\subsecno\the\subsubsecno. #1)}
\ifnum\lastpenalty>9000\else\bigbreak\fi
\noindent\quad{\secsym\the\subsecno.\the\subsubsecno. #1}
\writetoca{\string\qquad{\secsym\the\subsecno\the\subsubsecno.} {#1}}
\par\nobreak\medskip\nobreak}

\lockat


\def\figin{\epsfcheck\figin}\def\figins{\epsfcheck\figins}
\def\epsfcheck{\ifx\epsfbox\UnDeFiNeD
\message{(NO epsf.tex, FIGURES WILL BE IGNORED)}
\gdef\figin##1{\vskip2in}\gdef\figins##1{\hskip.5in}
instead
\else\message{(FIGURES WILL BE INCLUDED)}%
\gdef\figin##1{##1}\gdef\figins##1{##1}\fi}
\def\DefWarn#1{}
\def\figinsert{\goodbreak\midinsert}
\def\ifig#1#2#3{\DefWarn#1\xdef#1{fig.~\the\figno}
\writedef{#1\leftbracket fig.\noexpand~\the\figno}%
\figinsert\figin{\centerline{#3}}\medskip\centerline{\vbox{\baselineskip12pt
\advance\hsize by -1truein\noindent\footnotefont{\bf
Fig.~\the\figno:} #2}}
\bigskip\endinsert\global\advance\figno by1}

\def\inbar{\,\vrule height1.5ex width.4pt depth0pt}

\def\bA{{\bf A}}

\def\bF{{\IF}}
\def\bG{{\IG}}

\def\bO{{\IO}}

\def\cA{{\cal A}}

\def\cC{{\cal C}}
\def\cch{ {\cal \chi} }
\def\cD{{\cal D}}
\def\CD {{\cal D}}

\def\cF{{\cal F}}
\def\CF {{\cal F}}

\def\cgp {c_\Gamma^+}
\def\cgm {c_\Gamma^-}

\def\CM {{\cal M}}

\def\coker{{\mathop{\rm coker}}}

\def\eqdef{{\buildrel{\rm def}\over =}}
\def\eqsim{{\buildrel \sim \over =}}
\def\G {\Gamma}

\def\half{{\textstyle{1\over 2}}}
\def\holo{holomorphic}

\def\IA{{\bf A}}
\def\IB{\relax{\rm I\kern-.18em B}}
\def\IC{\relax\hbox{$\inbar\kern-.3em{\rm C}$}}
\def\ID{\relax{\rm I\kern-.18em D}}
\def\IE{\relax{\rm I\kern-.18em E}}
\def\IF{\relax{\rm I\kern-.18em F}}
\def\IG{{\relax\hbox{$\inbar\kern-.3em{\rm G}$}}}
\def\IGa{\relax\hbox{${\rm I}\kern-.18em\Gamma$}}
\def\IH{\relax{\rm I\kern-.18em H}}
\def\II{\relax{\rm I\kern-.18em I}}
\def\IK{\relax{\rm I\kern-.18em K}}
\def\IL{\relax{\rm I\kern-.18em L}}
\def\IM{\relax{\rm I\kern-.18em M}}
\def\IN{\relax{\rm I\kern-.18em N}}
\def\inbar{\,\vrule height1.5ex width.4pt depth0pt}
\def\index{{\mathop{\rm index}}}
\def\IO{\relax\hbox{$\inbar\kern-.3em{\rm O}$}}
\def\Iom{{\inbar\kern-3.00pt\Omega}}
\def\IOm{\relax\hbox{$\inbar\kern-3.00pt\Omega$}}
\def\IP{\relax{\rm I\kern-.18em P}}
\def\IPi{{\relax\hbox{${\rm I}\kern-.18em\Pi$}}}

\def\IQ{\relax\hbox{$\inbar\kern-.3em{\rm Q}$}}
\def\IR{\relax{\rm I\kern-.18em R}}
\def\ITh{\relax\hbox{$\inbar\kern-.3em\Theta$}}

\font\cmss=cmss10 \font\cmsss=cmss10 at 7pt
\def\IZ{\relax\ifmmode\mathchoice
{\hbox{\cmss Z\kern-.4em Z}}{\hbox{\cmss Z\kern-.4em Z}}
{\lower.9pt\hbox{\cmsss Z\kern-.4em Z}}
{\lower1.2pt\hbox{\cmsss Z\kern-.4em Z}}\else{\cmss Z\kern-.4em
Z}\fi}
\def\kG {\vec{k}_\Gamma}
\def\lieg{{\underline{\bf g}}}
\def\log {{\rm log}}
\def\mapdown#1{\Big\downarrow
        \rlap{$\vcenter{\hbox{$\scriptstyle#1$}}$}}

\def\mapleftu#1{\smash{
        \mathop{\longleftarrow}\limits^{#1}}}
\def\mapright#1{\smash{
        \mathop{\longrightarrow}\limits^{#1}}}
\def\maprightd#1{\smash{
        \mathop{\longrightarrow}\limits_{#1}}}
\def\maprightu#1{\smash{
        \mathop{\longrightarrow}\limits^{#1}}}
\def\mapse#1{\searrow
        \rlap{$\vcenter{\hbox{$\scriptstyle#1$}}$}}
\def\mapsw#1{\swarrow
        \rlap{$\vcenter{\hbox{$\scriptstyle#1$}}$}}
\def\mapup#1{\Big\uparrow
        \rlap{$\vcenter{\hbox{$\scriptstyle#1$}}$}}

\def\p {\partial}
\def\pb{\bar{\partial}}

\def\pseudo{co-}

\def\quarter{{\textstyle{1\over 4}}}
\def\rank{{\rm rank}}
\def\Sc {\Sigma_T^c}
\def\sdtimes{\mathbin{\hbox{\hskip2pt\vrule height 4.1pt depth -.3pt
width
.25pt
\hskip-2pt$\times$}}}
\def\sh{{h^{1/2}}}
\def\SG{{\Sigma_T}}
\def\Sh{{\Sigma_W}}
\def\Sw{{\Sh}}
\def\ST{{\SG}}
\def\Sym{{\mathop{\rm Sym}}}

\def\tbA{{\widetilde{\bA}}}

\def\tbX{{\widetilde{\bf X}}}
\def\tcF{{\widetilde{\cal F}}}

\def\tcM{{\widetilde{\cal M}}}

\def\top{topological}

\def\vskipabit{{\vskip0.25truein}}

\def\ymt{$YM_2$}

\Title{
\vbox{\baselineskip12pt\hbox{hep-th/9402107}\hbox{YCTP-P23-93}
\hbox{RU-94-20}}}
{\vbox{
\centerline{Large $N$ 2D Yang-Mills Theory}
\centerline{and}
\centerline{Topological String Theory} }}
\bigskip
\centerline{Stefan Cordes, Gregory Moore\foot{Currently visiting the
Rutgers
University Dept. of Physics}, and Sanjaye Ramgoolam}
\bigskip
\centerline{stefan@waldzell.physics.yale.edu}
\centerline{moore@castalia.physics.yale.edu}
\centerline{skr@genesis2.physics.yale.edu}
\smallskip\centerline{Dept.\ of Physics}
\centerline{Yale University}
\centerline{New Haven, CT \ 06511}
\bigskip
\bigskip
\centerline{\bf Abstract}
\noindent
We describe a topological string theory which reproduces
many aspects of the $1/N$ expansion of $SU(N)$
Yang-Mills theory in two spacetime dimensions
in the zero coupling ($A=0$) limit. The string
theory is a modified version of topological gravity coupled
to a topological sigma model with spacetime as target. The
derivation of the string theory relies on a new interpretation
of Gross and Taylor's ``$\Omega^{-1}$ points.''
We describe how inclusion of the area, coupling of
chiral sectors, and Wilson loop expectation values can
be incorporated in the topological string approach.

\Date{January 8, 1996}

\newsec{Introduction}

The possibility that the strong interactions might
be described by a theory of strings has been an
enduring source of fascination and frustration
to particle theorists for the past twenty-five years
\nref\veneziano{G.~ Veneziano, ``Construction of a
crossing-symmetric, Regge-behaved amplitude for
linearly rising trajectories,'' Nuovo Cim.
{\bf 57A} (1968) 190.}\nref\wilson{K.G.~ Wilson,
``Confinement of quarks", Phys.
Rev. {\bf D10} (1974) 2445.}\nref\thooft{G.~ 't Hooft,
``A planar diagram theory for string interactions", Nucl.
Phys. {\bf B72} (1974) 461.}\nref\migdal{A.A.~ Migdal,
`` Loop equations and 1/N expansion,'' Physics Reports
(Review section of Physics Letters) {\bf 102} (1983)
199-290.}\nref\polchinski{J.~ Polchinski, ``Strings and QCD?''
Talk presented at the Symposium on Black Holes,
Wormholes, Membranes and Superstrings,'' Houston,
1992; hep-th/9210045}\nref\gross{D.~ Gross, ``Some
new/old approaches to QCD,'' Published in String Theory
Workshop, Rome 1992: 251-268; hep-th/9212148.}\nref\gtrev{D.J.~
Gross and W. Taylor, ``Two-Dimensional QCD and Strings,''
Published in Strings '93, Berkeley, 1993: 214-225;
hep-th/9311072.}\nref\dougrev{M.R.~ Douglas, ``Conformal Field
Theory Techniques in Large N Yang-Mills Theory,''
hep-th/9303159. }\nref\bars{I.~ Bars, ``QCD and Strings in
2d", hep-th/9312018.}\refs{\thooft {--} \bars}.
In the early 80's some interesting progress on
this question was made in the case of large N
Yang-Mills theory in two dimensions ($YM_2$)
\ref\Mig{A.~ Migdal, ``Recursion equations in gauge theories",
Zh. Eskp. Teor. Fiz. {\bf 69} (1975) 810 (Sov. Phys. JETP.
{\bf 42} 413).},\ref\kazakov{V.A.~ Kazakov
and I.~ Kostov, ``Non-linear strings in two dimensional
$U(\infty)$ gauge theory,''  Nucl. Phys. {\bf B176} (1980) 199-215;
V.A. Kazakov, ``Wilson Loop average for an arbitrary
contour in two-dimensional $U(N)$ gauge theory,''
Nuclear Physics {\bf B179} (1981) 283-292.}.
Recently this work has been revived, considerably
extended, and deepened. Exact results are now available
for partition functions $Z(\CG,\Sigma_T)$ and Wilson loop
averages for a compact gauge group $\CG$  on two-dimensional
spacetimes $\Sigma_T$ of arbitrary topology\nref\Rus{B.~
Rusakov, ``Loop Averages and Partition Functions in $U ( N )$
gauge theory on two-dimensional manifolds", Mod. Phys. Lett.
{\bf A5}, 693 (1990).}\nref\witten{E.~ Witten, ``On gauge theories
in two dimensions,'' Commun. Math. Phys. {\bf 141}, 153
(1991).}\nref\blauthom{M.~ Blau and G.~ Thompson, ``Lectures
on 2d Gauge Theories: Topological Aspects and Path Integral
Techniques", published in Trieste HEP and Cosmology 1993: 175-244;
hep-th/9310144.}\refs{\Rus{--}\blauthom}
\foot{\ymt\ has area-preserving diffeomorphism
symmetry so $Z$ only depends on the gauge group,
topology and total area of $\ST$. For gauge group
$\CG=SU(N)$ and $\ST$ of genus $G$
we denote the partition function by  $Z(A,G,N)$.}.

Building on the  results \refs{ \Mig, \Rus{--}\blauthom}\
D. Gross and W. Taylor returned to the problem of strings and \ymt\
in a beautiful series of papers
\ref\GrTa{D.~ Gross, ``Two Dimensional QCD as a String Theory,"
PUPT-1356, LBL-33415; hep-th/9212149; D.~ Gross and
W.~ Taylor, ``Two-dimensional QCD is a String Theory," Nucl.
Phys. {\bf B400} (1993) 161-180; hep-th/9301068;
D.~ Gross and W.~ Taylor, ``Twists and Loops in the String
Theory of Two Dimensional QCD," Nucl. Phys. {\bf B403}
(1993) 395-452; hep-th/9303046.}
(See also \ref\Min{J.~ Minahan, ``Summing over inequivalent maps
in the string theory of QCD,'' Phys. Rev. {\bf D47} (1993) 3430.} ).
In particular, \GrTa\ derives the $N\to \infty$ asymptotic
expansion for the partition function $Z(A,G,N)$.
Moreover many aspects of the expansion in $1/N$ have a natural
geometrical explanation in terms of weighted sums over maps
from a worldsheet $\Sw$ to the spacetime $\ST$.
In some cases (e.g. when $\ST$ is a torus) Gross and Taylor
were able to write the weighted sum explicitly as a sum over
covering maps with weights given by symmetry
factors for the cover. Given the results of \GrTa\
no one could seriously doubt that \ymt\ is equivalent to a string
theory. Nevertheless, \GrTa\ left untied some loose ends, such
as the following two problems:

\noindent
(1.)  The problem of the true meaning of the
``$\Omega^{-1}$ points''.

\noindent
(2.)  The problem of finding the appropriate
string action.

With regard to problem (1),
the geometrical interpretation
of the $1/N$ expansion necessitated the introduction of
$\vert 2-2G\vert$ ``twist points,'' (``$\Omega$ points,''
or ``$\Omega^{-1}$ points''). In contrast to the clear
and natural
geometrical interpretation of all the other aspects of the
series, the nature of the $\Omega^{-1}$ points was
fraught with mystery.
 Related difficulties had already
presented themselves ten years earlier in the work of
Kazakov and Kostov \kazakov. Several authors have emphasized
the importance of a proper understanding of the $\Omega^{\pm 1}$
points.

As for problem (2),
one of the key motivations for Gross and Taylor's
work
was that the action for a string interpretation of
$YM_2$ might  have a natural generalization to
four-dimensional targets, or might
suggest essential features of a string interpretation
of $YM_4$. However, difficulties associated with
problem (1) presented a serious obstacle to finding
the action  for \ymt.
Indeed, after the appearance of the first papers of
\GrTa\ it was quickly noted in
\ref\djrdd{R.~ Dijkgraaf and R.~ Rudd, unpublished.}
\ref\bcov{M.~ Bershadsky, S.~ Cecotti, H.~ Ooguri, C.~ Vafa,
``Holomorphic anomalies in topological field theories," Nucl.
Phys. {\bf B405} (1993) 279-304; hep-th/9302103;
``Kodaira-Spencer theory of gravity and exact quantum string
amplitudes," Commun. Math. Phys. {\bf 165} (1994) 311-428;
hep-th/9309140. } that, for the case of the partition function of
a toroidal target, (where there are no $\Omega^{-1}$ points)
the interpretation in terms of covering maps naturally
suggests that the string action principle for \ymt\ will
involve a topological sigma model, with $\ST$ as
target, coupled to topological gravity. However, even in
the genus one case, the evaluation of Wilson loops
necessitates consideration of $\Omega^{-1}$-points.
No theory of \ymt\ can go very far without an understanding
of these factors.

In the present paper we will solve problem (1). The
solution of this problem allows us to make some
definite progress on problem (2).  The solution to
problem (1) is simple:
{\it there is no such thing as an ``$\Omega^{-1}$ point!''}
We have not completely solved problem (2) in the
sense that we have not reproduced {\it all} known
results on \ymt\ from the string approach.
Nevertheless, we have reproduced enough to
say that  (a.) a description in terms of topological
string theory is possible but (b.) the action is more
elaborate than the standard coupling of topological
gravity to the topological sigma model for $\ST$, and
(c) a careful analysis of contact terms is needed to reproduce the
$YM_2$
results.

In more detail, the paper is organized as follows.
We review some aspects of \GrTa\  and establish
notation in sec. 2. We will discuss both
 the ``chiral'' partition function $Z^+(A,G,N)$
(eq. (2.4))
as well as the ``nonchiral'' partition function
(eq. (2.2)),  both of which
 we view as asymptotic expansions in $1/N$.

In sections 3,4 we  review
some necessary background material from
mathematics, in particular, we describe the Hurwitz moduli
space $H(h,G)$ of holomorphic maps $\Sw\to \ST$
from a connected Riemann surface $\Sw$ of genus $h$
to a Riemann surface $\ST$  of
genus $G$ with fixed complex structure.
In sections 4.3, 4.4 we explain how $H(h,G)$ can be thought of
as the base of a principal fibre bundle for
$Diff^+(\Sw)\sdtimes Weyl(\Sw)$.

In section 5 we begin with the simplest quantity in
\ymt: the chiral  partition function at zero area:
$Z^+(A=0,G,N)$.  This expansion
can be interpreted as a sum over branched covers
\GrTa. Taking  proper account of
the $\Omega^{-1}$ factors leads to our first main
result, stated as Proposition 5.2 (section 5.2):
 $Z^+(A=0,G,N)$ is the generating function for
the orbifold Euler characteristic  $Z$ of the compactified
Hurwitz moduli space, $ \overline{H(h,G)}$of branched covers
 of $\ST$:
\eqn\mnresi{\eqalign {
Z^+(0,N,G) &=exp\Biggl[ \sum_{h=0}^{\infty}
\biggl({1\over N}\biggr)^{2h-2}
\chi \bigl( \overline{H(h,G)} \bigr)\Biggr]\cr
}}

A branched cover of surfaces $\Sw\to \ST$ can always
be interpreted as a holomorphic map for appropriate
complex structures on $\Sw,\ST$. Thus, the appropriate
category of maps with which to formulate the chiral $1/N$
expansion
of \ymt\ is the category of holomorphic maps. This is precisely the
situation best suited to an introduction of topological
sigma models. Accordingly, in section 6 we introduce a
topological string theory which  counts
holomorphic (and antiholomorphic) maps
$\Sw\to\ST$. Our central claim is that this string theory
is the underlying string theory of \ymt. The action is schematically
of the form:
\eqn\actn{
I_{\rm chiral\ YM_2\  string}=
I_{\rm tg} + I_{\rm t \sigma} +I_{c\sigma}}
The first two terms give the (standard) action of
2D topological gravity coupled to a topological
$\sigma$-model with $\ST$ as target.
The action $I_{c\sigma}$ turns out to be complicated
but can be deduced using a procedure which is
in principle straightforward.
This procedure is based on the point of view
that topological field theory
 path integrals are related to infinite-dimensionsl
generalisations of the
 Mathai-Quillen representative of equivariant Thom
 classes\nref\kanno{H. Kanno, ``Weil algebra
structure and geometrical meaning of BRST
transformation in topological quantum field theory,''
Z. Phys. {\bf C43} (1989) 477.}\nref\AJ{M.F.Atiyah
and L.Jeffrey, ``Topological Lagrangians and
Cohomology,'' Jour. Geom.  Phys. {\bf 7} (1990)
119.}\nref\kalkman{J.~ Kalkman, ``BRST
Model for Equivariant Cohomology and
Representatives for the Equivariant Thom Class",Commun. Math.
Phys. {\bf 153} (1993) 447.
 }
\refs{\kanno{--}\kalkman}. In section 7,
we use this point of view to construct  $I_{c\sigma}$ explicitly.

In sections 8 and 9 we show how many of the results
of  chiral \ymt\ can be derived from the topological string
theory \actn.

In section 8 we describe how the area can be restored
by a perturbation of the topological action \actn\ by the area
operator:
\eqn\area{
\CA = \int f^* \omega
}
where $\omega$ is the K\"ahler class of the
target space Riemann surface. This is, of course,
just the standard Nambu action, as one might well
expect. The novelty in the present context is that
the Nambu action is regarded as a {\it perturbation}
of a previously constructed theory. Calculating this
perturbation of the topological string theory is
rather subtle because of contact terms. These are
responsible for the polynomial dependence on the
area in \ymt.  In section 8.1 we isolate what we believe are
``the most important'' area polynomials and, after
some preliminary analysis of the contact terms between the
area operator \area\ and curvature insertions
arising from $I_{c\sigma}$ in \actn, as well as
those between area operators themselves,  we show in section 8.7
how these polynomials follow from the string picture.

In section 9 we discuss Wilson loop expectation
values in the case of nonintersecting Wilson loops.
Following the lead of \GrTa\ we show that these may
be incorporated in the string approach by computing
macroscopic loop amplitudes. The data of the representation
index on the Wilson line $\G$ is translated into covering
data of the boundary of the worldsheet over the lines
$\G$.

In section 10 we repeat the discussion of section 5 for
the partition function of the ``full nonchiral \ymt.''
We follow closely the geometrical picture
introduced in \GrTa.
In order to state the analog of \mnresi\ it is
necessary to introduce both holomorphic and
antiholomorphic maps, as well as
``degenerating coupled covers'' (see Definitions 10.3,10.4).
 We introduce a Hurwitz space for such maps, called {\it coupled
Hurwitz space}
 and, in Proposition 10.3,
 we state the result for the nonchiral
theory analogous to \mnresi.

In section 11 we explain how the nonchiral partition
function can be incorporated in topological
string theory. The path integral localizes on
both holomorphic and antiholomorphic maps. It
also localizes on singular maps
(``degenerated coupled covers'' ) and the
contributions from these singular geometries
must be defined carefully.
The answer for the partition
function of the string theory \actn\
depends on how we choose our contact terms for
these singular geometries. We discuss two
choices which lead to  two distinct answers:
\eqn\mnresii{
exp\Biggl\{\sum_{h\geq 0}\biggl({1\over N}\biggr)^{2h-2} Z_{\rm
string}(\Sw\to
\ST)\Biggr\}=
\cases{Z^+(A=0,N)  Z^-(A=0,N)\cr
Z(A=0,N)\cr}
}
In the first case we choose contact terms so that
singular geometries make no contribution
(we ``set all contact terms to zero''). This reproduces
answers of the chiral theory.
A more non-trivial choice of contact terms reproduces the full
zero-area theory.

 Some technical arguments are
contained in appendices.

Finally the reader should note that while we
were preparing this paper a closely related paper
appeared on hep-th
\ref\horava{P. Horava, ``Topological Strings and
QCD in Two Dimensions,''
EFI-93-66, hep-th/9311156. To appear in Proc. of
The Cargese Workshop, 1993. }. In this paper
P. Horava
proposes a formulation of \ymt\ in terms of
topological string theory. The theory in \horava\
 is based on counting of
{\it  harmonic maps}, rather than
holomorphic maps,
(or degenerated coupled covers)
 and, at least superficially,  appears to be different
from the proposal of this paper.

\newsec{The Gross-Taylor Asymptotic Series}
\subsec{ Partition Functions}
The partition function of two dimensional Yang-Mills theory
on an orientable closed manifold $\ST$
of genus $G$ is \refs{\Mig,\Rus}
\eqnn\pfun $$\eqalignno{
   Z(SU(N),\ST) &= \int [DA^{\mu}]exp[{{-1}\over {4 e^2}}
\int_{\ST} d^2x \sqrt {det G_{ij}} ~Tr F_{ij}F^{ij}]  \cr
 &= \sum_{R}(\dim R)^{2-2G}
e^{-{{\lambda A}\over {2N}} C_2(R)},&\pfun\cr }$$
where the gauge coupling
$\lambda =  e^2N$ is held fixed in the large $N$ limit,
  the sum runs over
all unitary irreducible representations $R$ of the gauge group
${\cal G}=SU(N)$,  $C_2(R)$ is the second casimir, and $A$ is the
area of the
spacetime in the metric $G_{ij}$. We will henceforth
absorb $\lambda$ into $A$.

Using the Frobenius relations between representations
of symmetric groups and representations of $SU(N)$
Gross and Taylor derived  an expression for the
$1/N$ asymptotics of \pfun\ in terms of a sum over
elements of symmetric groups. The result of   \GrTa\ is:
\eqn\fullgt{
\eqalign{
Z&( A,G, N)\cr
&\sim
\sum_{n^\pm,i^\pm=0}^{\infty}
\sum_{p^\pm_1,\ldots,p^\pm_{i^\pm} \in T_2\subset S_{n^\pm}}
\sum_{s^\pm_1,t^\pm_1,\ldots,s^\pm_G,t^\pm_G\in S_{n^\pm}}
\bigl({1\over N}\bigr)^{( n^+ + n^- ) (2G-2)+(i^+ + i^- )}\cr
&{{(-1)^{(i^++ i^-)}} \over  {i^+! i^-! n^+ ! n^- !}}
( A)^{(i^+ + i^-)}
e^{-\half  (n^+ + n^-) A}
e^{\half ((n^+)^2 + ( n^- )^2 - 2 n^+ n^- ) A/N^2}
\cr
&
\delta_{{\scriptscriptstyle S_{n^+} \times S_{n^-}}}
\biggl ( p^+_1\cdots p^+_{i^+} p^-_1\cdots p^-_{i^- }
\Omega_{n^+, n^-}^{2-2G}
\prod_{j=1}^G [ s^+_j, t^+_j ] \prod_{k=1}^G [ s^-_k, t^-_k ]
\biggr )
,}}
where $[s,t]= sts^{-1}t^{-1}$.
Here $\delta$ is the delta function on the group algebra
of the product of symmetric groups $S_{n^+}\times S_{n^-}$,
$T_2$ is the class of elements of $S_{n^\pm}$ consisting of
 transpositions, and $\Omega^{-1}_{n^+, n^-}$ are
certain elements of the group algebra of the symmetric
group $S_{n^+} \times S_{n^-}$ with coefficients in
$\IR ((1/N))$.
These will be discussed in detail below.

One of the striking features of \fullgt\ is that it
nearly factorizes, splitting into a sum over $n^+,i^+,\cdots $ and
$n^-,i^-,\cdots $.
Gross and Taylor interpreted the contributions of
the $(+)$ and $(-)$ sums  as arising from two ``sectors''
of a hypothetical worldsheet theory. These sectors
correspond to
orientation reversing and preserving maps, respectively.
One views the $n^+=0$ and $n^-=0$ terms as
leading order terms in a $1/N$ expansion.
At higher orders the two sectors are coupled via the
$n^+ n^-$ term in the exponential and via terms in
$\Omega_{n^+ n^-}$.
The latter are described by a simple set of
rules in \GrTa, and will be addressed in detail in
section 10 below.

The expression \fullgt\ simplifies considerably if
we concentrate on one chiral (or antichiral) sector.
In general we define {\it chiral} expectation amplitudes
in \ymt\ by  translating $SU(N)$ representation theory into
 representation theory of symmetric groups and
making the replacement:
\eqn\dfchsm{
\sum_{R\in Rep(SU(N))} \to \sum_{n\geq 0} \sum_{R\in Y_n}
}
where $Y_n$ stands for the set of Young Tableaux with $n$ boxes.
For example, in the  case of the partition function we may
define $\Omega_n  \equiv \Omega_{n,0}$ and write
the  ``chiral Gross-Taylor series'' (CGTS) as
\GrTa:
\eqn\Zchir {\eqalign{
Z^+(A,&G, N)\cr
{}~& =~
\sum_{n,i,t,h=0}^{\infty} e^{-n A/2} (-1)^i {{ (
A)^{i+t+h}} \over{i!t!h!}}
 \bigl({1\over N}\bigr)^{n(2G-2)+2h+ i+ 2t} {{n^h(n^2-n)^t }\over
{2^{t+h}
}}\cr
&
\sum_{p_1,\ldots,p_i \in T_2}
\sum_{s_1,t_1,\ldots,s_G,t_G\in S_n}
\biggl[ {1\over {n!}}\delta (p_1\cdots p_i \Omega_n^{2-2G}
\prod_{j=1}^G
s_jt_js_j^{-1}t_j^{-1}) \biggr]
.\cr
 }}

\subsec{ $\Omega$ factors}
Let us now define $\Omega$. We postpone describing $\Omega_{n^+,n^-}$
to section 10.4 and concentrate on the "chiral $\Omega$ factors"
$\Omega_n$.
This is the element of the group algebra of $S_n$
defined by the equation
$$ \hbox{dim }R_Y = {N^n\over n!} \chi_{\scriptscriptstyle Y}
(\Omega_n)$$
Here $R_Y$ is an $SU(N)$ representation associated to
a Young tableaux $Y$ with $n$ boxes, and $\chi_{\scriptscriptstyle
Y}$ is the
character in  the corresponding representation of $S_n$.
Explicitly, $\Omega_n$ is
given by
\eqn \Omeg {
 \Omega_n = \sum_{v\in S_n}
\bigl({1\over N}\bigr)^{n-K_v} v \qquad .}
 Here,  $K_{v}$ is the number of cycles
in the permutation $v$.

When $G>1$ one must introduce $\Omega^{-1}$ which is the inverse of
$\Omega$ in the group algebra
\foot{Note that for some values of $N$, $\Omega_n$ will fail to have
an
inverse. This does not happen when $N>n$. Hence $\Omega$ may always
be
considered
as invertible in the $1/N$ expansion. }
For example,
we have
\eqn\example{
\eqalign{
\Omega_2&=1+{1\over N} v\cr
\Omega_2^{-1} &= \sum_{i=0}^{\infty} (-1/N)^i v^i\cr
       &= \sum_{i=0}^{\infty} (1/N)^{2i} - v \sum_{i=0}^{\infty}
        (1/N)^{(2i+1)}\cr
          &={1\over (1-1/N^2)} \cdot 1-{1/N\over (1-1/N^2)}
 \cdot v \cr}
}
for $n=2$. In the first line we have written it as an element of the
free
algebra generated by elements of $S_2$. In the second we have reduced
it to
to an element of the group algebra of $S_2$ whose
coefficients are expansions in $1/N$.

\newsec{Maps and Coverings}

We would like to interpret the terms
in the $1/N$ expansion as weighted sums of maps
$\Sigma_W\to \Sigma_T$ between compact orientable surfaces  without
boundary,
of genus $h,G$,
respectively.  In the next two sections we
summarize some relevant
mathematical results pertaining to such maps.

\subsec{Homotopy groups}

We will  use heavily the properties of homotopy groups of
 punctured Riemann surfaces.
 As abstract groups these
are ``F-groups.''   The group $F_{G,L}$ may be described in
terms of generators and relations by
\eqn\fgp{
F_{G,L}\equiv  \biggl \langle \{
\alpha_i,\beta_i\}_{i=1,G},\{\gamma_s\}_{s=1,L}\vert
\prod_{i=1}^G [\alpha_i,\beta_i]\prod_{s=1}^L \gamma_s=1\biggr
\rangle
}
(The product is ordered, say, lexicographically.)

Consider a compact orientable surface $\ST$ of genus $G$. If we
remove
$L$ distinct points, and choose a basepoint $y_0$,  then there is an
isomorphism
\eqn\fgpiso{
F_{G,L}\cong \pi_1(\Sigma_T-\{P_1,\dots P_L\},y_0)
.}
This isomorphism is not canonical. The choices are
parametrized by the infinite group $Aut(F_{G,L})$.

On several occasions we will make use of a set of generators
$\alpha_i, \beta_i,\gamma_i$ of $\pi_1$ so that, if we cut along
curves in the
homotopy class the surface looks like

\ifig\ezfg{ A choice of generators for the homotopy group
of a punctured surface. The curves $\gamma(P)$ become
trivial if we fill in the puncture $P$.}
{\epsfxsize3.0in\epsfbox{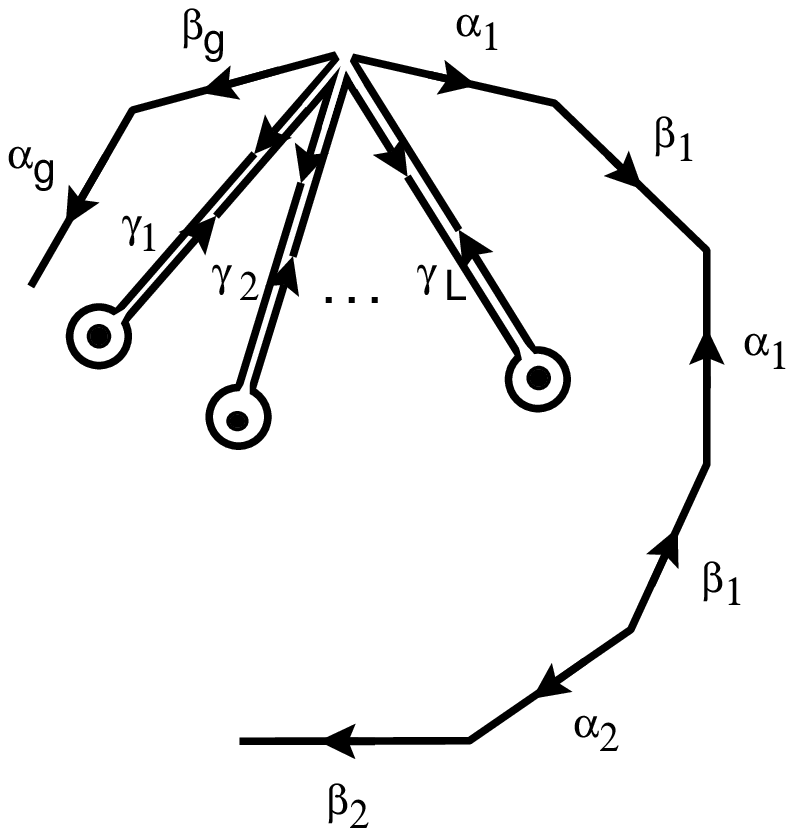}}

The conjugacy class of the curves $\gamma(P)$ can be characterised
intrinsically as follows.
The process of filling in a point $P_1$ defines an inclusion
\eqn\inclu{
i: \ST - \{P_1,P_2, \cdots P_L\} \rightarrow \ST - \{ P_2,\cdots
P_L\}
                    }
with an induced map $i_*$ on $\pi_1$. $[\gamma(P_1)]$ is
 the kernel of $i_*$.

\subsec{Branched coverings}

Maps of particular importance to us are branched coverings.

\noindent
{\bf Definition 3.1}.

a.) A continuous map $ f:\Sigma_W \rightarrow
\Sigma_T $ is a  {\it branched cover} if  for any open set $U \subset
\Sigma_T$,
the inverse $f^{-1}(U)$
is a union of disjoint open sets on each of which $f$ is
topologically
equivalent to the complex map $z\mapsto z^n$ for some $n$.

b.) Two branched covers $f_1$ and $f_2$ are
said to be {\it equivalent} if there exists a homeomorphism $\phi :
\Sigma_W
\rightarrow \Sigma_W $ such that $f_1\circ \phi = f_2 $.

For $Q \in \Sw$, the integer $n$ will be called the
{\it ramification index} of $Q$ and will be denoted
$Ram(f,Q)$. For any $P \in \ST$ the sum
$$\sum_{Q\in f^{-1}(P) } Ram(f,Q)$$
is independent of $P$ and will be called the index of $f$ (sometimes
the
degree).
Points $Q$ for which the integer $n$ in condition $(a)$  is bigger
than $1$
 will
be called {\it ramification points}. Points $P\in \ST$ which
are images of ramification points will be called
{\it branch points}.  The set of branch points is the branch locus
$S$.
The branching number at $P$ is
$$B_P=\sum_{Q\in f^{-1}(P)} [Ram(f,Q)-1] $$
The branching number of the map $f$ is $B(f)= \sum_{P\in S(f) } B_P$.
A branch point $P$ for which the branching number is
$1$ will be called a {\it simple branch point}.
Above a simple branch point all the inverse images
have ramification index $=1$, with the exception of
one point $Q$ with index $=2$.
\foot{Unfortunately, several authors use these terms in
inequivalent ways.}

We will often use the Riemann-Hurwitz formula. If $f:\Sw\rightarrow
\ST$
is a branched cover of index $n$ and branching number $B$, $\Sw$ has
genus $h$,
$\ST$ has genus $G$, then :
\eqn\Rhurw{ 2h-2 = n(2G-2)+B .}

Equivalence classes of branched covers may be
related to  group homomorphisms through the
following construction. Choose a point $y_0$ which is not a branch
point
and label the inverse images $f^{-1}(y_0)$ by
the ordered set
$\{x_1, \dots x_n \}$.
Following the lift of elements of $\pi_1(\Sigma_T - S ,y_0)$
the map $f$
induces a homomorphism
$$f_\#:\pi_1(\Sigma_T - S ,y_0)\to S_n\qquad . $$

Suppose $\gamma(P)$ is a curve
surrounding a branch point P as in \ezfg.
There is a close relation between the cycle structure of $v_P=f_\#
(\gamma(P))$
and the topology of the covering space over a neighbourhood of $P$.
If the
cycle
decomposition of $v_P$ has $r$ distinct cycles then $f^{-1}(P)$ has
$r$
distinct points.
 Moreover, a cycle of length $k$ corresponds to a ramification point
$Q$ of
index $k$.

With an appropriate notion of equivalence
the  homomorphisms are in 1-1 correspondence
with equivalence classes of branched covers.

\noindent
{\bf Definition 3.2}.
Two homomorphisms
$\psi_1,\psi_2:\pi_1(\Sigma_T - S ,y_0)\to S_n$ are said to be
{\it equivalent} if  they differ by an inner automorphism of $S_n$,
i.e., if $\exists g$ such that  $\forall x$, $\psi_1(x)=g \psi_2(x)
g^{-1}$.

\noindent
{\bf Theorem 3.1}. \ref\Fulton{W.~ Fulton, ``Hurwitz schemes and
irreducibility of moduli of algebraic curves," Annals of Math.
{\bf 90} (1969) 542.}
\ref\Ezell{C.L.~ Ezell, ``Branch point structure of covering maps onto
nonorientable surfaces," Transactions of the American Mathematical
Society {\bf 243}, 1978. }.
Let $S\subset \Sigma_T$ be a finite set and $n$ a positive
integer.
There is a one to one correspondence between equivalence
classes of homomorphisms
$$\psi:\pi_1(\Sigma_T - S ,y_0)\to S_n$$
and equivalence classes of $n$-fold branched coverings of
$\Sigma_T$ with branching locus $S$.

{\it Proof. }
We outline the proof which is described in \Ezell.
The first step shows that equivalent homomorphisms determine
equivalent
branched coverings.  Given a branched cover,
we can delete  the branch points from $\ST$ and the inverse images of
the
branch points from $\Sw$ giving  surfaces $\overline {\Sigma}_W$ and
$\overline
{\Sigma}_T$ respectively.
The branched cover restricts to a topological (unbranched)  cover of
$\overline {\Sigma}_T$ by
 $\overline {\Sigma}_W$. To this map we can apply the theorem
\ref \Mas{W.S.~ Massey, {\it A basic course in Algebraic Topology},
Springer-Verlag, 1991.} which establishes a one-to-one correspondence
between conjugacy classes of subgroups of
$\pi_1(\overline {\Sigma}_T  )$ and equivalence classes of topological
coverings of $\overline {\Sigma}_T$.

Similarly the second step proves that equivalent covers
determine  equivalent homomorphisms. The restriction of $\phi$ to the
inverse
images of $y_0$  determines the permutation which conjugates one
homomorphism
into the
other.

Finally one proves that the map from equivalence classes of
homomorphisms to
equivalence classes of branched covers is onto. We cut  $n$ copies of
$\ST$
along
chosen generators of $\pi_1(\ST-S)$  (illustrated in Figure 1), and
we glue
them
together according to the data of the homomorphism.
$\spadesuit$

\bigskip
This theorem goes back to Riemann.
Since the \ymt\ partition function sums over covering surfaces which
are not necessarily connected we do not restrict to homomorphisms
whose images are transitive subgroups of $S_n$.

\noindent
{\bf Definition 3.3}.
An {\it automorphism} of a branched covering $f$ is a homeomorphism
$\phi$ such
that
$f\circ \phi = f $.

It follows from the above that the number of such
automorphisms of a given equivalence class of branched coverings is
equal
to the order of the centraliser of the subgroup generated by the
image of
$\pi_1(\Sigma_T -  S, y_0)$ in $S_n$ . For a homomorphism $\psi$ we
call
this
$\vert C(\psi)\vert$.
 Then $n!/\vert C(\psi)\vert$, the number of cosets of this
subgroup, is the number of distinct homomorphisms related to the
given
homomorphism by conjugation in $S_n$.

\subsec{Continuous maps}

The space of branched coverings is not the only space of
maps of surfaces one might wish to consider in rewriting
\ymt\ as a string theory. Another natural choice of
category is the category of continous maps. We mention
here two theorems concerning the classification of
these maps, and their relation to the category of
branched coverings.

Since Riemann surfaces have a contractible universal
cover, continuous maps between Riemann
surfaces are topologically classified
by their action on homotopy groups:

\noindent
{\bf Theorem 3.2}. \ref\Spanier{E.H.~ Spanier, {\it Algebraic Topology},
Mc. Graw-Hill, 1966.}
Homotopy classes of continuous
maps $f:(\Sigma_W,x_0) \to (\Sigma_T,y_0)$ of fixed degree
are in 1-1 correspondence with homomorphisms
$f_*:\pi_1(\Sigma_W,x_0)\to \pi_1(\Sigma_T,y_0)$.

 A map $\Sigma_W \rightarrow
\Sigma_T$ is said to be a pinch map if there is a compact connected
submanifold
$H \subset \Sigma_W$, with boundary consisting of a simple closed
curve in the
interior of $\Sigma_W$, such that the $\Sigma_T$ is
$\Sigma_W/H$, the quotient of $\Sigma_W$ with $H$ identified to a
point,
and such that $f$ is the quotient map. Pinch maps can
collapse entire regions of surface to a single point as in

\ifig\fhhpiv{Example of a pinch map.}
{\epsfxsize2.0in\epsfbox{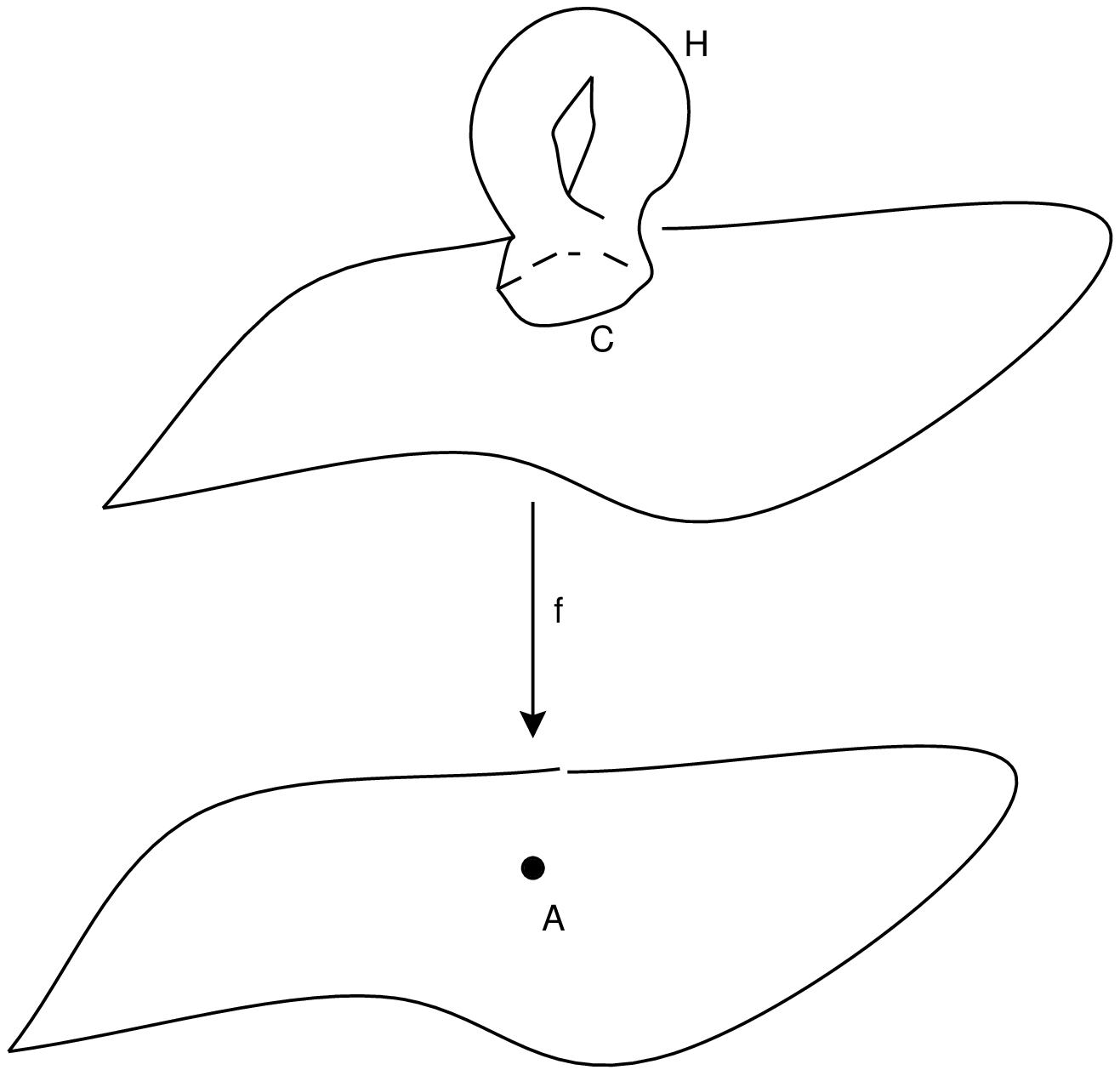}}

\noindent
{\bf Theorem 3.3}. \ref\Edmonds{A.L.~ Edmonds, ``Deformation of maps
to branched coverings in dimension two", Annals of Mathematics {\bf 110}
(1979) 113-125. }.
In each homotopy class of maps $f:\Sigma_W\to \Sigma_T$
there is a representative $f=p\circ \pi$
where $p$ is a  pinch map and $\pi$ is a
branched covering.

Notice that  pinch maps can only decrease the
Euler character of $\Sw$. It therefore follows from theorem 3.3 and
\Rhurw\
that the existence of a nonconstant  $f:\Sw\to \ST$
implies that $2h-2\geq n(2G-2)$. This is the now-famous Kneser bound.

\newsec{The Hurwitz space $H$ of branched coverings}

\subsec{Definition}

The Hurwitz space of branched coverings is nicely
described in  \Fulton
\ref\harris{J.~ Harris and D.~ Mumford, ``On the Kodaira dimension of
the Moduli Space of Curves," Invent. math. {\bf 67} (1982) 23-86.}.
Let $H(n,B,G;S)$ be the set
of equivalence classes of branched coverings
of $\Sigma_T$, with degree $n$, branching number $B$, and
branch locus $S$, where $S$ is a set of distinct points on
$\Sigma_T$.  $H(n,B,G;S)$ is a finite set.
The union of these spaces over sets $S$ with
$L$ elements is the Hurwitz space
 $H(n,B,G,L)$ of equivalence classes of branched coverings of
$\Sigma_T$ with
degree $n$,
branching number $B$ and  $L$ branch points.
Finally let $C_L(\Sigma_T)$ be the configuration space
of ordered $L$-tuples of distinct points on $\Sigma_T$, that is
$$C_L(\Sigma_T)=\{(z_1,\ldots,z_L)\in
\Sigma_T^L\vert z_i \in \Sigma_T, z_i \ne z_j { }\hbox{ for } i \ne
j\}.
$$
The permutation group $S_L$ acts naturally on $C_L$ and
we denote the quotient $\CC_L(\Sigma_T)=C_L(\Sigma_T)/S_L$.
There is a map
\eqn\proj{
p: H(n,B,G,L)\to \CC_L(\Sigma_T)
}
which assigns to each covering its branching locus.
This map can be made a
topological (unbranched) covering map \Fulton\ with
discrete fiber $H(n,B,G;S)$
over $S\in \CC_L$.

The lifting of closed curves in $\CC_L$ will in
general permute different elements of the
fibers $H(n,B,G,S)$. Note however that
$Aut f $ is invariant along any lifted curve
so that $Aut f $ is an invariant of the different
components of $H(n,B,G,L)$.

\subsec{H as an analytic variety}

One great advantage of branched covers is that they allow
us to introduce the powerful methods of complex analysis,
which are crucial to introducing ideas from topological field theory.

Choose a complex structure $J$ on $\ST$.
Then given a branched cover $f:\Sw\longrightarrow \ST$ there is a
 unique complex structure on $\Sw$ making $ f$ holomorphic.
 (use the complex structure $f^*(J)$ on $\Sw$ )
 \ref\Ahl{L.V.~ Ahlfors and L.~ Sario,
{\it Riemann Surfaces}, Princeton University Press, 1960.}. (This is far
from true for pinch maps).  Conversely, any  nonconstant
holomorphic map $f:\Sw \longrightarrow \ST$ defines a branched cover.
It follows that we can consider the Hurwitz space $H(n,B,G,L)$ as
 a space of holomorphic maps. The complex structure $J$ on $\ST$
 induces a complex structure on $H(n,B,G,L)$ such that $p$ is a
holomorphic
fibration. Moreover, the induced complex structure on $\Sw$ defines a
holomorphic map
$m: H(n,B,G,L) \longrightarrow  \CM_{h,0}$ where $\CM_{h,0}$ is
the Riemann moduli space of curves of genus $h$, where $h$ is given
by \Rhurw .
The image of $H$ is a subvariety of $\CM_{h,0}$.


\subsec{Fiber Bundle approach to Hurwitz space}

For comparison with topological field theory we will
need another description of Hurwitz space as the base
space of an
infinite-dimensional fiber bundle.

Let $\Sw$ be an orientable surface, and suppose
$\ST$ is a Riemann surface with a choice
of K\"ahler metric and complex structure $J$.
Let us begin with the configuration space
\eqn\CurlyM{
\tcM ~=~ \left \{ (f,h) \vert~ f \in \CC^\infty( \Sw, \ST ), h \in
{\rm
Met} (\Sw) \right\}.}
where $\CC^\infty( \Sw, \ST )$ is the space of  smooth
($\CC^\infty$) maps,
$f \colon \Sw \to \ST$ and ${\rm Met} ( \Sw )$ is the space of smooth
metrics on $\Sw$.
A choice of metric $h$ induces a complex
structure: $\epsilon(h)\in \Gamma[End(T\Sw)]$,
$\epsilon^2=-1$.  If we choose a basepoint $h_0$ in
the space of metrics, and choose oriented isothermal coordinates
relative to $h_0$ then we can define a basepoint complex
structure to be  the standard antisymmetric
tensor $\hat\epsilon_{\alpha\gamma}$,
\eqn\epsi { \hat\epsilon_{\alpha\gamma} = \pmatrix { 0 \qquad 1 \cr
-1 \qquad
0}   }
in the isothermal coordinates. In these terms  we define
$\epsilon_\alpha^{~ \beta} ( h ) = \sh \hat\epsilon_{\alpha\gamma}
h^{\gamma\beta}$.

The subspace of
pairs defining a holomorphic map
$\Sw\to\ST$ is then given by
\eqn\bigfam{
\tilde\CF=\{(f,h): df \epsilon = J df\}\subset \tcM
}
The defining equation $df \epsilon = J df$ is an equation in
$\Gamma \bigl[ End (T_x\Sw,T_{f(x)}\ST) \bigr]$.

Let  $Diff^+(\Sw)\sdtimes Weyl(\Sw)$ be   the semidirect product of
the group of orientation preserving diffeomorphisms of
$\Sw $ and the group of Weyl transformations on $\Sw$.
There is a natural action of this group on $\tcM$.
 There is an action
of $Diff^+ (\Sw)\sdtimes Weyl(\Sw)$ on $\tcF$. The quotient space
\eqn\dfofeff{
\CF\equiv \tilde\CF/\bigl( Diff^+ (\Sw)\sdtimes Weyl(\Sw)\bigr)
\qquad
}
parametrizes holomorphic maps $\Sw\longrightarrow \ST$.

We have now provided two descriptions of the space of
 holomorphic maps: Hurwitz space $H$ and \dfofeff.
Let  $H(h,G)=\amalg_{n(2-2G)-B=2-2h} H(n,B,G,L)$
where the disjoint union runs over $n,B,L\ge 0$ compatible with
\Rhurw.
There is a map  $\sigma: H(h,G,L)\to \CF$
which is generically smooth and one-one. The space $\CF$ has orbifold
singularities
where the
group $Diff^+ (\Sw)$ fails to act freely. Because we divide by
$Diff^+(\Sw)$,
the local orbifold group at $(f,h) \in \CF$ is $Aut f$.
This will be important when we introduce the orbifold Euler character
of $H$.
The map  $ \tcM \longrightarrow  Met (\Sw) $ obtained from $(f,h)
\rightarrow
h$
induces the map
$m: H(h,G) \rightarrow   {\cal M}_{h,0} $
of sec. 4.2 and  relates  the bundle description of
Hurwitz space to the bundle description of   ${\cal M}_{h,0}$.
Since $Aut f \subset Aut (\Sw)$ , orbifold points of $\CF$ map  to
orbifold points of $ {\cal M}_{h,0}$.

\subsec{Geometry of $\CF$}

Our discussion of the topological string theory
approach to \ymt\ requires a brief discussion
of the geometry of $\CF$. In particular, we need
to define a connection on $T\CF$ and compute its
curvature.

Let us first make the tangent space  to $\tcF$ more explicit.
The tangent space to $\tcM$ is
\eqn\tnmt{
T_{{\scriptscriptstyle f,h}}\tcM=\Gamma[f^\ast (T \ST)]\oplus
\Gamma[S^{\otimes
2}(T^*\Sw)],}
where $\Gamma$ is the space of $\CC^\infty$ sections and $S^{\otimes
n}$ is the
${\rm n^{th}}$ symmetric power.
The tangent space to $\tcF$ is the subspace of pairs $(\delta f,
\delta h)$
which
preserve the equation $df  \epsilon = J df$.
In order to characterize this subspace by a differential equation we
identify
the differential $df$ with a section of $T^\ast \Sw  \otimes f^\ast
(T\ST)$
($\cong \Gamma \bigl[ End (T_x\Sw,T_{f(x)}\ST) \bigr]$).
Then, in order to vary with respect to $f$ we must compare $T \ST$ at
different
points.
We do this using the K\"ahler metric on $\ST$ to define a pullback
connection
\foot{See appendix B for a careful derivation of this connection.}
on $f^\ast (T \ST)$:
$$
\nabla:\Gamma[ f^\ast (T \ST)]\to \Gamma[T^\ast \Sw \otimes f^\ast (T
\ST)].
$$
Finally, let $k(\delta h)=\delta \epsilon$ be the variation of
complex
structure on
$\Sw$ induced from a variation of metric $\delta h \in S^{\otimes
2}(T^*\Sw)$.
The tangent space  $\tcF$ at $(f,h)$ is the subspace of
$T\tcM$ defined by
\eqn\taneq{
T_{{\scriptscriptstyle f,h}} \tcF=\{ (\delta f,\delta h):
\nabla (\delta f)+ J \nabla (\delta f) \epsilon +J df k (\delta h)
=0\}.}
See appendix B for the proof.

We now separate out  ``pure gauge''  deformations.
The action of the gauge group on $\tcF$ defines a subbundle
$T^{vert} \tcF \subset T \tcF$ with fibers isomorphic to $im C $
where
\eqn\defgge{\eqalign{
C \colon T [ Diff^+ (\Sw)\sdtimes Weyl(\Sw) ] ~&\longrightarrow~
T^{vert}
\tcF\cr
C  \pmatrix{\xi^\alpha\cr \delta \sigma\cr}
{}~&\longmapsto~ \pmatrix{\xi^\alpha \p_\alpha f^i \cr
(P \xi)_{\alpha \beta} + (\delta \sigma+ \nabla\cdot \xi)h_{\alpha
\beta}\cr}\cr}}
and
\eqn\orlpi{
(P \xi)_{\alpha \beta}
{}~=~ \nabla_{(\alpha}\xi_{\beta)} - h_{\alpha\beta} \nabla\cdot
\xi,}
as is familiar from string theory.
In general $\ker C=0$ and we have an isomorphism
$T_{{\scriptscriptstyle [f,h]}} \cF \cong T_{{\scriptscriptstyle
f,h}}
\tcF/im C $.

To go further we use the natural $Diff^+ (\Sw)$-invariant metric on
$\tcM$
(hence on $\tcF$), given by
\eqn\natmetric{\eqalign{
&\left\langle ( \delta f_1, \delta h_1 ), ( \delta f_2, \delta h_2 )
\right\rangle_{{\scriptscriptstyle T \tcM}}\cr
&\qquad\qquad=~ \int d^2 z~ \sh~ \left \{
G_{ij} \delta f_1^i \delta f_2^j
+ ( h^{\alpha\gamma} h^{\beta\delta} + c~ h^{\alpha\beta}
h^{\gamma\delta} )
\delta h_{1~\alpha\beta} \delta h_{2~\gamma\delta} \right\}.\cr}}
with $c \in \IR^+$ arbitrary.
This allows us to define adjoints and orthogonal projections.
We now introduce the operator:
\eqn\defO{
\bO ~=~ \pmatrix{
\CD & J df k \cr
\p f & P^\dagger\cr}\colon
T_{{\scriptscriptstyle f,h}}\tcM \to
\Gamma[T \Sw \otimes f^\ast (T\ST)] \oplus \Gamma[T\Sw].}
where the components are given by:
\eqn\dfcomps{\eqalign{
\CD \chi^i &= \nabla \chi^i + J (\nabla \chi^i)\epsilon\cr
\p f \chi^i & = h^{\alpha \beta} (\p_\beta f^i) G_{ij} \chi^j \cr}}
Consider deformations $(\delta f,\delta h)$
in the kernel of  $\bO$.
The first line of \defO\ ensures that $(\delta f,\delta h)\in T \tcF$
and the second ensures that $( \delta f, \delta h ) \not\in T^{vert}
\tcF$.
An index theorem shows that
$\dim_{\IC} \ker \bO - \dim_{\IC}  {\rm coker }\bO=B=2h-2-n(2G-2)$
is the total branching number.
On the other hand, a generalisation of Kodaira-Spencer theory
described
in appendix A shows that $\dim_{\IC} \CF =B$.
In the generic case ($G, h > 2$),  $\dim_{\IC}  {\rm coker }\bO=0$.
Moreover we have an {\it orthogonal} decomposition
\eqn\prtng{
T_{{\scriptscriptstyle f,h}} \tcF
{}~=~ T( Diff^+ (\Sw) \sdtimes Weyl(\Sw) ) \oplus \ker\bO.}
Even though the metrics are not $Weyl(\Sw)$ invariant, the orthogonal
decomposition \prtng\  is invariant.
Therefore $\ker\bO$ is isomorphic to the tangent space
$T_{[f,h]}\CF$.

The natural metric  \natmetric\ on $\tcF$ also defines connections
on the principal $Diff^+ (\Sw)\sdtimes Weyl(\Sw) $ bundle
$\pi: \tcF\to \CF$ as well as on the tangent $T\tcF\to \tcF$.
In the first case we define a lift of a curve $\gamma(t)\subset \CF$
to be $\tilde \gamma(t)\subset \tcF$ defined by the conditions:
\eqn\conds{\eqalign{
{d \over dt}{\tilde \gamma} &\in \ker\bO,\cr
d \pi \biggl({d \over dt}{\tilde \gamma}\biggr)& = {d \over
dt}{\gamma}.\cr}}
In the second case we define $\tilde \nabla$ on
$T\tcF$ by declaring $\ker \bO$ to be the horizontal subspace
of the fiber.
Finally we use these connections to define a connection $\nabla$ on
$T\cF\to \cF$.
It suffices to define the parallel transport $ X(t)$ of
$ X(0)\in T_{\gamma(0)}\cF$ along a path $\gamma(t)\subset \cF$.
Choose lifts $\tilde \gamma(t)\subset \tcF$ and
$\tilde X(0)\in T_{\tilde \gamma(0)}\tcF$.
Use $\tilde \nabla$ to define the parallel-transported
$\tilde X(t) \in T_{\tilde \gamma(t)} \tcF$ and define
$X(t)=d \pi \bigl(\tilde X(t)\bigr)$.
Since $\tilde \nabla$ preserves the orthogonal
decomposition \prtng\ our definition is independent of the choices
made
in lifting.
See Figure 3.

\ifig\fhhpiv{ Construction of a connection on $T \cF \to \cF$ :
$\delta ( t )$ is the lift of $\gamma ( t )$, determined by
choosing as initial point a lift
$Y(0)$ of $X(0)$.}
{\epsfxsize3.0in\epsfbox{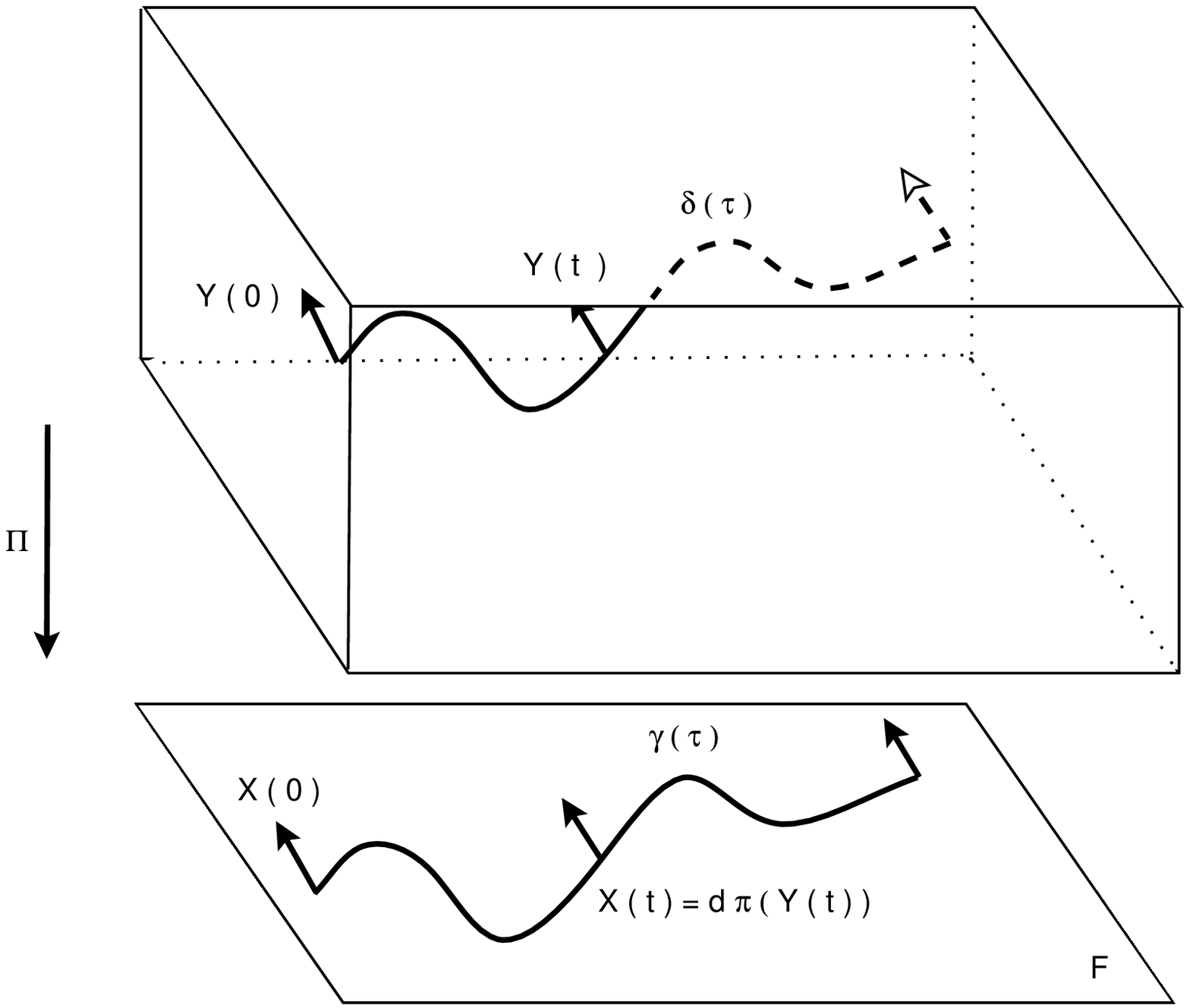}}

Finally, let us describe the curvature of the connection
$\tilde \nabla$ on $T\tcF$ more explicitly.
In local coordinates we may describe the connection
on $T_{f,h}\tcF$
as follows. Introduce local tangent vector fields
$$
{\bf X}=\pmatrix{\chi^i \cr \psi_{\alpha\beta}\cr},
\qquad\qquad
\tilde {\bf X} =\pmatrix{\tilde \chi^i \cr \tilde
\psi_{\alpha\beta}.\cr}
$$
of as elements of $\ker \bO$.
Then
\eqn\explcon {
\tilde\nabla_X \tilde X =\delta_X  \tilde X
 + \bO^\dagger {1\over \bO\bO^\dagger}\delta_X \bO \tilde X.}
where $\delta \tilde X= X\circ \tilde X$ and
$\delta\bO=X\cdot \bO$. Equation \explcon\ makes sense since $
\bO\bO^\dagger$
is invertible. A simple calculation,
using repeatedly the fact that $\bO \tilde X=0$ shows
that the curvature on horizontal vectors is given by
\eqn\xRx{\eqalign{
( \tbX_1,& {\bf R}[{\bf X}_1, {\bf X}_2] \tbX_2 )\cr
&=~ \left ( \tbX_1, (\delta_1 \bO^\dagger ) ( \bO \bO^\dagger )^{-1}
(  \delta_2 \bO ) \tbX_2 \right )
- \left ( \tbX_1, (\delta_2 \bO^\dagger ) ( \bO \bO^\dagger )^{-1}
(  \delta_1 \bO ) \tbX_2 \right )\cr}}

When we descend to $\CF$ we are working on an analytic space, and,
from \natmetric\ we see that $T\CF$ is a holomorphic Hermitian vector
bundle.
Thus if we choose a local holomorphic framing $\bG_I$ we may
use \natmetric\ to form the
positive definite matrix of inner products $h_{IJ}=\langle \bG_I,
\bG_J \rangle$.
In these terms the connection and curvature are given by
\ref\wells{R.O. Wells, {\it Differential Analysis on Complex
Manifolds},
Springer
1980}:
\eqn\hvb{
\nabla=\p \log h \qquad \qquad  R=\p\pb \log h
}
We will return to this formula in our discussion of contact terms.

\subsec{Compactification of Hurwitz Space}

Consider $H(n,B,G)$ the space of branched coverings with  a given
degree
and branching number of a surface of genus $G$.
A $B$-dimensional open subset of this space is $H(n,B,G,L=B)$, which
consists of maps where all the branch points are simple.
We will refer to this space as {\it simple Hurwitz space}.

Simple Hurwitz space can be (partially) compactified to form
the Hurwitz space $H(n,B,G)$ by adding $L$-dimensional
compactification varieties of the form $H(n,B,G,L<B)$:
 \eqn\unionH{ H(n,B,G) = \bigcup_{L} H(n,B,G,L).}

We thus have in mind the following schematic description
of Hurwitz space:

\ifig\fhhpiv{Simple Hurwitz space with
other Hurwitz spaces as compactification varieties.}
{\epsfxsize2.0in\epsfbox{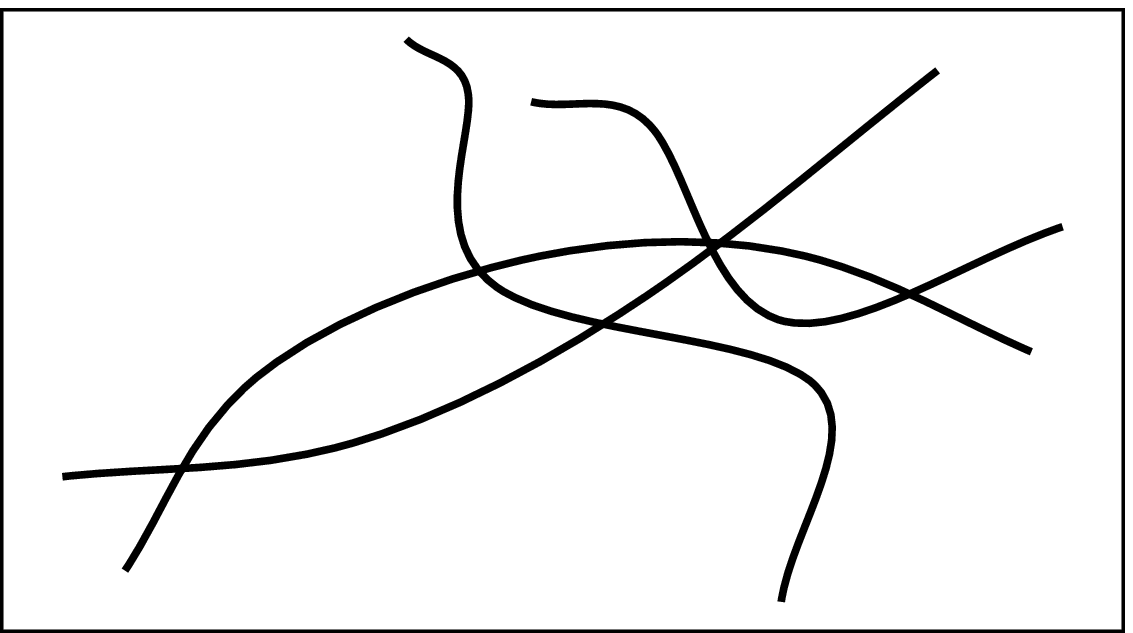}}

Compactified Hurwitz space is a bundle over the compactified
configuration
space where the branch points on the target are allowed to collide.
The
compactification subvarieties may be described in terms of
basic degenerations of branched coverings.
We will use the facts in sections 3.1 and 3.2 to describe some
properties of
degenerations that happen when two branch points collide.
The following observation is basic . Let $f\in H(n,B,G;S) $.
Choosing a set of generators for
$\pi_1$ we may then associate a cut surface as in the
proof of Theorem 3.1, and as in the LHS of Fig. 5.

\ifig\ezfgii{A collision of neighboring branch cuts
(for an appropriate choice of generators) produces
the product of the monodromy data.}
{\epsfxsize4.0in\epsfbox{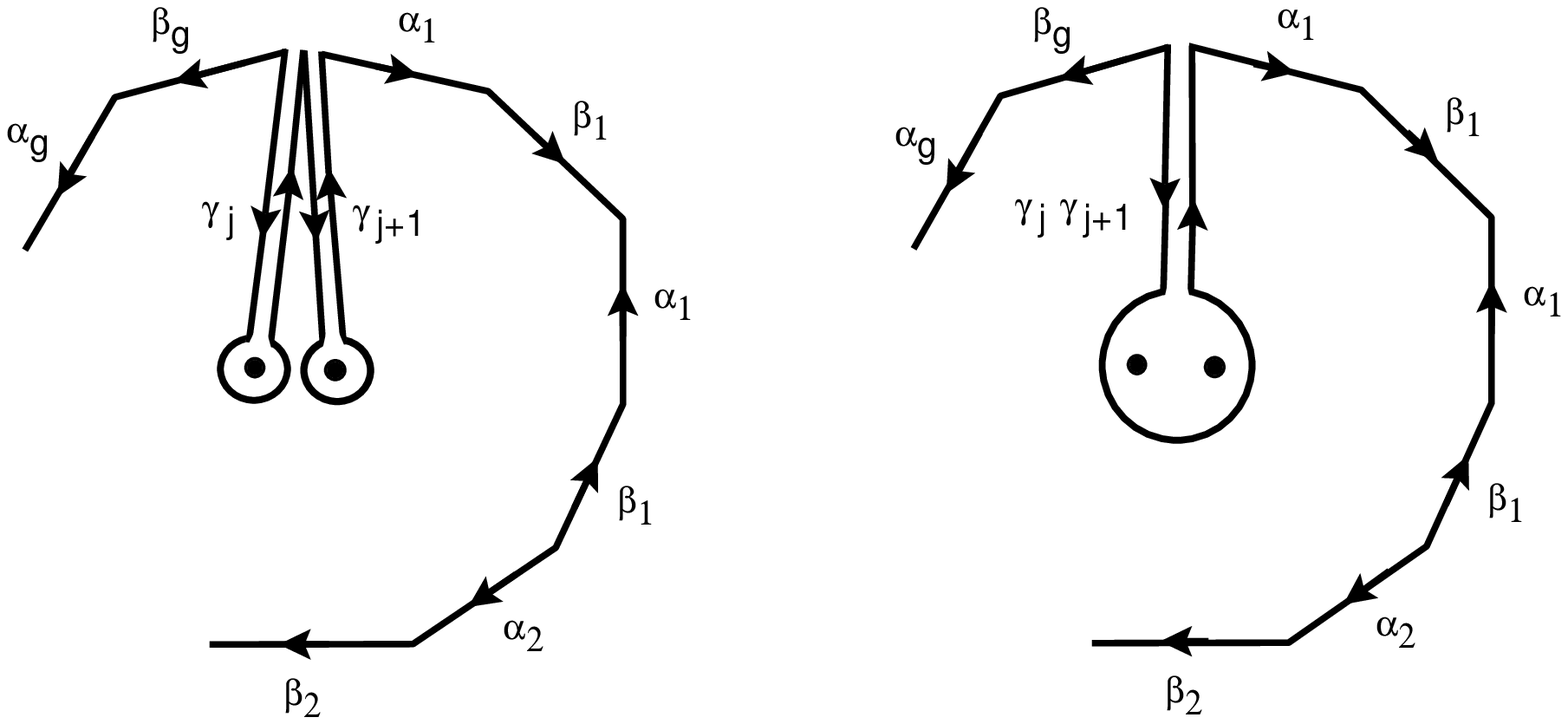}}

Choose two generators of $\pi_1(\Sigma_G -S,y_0)$,
 $\gamma_j$ and $\gamma_{j+1}$
(see \fgp). Consider a closed path
$\gamma^*$ homotopic to the product $\gamma_j\gamma_{j+1}$.
As we let the points enclosed by $\gamma_j$ and $\gamma_{j+1}$
approach
each other, the image of $\gamma^*$ in $S_n$ does not change and
remains
the product $u_ju_{j+1}$ where $u_j$ is the image of
$\gamma_j$ and $u_{j+1}$ is the image of  $\gamma_{j+1}$ under
$f_\#$.

Using the above rule we see that there are three types
of  collisions of simple branch points. They are classified by the
behaviour of
the inverse ramification points, and illustrated in Figure 6.

\noindent
{\it Type 1}. A collision of type 1  produces a single ramification
point of
index $3$. For example if $u_j= (12)$, $u_{j+1}= (23)$ then collision
of
$P_j,P_{j+1}$ produces a ramification point with $u=(123)$. Starting
from the
space of generic branched covers successive
collisions
can lead to multiple branch points; a collision of $\ell$ simple
branch
points
can lead to a ramification point of index $\ell + 1$.

\noindent
{\it Type 2}. A collision of type $2$
produces {\it two } ramification points.
This occurs when $u_{j}$ and $u_{j+1}$
are disjoint transpositions.

\noindent
{\it Type 3}.
A collision of type $3$ produces {\it no } ramification point but
instead
produces a double point. This occurs when $u_j$ and $u_{j+1}$ are the
same
transposition. Collisions  of any two branch points in a
hyperelliptic curve
produce degenerations of the third type.

Collisions of types $1$,  $2$  and
 their generalisations explain why $H(n,B,G,L<B) $ are used as some
of the
compactification
 varieties in the compactification of Hurwitz space.

\ifig\fhhpiv{Three types of Collisions}
{\epsfxsize3.5in\epsfbox{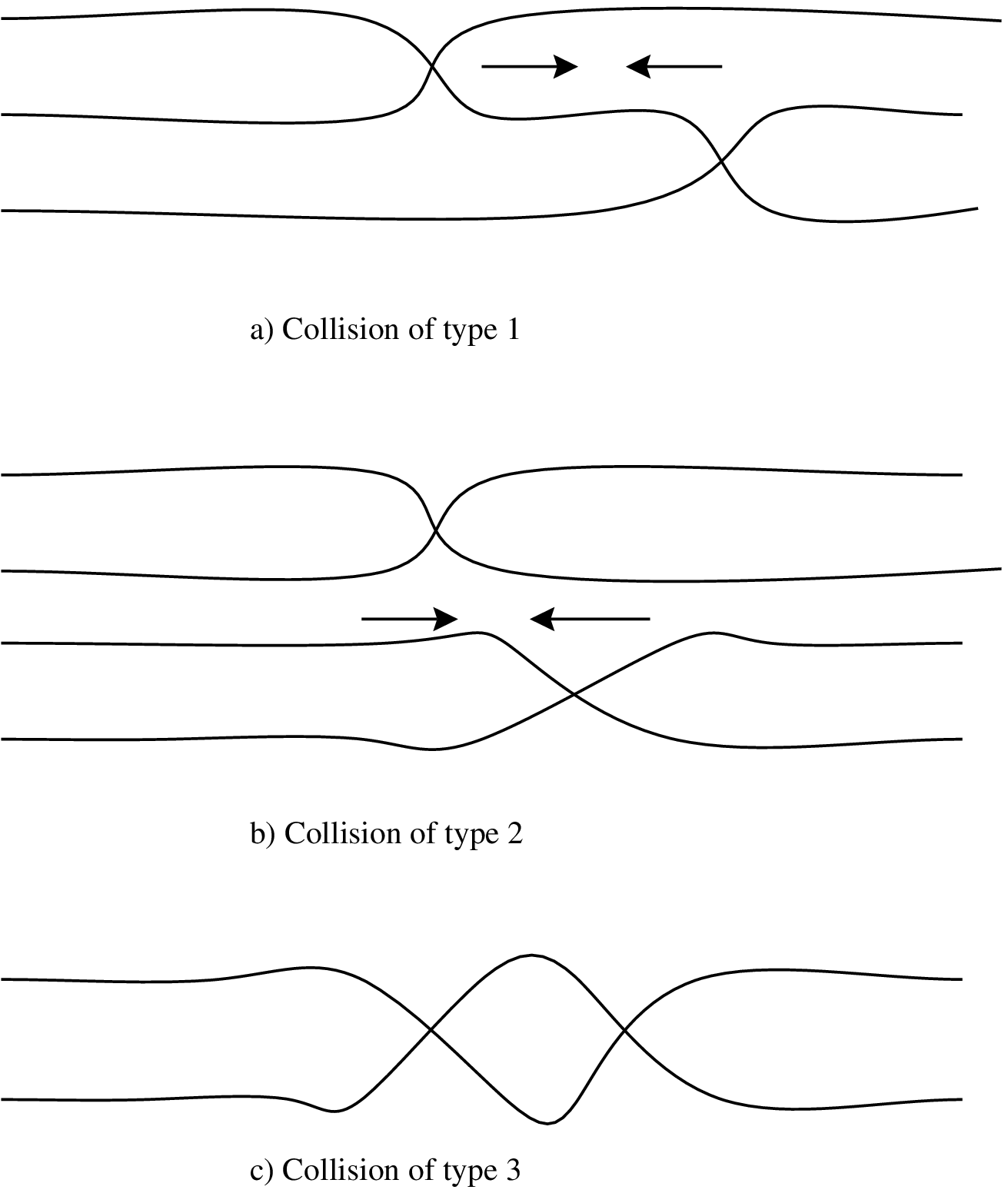}}

We now describe a class of collisions of branch
points  generalising collisions of type 3, which will be useful
in the discussion of the nonchiral theory. Let
$$u_j= (1,2\cdots k,k+1) (k+2)\cdots (n)$$ and
$$u_{j+1} = (1)\cdots (k-1) (k, k+1, \cdots, 2k) (2k+1)\cdots (n).$$
The monodromy $u^*$ around $\gamma^*$ is
$$(1,2,\cdots,k-1,k+1) (k, k+2,\cdots,2k) (2k+1)\cdots(n)\qquad .$$
The product permutation has two cycles of length $k$ and
remaining cycles
of length $1$.
Before collision the total branching number  at the ramification
points
is $(n - K_{u_j}) + (n-K_{u_{j+1}} ) = 2k$. After collision
the branching occurs at a single point so the branching number is
$K_{u^*}-n = 2k-2$. But the genus of the worldsheet does not change
during the collision. So there is a collapsed tube connecting the two
cycles of length $k$, the other sheets labelled $2k+1, \cdots,  n$ do
not
participate in the collision. The fact that collision of branch
points
can produce tubes connecting ramification points of equal index will
be
used in section 11.3  (More complicated collisions can occur but do
not seem
to contribute to the $YM_2$ partition function).
Note that the deficiency in total branching number
can only be even when branch points collide, which is clear
geometrically.
 This also follows from the fact
that $(-1)^{K_u - n}$ is equal to the parity of the permutation
$u$.

\noindent
{\bf Remark}. Beware.
Compactifications of moduli spaces are not unique.
Moreover, the construction of compactifications is a tricky
business. Compactifications of the base of Hurwitz space
are described in
\ref\blgz{A. Beilinson and V. Ginzburg, ``Infinitesimal structure of
moduli
spaces of G-Bundles,'' International Mathematics Research Notices
1992, No.
4. }.
The complete  mathematical description
of families of maps associated
with collisions of arbitrary branch points is rather complicated.
A compactification of Hurwitz spaces and its relation to
the Deligne-Mumford compactification of moduli spaces of complex
structures
 \ref\Mumford{D.~ Mumford, ``Towards an enumerative geometry
of the moduli space of Curves'' in {\it Arithmetic and Geometry}
(Birkhauser). }  and to the   compactification of
configuration spaces of \ref\knud{F.Knudsen, ``The projectivity
of the moduli space of stable curves,''   Math. Scand.
{\bf 52} (1983) 161.} is discussed in detail in
\harris .

\newsec{The CGTS and
 the space of branched coverings}

In this section we make our first connection between the topology of
Hurwitz space and $YM_2$ amplitudes. Consider the CGTS \Zchir.
As in 2D gravity, relations to topological field theory become most
transparent
in the
limit $A\rightarrow 0$.
Accordingly in this section we will study the series
\eqn\Zchiri{
Z^+(0,G, N) = \sum_{n=0}^{\infty}
N^{n(2-2G)}
\sum_{s_1,t_1,\ldots,s_G,t_G\in S_n}
\biggl[ {1\over {n!}}\delta ( \Omega_n^{2-2G} \prod_{j=1}^G
s_jt_js_j^{-1}t_j^{-1}) \biggr].}

\subsec{Recasting the CGTS as a sum over branched coverings}

The first step in rewriting \Zchiri\
is to count the weight of a given power of $1/N$. To this end
we expand the
$\Omega^{-1}$ point as an
element of the free algebra generated by elements of
the symmetric group,
\eqn \Omeginv {
 \Omega_n^{-1} =  1 + \sum_{k=1}^{\infty}
{\sum}^\prime _{v_1\cdots v_k\in S_n}
\biggl({1\over N}\biggr)^{\sum_{j=1}^k n-K_{v_j}  }
 (v_1v_2\cdots v_k ) (-1)^k .}
where the primed sum means no $v_i=1$.
We could rewrite \Zchiri\ by  imposing relations
 of the symmetric group  of $S_n$ as in \example.  However,
we decline to do this and rather substitute
the expansion \Omeginv\ into \Zchiri\
to obtain
\eqn\Zchirii {\eqalign{Z^+(0,G,N) =
\sum_{n=0}^{\infty}\sum_{L=0}^{\infty}
&N^{n(2-2G)}
\sum_{s_1,t_1,\ldots,s_G,t_G\in S_n}
\sum_{v_1,v_2, \ldots,v_L \in S_n}^{{}\prime}
\cr
N^{\sum_{j=1}^L (K_{v_j}-n)} &
\biggl[ { d(2-2G,L)\over {n!}}
        \delta(v_1v_2 \cdots v_L \prod_{j=1}^G
s_jt_js_j^{-1}t_j^{-1}) \biggr].\cr}}
 where $d(m,L)$ is defined by
\eqn\genfun {
 (1+x)^m = \sum_{L=0}^{\infty} d(m,L) x^L}
Explicitly we have
\eqn\dex{ \eqalign{
d(2-2G,L) &= (-1)^L { { (2G+L-3)!}  \over {(2G-3)! L! } } ,\qquad
 \hbox{ for} \qquad G>1 \cr
d(0,L) &= 0 , \hbox {unless}~ L=0 \cr
d(2,L) &= 0 , \hbox {unless}~ L=0,1 ,2. \cr }}
For $G>1$, $d(2-2G,L)$  is the number of ways of  collecting $L$
objects into
$2G-2$ distinct sets. The equation \Zchirii\ correctly gives the the
partition
function for any $G$ including  zero and one. For example the
vanishing of
$d(0,L)$ for $L>0$ means that in the zero area limit only
maps with no branch points contribute to the torus partition
function. And for
genus zero
the vanishing of $d(2,L)$ for $L>2$ means that only maps with no more
than two
branch points contribute to the CGTS for the sphere.

To each nonvanishing term in the sum \Zchirii\ we may
associate a homomorphism $\psi:F_{G,L}\to S_ n$,
where $F_{G,L}$ is an $F$ group \fgp,  since, if
the permutations
 $v_1,\dots,v_L, s_1,t_1, \dots, s_G,t_G$
 in $S_n$
satisfy $v_1\cdots v_L \prod_{i=1}^{G}
s_it_is_i^{-1}t_i^{-1} =1$  we may define
\eqn\homomorph{
\psi: \alpha_i\to s_i \qquad \psi:\beta_i \to t_i \qquad
\psi:\gamma_i \to v_i
}
Moreover, if
there exists a
$g \in S_n$ such that
$g \{
v_1, \cdots ,v_L; s_1  , t_1 \cdots s_G ,t_G\} g^{-1}=
 \{
v_1^{'} ,\cdots v_L^{'}; s_1^{'}, t_1^{'} ,
\cdots s_G^{'}, t_G^{'}\}$
as ordered sets
then by definition 3.2 the induced homomorphisms are equivalent.
Evidently the class of $\psi$ will appear in the sum $n!/\vert
C(\psi)\vert$
times in \Zchirii.
Therefore, we may  write \Zchirii\ as
\eqn \Zchiii {
Z^+(0,G,N)  =
\sum_{n=0}^{\infty} \sum_{B=0}^{\infty}
N^{n(2-2G)-B} \sum_{L=0}^{B} d(2-2G,L)
\sum_{\psi \in \Psi (n,B,G,L)}
       {1\over {\vert C(\psi)\vert} } }
where $\Psi (n,B,G,L)$ is the set of  equivalence classes of
homomorphisms  $F_{G,L}\to S_n$, with the condition that the
$\gamma_i$
all map
to elements of $S_n$ not equal to the identity.
We have
collected terms with fixed value of:
\eqn\dfb{
B\equiv \sum_{i=1}^{L} (n-K_{v_i}) \qquad .
}

Now we use Theorem 3.1 to rewrite the sum \Zchiii\ as a sum over
branched coverings.  To do this we must make several
choices.
We choose a point $y_0\in \ST$
and for each $n,B,L,\psi$ we also
make a choice of :

1. some set $S$ of $L$ distinct
points on $\ST$.

2. an isomorphism \fgpiso.

To each $\psi,S$  we may then associate
a homomorphism
$\pi_1(\ST - S,y_0 )\to S_n$. By theorem 3.1
we see that, given a choice of $S$,  to each class $[\psi]$
we associate the equivalence class of a branched covering
$f \in H(n,B,G;S)$, where $f:\Sw\to \ST$.
The genus of the covering surface $h=h(G,n,B)$ is given by
the Riemann-Hurwitz formula \Rhurw.
Note that the power of ${1\over N}$ in
\Zchiii\ is simply $2h-2$.
Finally, the centralizer
$C(\psi)\subset S_n$ is isomorphic to
the automorphism group of the associated
branched covering map $f$. The order of
this group, $\vert Aut(f)\vert$, does not depend on the choice
of points $S$ used to construct $f$.
Accordingly,  we can write $Z^+$ as a sum over
equivalence classes of branched coverings:
\eqn\Zii{
Z^+(0,G,N)  =
\sum_{n=0}^{\infty} \sum_{B=0}^{\infty}
\sum_{L=0}^{B} \biggl({1\over N}\biggr)^{2h-2}  d(2-2G,L)
 \sum_{f \in H(n,B,G;S) }
{1\over {\vert Aut f\vert} } }

\subsec{Euler characters}

We have now expressed the CGTS as a sum over
equivalence classes of branched coverings. We
now interpret the weights in terms of the Euler characters of
the Hurwitz space $H$.

To begin we write
\eqn\euli{
d(2-2G,L) =
{{(\chi_G ) (\chi_G-1) \cdots (\chi_G -L+1)}\over {L!}},}
where $\chi_G = 2-2G$.
The RHS of \euli\ is the Euler character of
the  space $\CC_L(\ST)=C_L(\ST)/S_L$.
This may be
easily proved in two ways. Recall that
 it is general property of fibre bundles
with connected base that their Euler character is
the product of Euler characters of base and
fibre \Spanier \ref\BottTu{R.~ Bott and L.~ Tu, {\it Differential
Forms in
Algebraic Topology}, Springer Verlag, New York, 1982.}.
 Let  ${\cal{M}}_{G,L}$ be the uncompactified
moduli space of complex
structures of a surface of genus $G$ with $L$ punctures. The
fibration:
\eqn\fibration{
\matrix{ \CC_L(\ST)&\mapright{}  & \CM_{G,L} \cr
                   &             & \mapdown{}\cr
                   &             & \CM_{G,0}\cr}}
together with the celebrated Harer-Zagier-Penner formula:
\ref\Penner{R.C.~ Penner, ``Perturbative Series and the moduli space
of Riemann surfaces," J. Differential Geometry, {\bf 27} (1988) 35-53. }
\eqn\Pen { {\cal{\chi}}
({\cal{M}}_{G,L}) = (-1)^L {(2G-3+L)!(2G-1) \over {L!
(2G)! }} B_{2G} }
(where  $G \ge 0, 2G -2 + L > 0$, and
$B_{2G}$ is a Bernoulli number)
gives the result:
\eqn\euc{\eqalign{
 \chi({\cal C}_L(\ST)) &=
{ {{\cal{\chi}} ( {\cal{M }}_{G,L})
   }
     \over { {\cal{\chi}} ({\cal{M}}_{G,0})
            }
 } \cr
              &= (-1)^L { {(2G+L-3)!} \over {(2G-3)! L!}
                         },  \cr}}
An alternative proof, (which also covers the cases of interest at
genus zero) uses the fibration of configuration spaces described
in \ref\Birman{J.S.~ Birman, Braids, Links and Mapping Class Groups,
pages 11, 12, Princeton University Press, 1975.}. Let $C_{m,n}
(\Sigma_T)$ be
the configuration space of $n$ labelled  points on a surface
$\Sigma_T$ of
genus $G$ with $m$  {\it fixed} punctures.
There is a fibration
\eqn\fibc{\matrix{
C_{L-1,1} (\Sigma_T) & \mapright{}  &C_{0,L} (\Sigma_T)\cr
                     &              &\mapdown{} \cr
                     &              &C_{0,L-1}(\Sigma_T).  \cr }}
Using the product formula for Euler characters of a fibration we get
\eqn\fibrec{
\chi ( C_{0,L}(\Sigma_T) = (2-2G-(L-1) ) \chi ( C_{0,L-1}(\Sigma_T) )
}
This recursion relation together with $\chi ( C_{0,1}(\Sigma_T))
 =\chi(\Sigma_T) $ gives
$ \chi ( C_{0,L}(\Sigma_T)) =  (\chi_G ) (\chi_G-1) \cdots
(\chi_G-L+1)$.
But  $C_{0,L}(\Sigma_T)$ is a topological covering space of
$\CC_L(\Sigma_T)$
of degree
$L!$ so this leads to
\eqn\euci { \chi (\CC_L(\ST)) = d(2-2G,L) . }

Using \euci,  we can further rewrite the CGTS as
\eqn\euchar {\eqalign {
Z^+(0,G,N) &= \sum_{n=0}^{\infty} \sum_{B=0}^{\infty}
N^{n(2-2G)-B} \sum_{L=0}^{B} \chi ( {\CC_L (\ST)) }
     \sum_{ f \in  H( n,B,G,S )  }
{1\over { \vert Aut f\vert } } \cr
}}

Let us now return to the fibration \proj. Consider first
the case where the covering surface $\Sw$ has
no automorphisms.   From the results of
\ref\Edtwo {I.~ Berstein and A.L.~ Edmonds, ``On the
classification of generic branched coverings of surfaces," Illinois
Journal of Mathematics, Vol {\bf 28}, number 1, (1984).}, it follows that
this willhappen for primitive branched coverings of
surfaces with $G>2$ with  $B>n/2$ simple branch points.
In such cases we can identify
\eqn\eulhur{\eqalign{
\chi ( \CC_L (\ST)  )\sum_{ f \in  H( n,B,G,S )  }
{1\over { \vert Aut f\vert } }&=\chi ( \CC_L (\ST)  )  \vert
H(n,B,G;S) \vert\cr
&=\chi(H(n,B,G,L))\cr}}
where we have again used the fact that the Euler character of a
bundle is the product of that of the base and that of the fibre
\Spanier\ (the Euler character of the fiber is
$ {\cal \chi}  (H(n,B,G;S )) = \vert H(n,B,G;S) \vert$).

When $H(n,B,G,L)$ contains coverings with automorphisms
the corresponding space $\CF$ has orbifold singularities.
We introduce the orbifold Euler character
$\chi_{ \rm orb} (H)$ as the Euler character of $\chi(\CF)$
calculated
by resolving its orbifold
singularities.
The division by the factor $\vert Aut(f)\vert$ is the correct factor
for calculating the orbifold Euler characteristic of the subvariety
$H(n,B,G,L)$ since $Aut(f)$ is the local orbifold group of the
corresponding point in $\CF$.
With this understood we naturally define:
\eqn\orbeul{
\chi_{\rm orb} ( (H( n,B,G,L) ))
\equiv \chi ( {\cal C}_G (L) )\sum_{ f \in  H( n,B,G,S )  }
{1\over { \vert Aut f\vert } }
}
in the general case.
Thus we finally arrive at our first main result:

\noindent
{\bf Proposition 5.1}.
The CGTS is the
generating functional for the orbifold Euler characters of the
Hurwitz spaces:
\eqn\euchario {\eqalign {
Z^+(0,G,N) &= \sum_{n=0}^{\infty} \sum_{B=0}^{\infty}
\biggl({1\over N}\biggr)^{2h-2} \sum_{L=0}^{B}
\chi_{\rm orb} ( H( n,B,G,L) )\cr
}}
where $h$ is determined from $n,G$ and $B$ via the
Riemann-Hurwitz theorem.
\par
The $L=B$ contribution in the sum is  the Euler character
of the space of generic branched coverings.
 As described in section 4.4
compactification of this space involves addition of   boundaries
corresponding to the space of
 maps with higher branch points, i.e.,  where $L < B$.

Quite generally, \foot{ We thank E. Getzler for these clarifying
remarks.}
suppose $X$ is a closed manifold with boundary $\p X$. The inclusion
$\p X \hookrightarrow X $ gives rise to a long exact sequence in
homology
\eqn\Lexseq{ \cdots\rightarrow H_i(\p X) \rightarrow
H_i(X)\rightarrow H_i(X,
\p X)
\rightarrow H_{i-1} (\p X) \rightarrow \cdots }
By Lefschetz duality we may write the relative cohomology groups in
terms of
the homology of the interior $X^0$:
\eqn\LD{H^i(X, \p X) = H_{n-i} (X^0),}
where $n$ is the dimension of X
\ref\vick{J. Vick, {\it Homology Theory}, Academic Press, 1973}.
 If $n$ is even then $\chi(X,\p X)= \chi(X^0)$.
Applying the above discussion to Hurwitz space we see that we can
interpret
$$\sum_{L=0}^{B} \chi_{\rm orb} ( H( n,B,G,L) )$$
as the Euler character of a
partially compactified Hurwitz space $\overline{(H( h,B,G))}$
obtaining by
adding
degenerations of type 1 and 2 and their generalizations.

\noindent
{\bf Proposition 5.2}: The CGTS is the
generating functional for the orbifold Euler characters of the
analytically compactified Hurwitz spaces:
\eqn\eucharit {\eqalign {
Z^+(0,G,N) &= \sum_{n=0}^{\infty} \sum_{B=0}^{\infty}
\biggl({1\over N}\biggr)^{2h-2}
\chi_{\rm orb} ( \overline{(H( n,B,G))})\cr
&= \exp \biggl[  \sum_{h=0}^{\infty} \biggl({1\over N}\biggr)^{2h-2}
\chi_{\rm orb} ( \overline{(H( h,G))})\biggr]
 \cr}}

\noindent
{\bf Remarks}:

1. Recall that we allow for the possibility of disconnected
worldsheets. Expressing the result in terms of connected coverings
leads
to the final equation.

2. The importance of the high-codimension compactification
varieties in the \ymt\ partition sum ( Hurwitz spaces with $L<B-1$)
is extremely intriguing from
the point of view of the topological field theory discussed  in the
remainder of this paper. \ymt\ appears to be an example
where the ``higher contact terms'' of the toplogical field theory
are extremely important in getting the correct answer.
This will be a recurring theme throughout this paper.

3. There is more than one way to interpret
the expansion $Z^+$ of \Zchiri\  as
a sum over maps. The ambiguity arises from the
treatment of the $\Omega^{-1}$ terms. Had we used
the relations of the symmetric group in writing a formula
for $\Omega^{-1}$ we would have found that
the coefficient of any permutation $v$ multiplies
an infinite series  $(1/N) ^{n-K_v} + \cdots $. All powers
in the series, but the leading one,  would be
too large to be accounted for by branching alone.
We could still describe the \ymt\ partition function
in terms of maps but we would need to
invoke the pinch maps of section 3.3.
As an example, consider the expansion of
$\Omega^{-1}$ in  \example .
In  the first line  we have written the inverse omega
point as
a sum where $i$ transpositions come with a factor $N^{-i}$.
Interpreting each factor $v$ as the data of some branch
point leads to a description in terms of branched covers.
Using the relations of the symmetric group we obtain
the last line of \example. The higher powers of $1/N$
must be accounted for by collapsed handles, for example, by
pinch maps. The advantage of excluding
pinch maps and associating  each term in the CGTS with
branched coverings is that, as in sec. 4.2, when the  target is
equipped with a
complex structure such maps can always be
interpreted as  holomorphic or antiholomorphic.
This is an encouraging sign because
topological sigma models count {\it  (anti)-holomorphic }
maps
\ref\Witten{E.~ Witten,
``Topological sigma models, ''Commun. Math. Phys. {\bf 118}
(1988) 411-419. }.
This remark is the first step on the road to the
construction of the equivalent string theory in
section 6.

4. The paper of Gross and Taylor already showed that
by expanding  $\Omega^{-1}$ one could interpret all
contributions to $Z^+$ in terms of branched covers.
However, in the picture of \GrTa\ there are
$ \vert 2-2G \vert $ special points: ``twist points,'' on
the target space, and one imagines that all the branch
points $v_1 \cdots v_L$ are somehow ``anchored''
to  these special twist points, for all values of $L$.
In this paper we allow the branch points
to ``sail'' over the entire target space $\ST$. From
the latter point of view the combinatorial factors
$d(2-2G,L)$ are more natural.

\newsec{Synopsis of Topological Field Theory}

In this section we briefly review topological field theory.
A number of reviews of this subject already
exist\refs{\Van,\Witr,\bbrt,\blau,\DiVV,\CMRLH} and it is to these
that we refer the interested reader for more detail and further
references.

Topological Field Theories (TFT) study the topology of moduli spaces.
There are usually a number of different descriptions of a moduli space,
${\cal M}$.
For the purpose of formulating a TFT, it is convenient to characterize
${\cal M}$ as a subspace within a ${\cal G}$-manifold\foot{Recall that for any
group ${\cal G}$,  a ${\cal G}$-manifold is a manifold on which the
action of ${\cal G}$ is defined at every point.
In the context of TFTs, $\widetilde {\cal C}$ and ${\cal G}$ are
typically infinite dimensional.
We will remain formal and largely ignore this fact in our discussion.},
$\widetilde {\cal C}$:
$$\eqalign{
{\cal M} ~=&~ \widetilde {\cal M} / {\cal G}\cr
\widetilde {\cal M} ~=&~ \{ \varphi \in \widetilde {\cal C}~ \vert~
s ( \varphi ) = 0 \}\cr}
$$
Here $s$ is a ${\cal G}$-equivariant map between $\widetilde {\cal C}$
and an auxilliary ${\cal G}$-manifold, $\widetilde {\cal V}$:
$$
s\colon \widetilde {\cal C} \longrightarrow \widetilde {\cal V}.
$$
The precise natures of the map, $s$, and the auxilliary
${\cal G}$-manifold, $\widetilde {\cal V}$, depend on the
moduli space under consideration.
\medskip\noindent
{\bf Remarks:}
\itemitem{6.1.1.}
We may regard
$$\matrix{
\widetilde {\cal E} & \longleftarrow & \widetilde {\cal V}\cr
\mapdown{\widetilde {\cal \pi}} & & \cr
\widetilde {\cal C} & & \cr}
$$
as an equivariant ${\cal G}$-bundle which may be either trivial (as in
case of topological Yang-Mills theory\refs{\donaldson}) or non-trivial
(as in case of topological sigma models\refs{\Witten}).
$s$ induces a section, $\widetilde s$, of $\widetilde {\cal E}$.
\itemitem{6.1.2.}
The action of ${\cal G}$ on $\widetilde {\cal E}$ may fail to be free, in
which case the quotient $\widetilde {\cal E} /{\cal G}$ is not a manifold.
For example, let $\widetilde {\cal E} = {\cal A}$ be the space of gauge
connections on a principal $G$-bundle, $P$.
The group of gauge transformations acts on ${\cal A}$.
For a reducible connection, $A \in {\cal A}$, ${\cal G}$ possesses a
non-trivial isotropy group.
\itemitem{6.1.3}
${\cal G}$ may, of course, be trivial, in which case
${\cal M} = \widetilde {\cal M}$.
This is the case for the topological sigma model\refs{\Witten}.
\medskip\noindent
Since the quotient space, $\widetilde {\cal E} / {\cal G}$, is potentially
a singular space, we cannot in general study its cohomology directly.
Instead one must examine the ${\cal G}$-equivariant cohomology
of $\widetilde {\cal E}$.

\subsec{Topological Description of Equivariant Cohomology}

A sketch of equivariant cohomology necessitates a brief discussion
of universal bundles.
Imagine that we are given a {\it contractible} ${\cal G}$-manifold,
$X$, on which ${\cal G}$ acts {\it freely}, then
\item{1.}
Since $E {\cal G}$ is contractible, $\widetilde {\cal E}$ and
$\widetilde {\cal E} \times X$ have the same homotopy type
and thus have identical de Rham cohomology\refs{\BottTu}.
\item{2.}
Since ${\cal G}$ acts freely on $X$, $\widetilde {\cal E} \times X$
inherits a free ${\cal G}$ action:
\eqn\GrpAct{
g \cdot ( e, x ) = ( g e, x g^{-1} ).}
Then the quotient
\eqn\AssocBund{
\widetilde {\cal E} \times_{\cal G} X
\quad\eqdef\quad {{\widetilde {\cal E} \times X} \over {\cal G}}}
defines a manifold.
\medskip\noindent
The case when $X$ is a principal ${\cal G}$-bundle is of particular
importance.
Such spaces are to a large degree unique:
\medskip\noindent
{\bf Definition 6.1.4}:
To a compact, finite dimensional group, ${\cal G}$ we can associate
a {\it universal} ${\cal G}$-{\it bundle}, $E {\cal G}$,
\eqn\UnivBund{\matrix{
E {\cal G} & \mapleftu{} & {\cal G}\cr
\mapdown{\pi_{\cal G}} & & \cr
B {\cal G} & & \cr}}
$E {\cal G}$ is a contractible principal ${\cal G}$-bundle over the
{\it classifying space}, $B {\cal G} \eqdef E {\cal G} / {\cal G}$.
$E {\cal G}$ is unique up to equivariant homotopy type, while $B {\cal G}$
is unique up to homotopy type.
\medskip\noindent
Note that if ${\cal G}$ acts freely on $\widetilde {\cal E}$, then
$\widetilde {\cal E} \times_{\cal G} E {\cal G}$ and
$\widetilde {\cal E} / {\cal G}$ have the same homotopy type and
therefore have identical de Rham cohomologies.
More generally,
\eqn\AssocUB{\matrix{
\widetilde {\cal E} \times E {\cal G} & \longleftarrow & {\cal G}\cr
\mapdown{\widetilde \pi_{\cal G}} & & \cr
\widetilde {\cal E} \times_{\cal G} E {\cal G} & & \cr}}
may be viewed as a ${\cal G}$-bundle over $B {\cal G}$.
These properties motivate the following
\medskip\noindent
{\bf Definition 6.1.5}:
The ${\cal G}$-equivariant cohomology of $\widetilde {\cal E}$ is defined to be
\eqn\ECTopDefi{
H^\bullet_{{\cal G}, {\rm top}} ( \widetilde {\cal E} )
\quad\eqdef\quad H^\bullet \left (
\Omega ( \widetilde {\cal E} \times_{\cal G} E {\cal G} ), d \right ).}
\medskip\noindent
Though this definition offers a perfectly sensible way in which to
define the topology of ``$\widetilde {\cal E} / {\cal G}$", the manifold
$\widetilde {\cal E} \times_{\cal G} E {\cal G}$ is generally quite
difficult to study directly.
In certain cases, e.g. ${\cal G}$ finite dimensional and compact,
the cohomology of this space may be studied very effectively
via indirect means.
\par
The pullback by the projection, $\widetilde \pi_{\cal G}$, induces an
injective homomorphism:
\eqn\PullBpig{
\widetilde \pi_{\cal G}^\ast\colon
\Omega^\bullet ( \widetilde {\cal E} \times_{\cal G} E {\cal G} )
{}~\longrightarrow~ \Omega^\bullet ( \widetilde {\cal E} \times E {\cal G} )}
Since $\widetilde {\cal E} \times E {\cal G}$ is topologically a simpler
space, it is very useful to characterize $\widetilde \pi_{\cal G}^\ast
H^\bullet ( \widetilde {\cal E} \times_{\cal G} E {\cal G} )$ as a subset
of $H^\bullet ( \widetilde {\cal E} \times E {\cal G} )$.
Since ${\cal G}$ acts freely on $\widetilde {\cal E} \times E {\cal G}$,
there is at each point $x \in \widetilde {\cal E} \times E {\cal G}$
a map of ${\cal G}$ into the fiber over $y = \widetilde \pi_{\cal G} ( x )$,
\eqn\GroupMap{
R_x\colon {\cal G}
{}~\longrightarrow~ \widetilde {\cal E} \times E {\cal G} \vert_y.}
where $\widetilde {\cal E} \times E {\cal G} \vert_y
\eqdef \widetilde \pi_{\cal G}^{-1} ( y )$.
The differential of $R_x$ defines a map from $\lieg = {\rm Lie}\ {\cal G}$
into the vertical tangent space
\eqn\DiffGM{\eqalign{
C_x \eqdef d R_x \colon& \lieg
\longrightarrow T_x ( \widetilde {\cal E} \times E {\cal G} )^{\rm vert}\cr
C_x\colon& g \longmapsto X_g \vert_x ~\eqdef~ C_x g\cr}}
This defines two actions of $\lieg$ on
$\Omega^\bullet (\widetilde {\cal E} \times E {\cal G} )$:
\item{1.}
{\it Contraction}: For all $g \in \lieg$,
$$\eqalign{
i ( g )\colon \Omega^k ( \widetilde {\cal E} \times E {\cal G} )
\quad\longrightarrow&\quad
\Omega^{k-1} ( \widetilde {\cal E} \times E {\cal G} )\cr
i ( g )\colon \omega
\quad\longmapsto&\quad
i_{X_g} \omega\cr}
$$
where $i_{X_g}$ is the usual interior product with a vector field $X_g$.
\item{2.}
{\it Lie Derivative}: For all $g \in \lieg$,
$$\eqalign{
{\cal L} ( g )\colon \Omega^k ( \widetilde {\cal E} \times E {\cal G} )
\quad\longrightarrow&\quad
\Omega^k ( \widetilde {\cal E} \times E {\cal G} )\cr
{\cal L} ( g )\colon \omega
\quad\longmapsto&\quad
{\cal L}_{X_g} \omega\cr}
$$
where ${\cal L}_{X_g} \equiv [ i_{X_g}, d ]_+$ is the usual Lie
derivative with the vector field $X_g$.
\medskip\noindent
These derivations characterize a special subcomplex of
$\Omega^\bullet ( \widetilde {\cal E} \times E {\cal G} )$.
\medskip\noindent
{\bf Definition 6.1.6}:
Let $\{ T_i \}$ be a basis for $\lieg$.
A form $\eta \in \Omega^\bullet ( \widetilde {\cal E} \times E {\cal G} )$
is called $\lieg$-{\it basic}, if it is {\it both}
\item{1.}
{\it Horizontal}, i.e. $\eta \in \bigcap_{i=1}^{\dim\ {\cal G}}
\ker\ i ( T_i )$, {\it and}
\item{2.}
{\it Invariant}, i.e. $\eta \in \bigcap_{i=1}^{\dim\ {\cal G}}
\ker\ {\cal L} ( T_i )$.
\par\noindent
The $\lieg$-{\it basic subcomplex} will be denoted by
$\Omega ( \widetilde {\cal E} \times E {\cal G} )_{\lieg-{\rm basic}}$.
\medskip\noindent
{\bf Theorem 6.1.7}:
In the case that ${\cal G}$ is finite dimensional and compact, the de Rham
cohomology of $\widetilde {\cal E} \times_{\cal G} E {\cal G}$ is precisely
the basic cohomology of $\widetilde {\cal E} \times E {\cal G}$.
\medskip\noindent
{\it Proof}: See Mathai and Quillen\refs{\MatQui}
\medskip\noindent
It follows from Definition 6.1.5 and Theorem 6.1.7, that the
${\cal G}$-equivariant cohomology of $\widetilde {\cal E}$ can be
computed as follows:
\eqn\ECTopDefi{\eqalign{
H^\bullet_{{\cal G}, {\rm top}} ( \widetilde {\cal E} )
\quad\eqsim&\quad
H^\bullet \left ( \Omega ( \widetilde {\cal E} \times_{\cal G} E {\cal G} ),
d \right )\cr \quad\eqsim&\quad
H^\bullet \left (
\Omega ( \widetilde {\cal E} \times E {\cal G} )_{\lieg-{\rm basic}},
d \right )\cr}}
\medskip\noindent
{\bf Remarks}:
\itemitem{6.1.8}The local gauge groups, ${\cal G}$, encountered in TFT are
neither finite
dimensional nor compact.
Nevertheless the cohomologies found are remarkably similar to those of
related compact groups.
\itemitem{6.1.9}
If $\widetilde{\cal E} \maprightu{\widetilde \pi} \widetilde {\cal C}$ is a
non-trivial bundle
with standard fiber $\widetilde {\cal V}$, and fiber metric,
$( \cdot, \cdot )_{\widetilde {\cal V}}$, then $\widetilde {\cal E}$ is
associated
to a principal $SO ( \widetilde {\cal V} )$ bundle, $\widetilde {\cal F}$, the
bundle of all orthonormal frames on $\widetilde {\cal E}$:
\eqn\SOVAssoc{
\widetilde {\cal E}
{}~=~ \widetilde {\cal F} \times_{SO ( \widetilde {\cal V} )}
\widetilde {\cal V}}
It is then convenient to express the cohomology of $\widetilde {\cal E}$ in
terms of the basic cohomology of
$\widetilde {\cal F} \times \widetilde {\cal V}$, i.e. from Theorem 6.1.7
\eqn\SOVCoho{\eqalign{
H^\bullet \left ( \Omega ( \widetilde {\cal E} ), d \right )
{}~\eqsim&~
H^\bullet \left ( \Omega ( \widetilde {\cal F} \times_{SO ( \widetilde {\cal V}
)}
\widetilde {\cal V} ), d \right )\cr
{}~\eqsim&~ H^\bullet \left ( \Omega  ( \widetilde {\cal F}
\times \widetilde {\cal V} )_{so ( \widetilde {\cal V} )-{\rm basic}},
d \right )\cr}}
Note also that in this case the ${\cal G}$-equivariant cohomology of
$\widetilde {\cal E}$ is given by
\eqn\GSOVCoho{
H_{{\cal G}, {\rm top}}^\bullet ( \widetilde {\cal E} )
\quad=\quad H^\bullet \left (
\Omega ( \widetilde {\cal F} \times \widetilde {\cal V}
\times E {\cal G} )_{\lieg \oplus so ( \widetilde {\cal V} )-{\rm basic}},
d \right )}
so that effectively there are {\it two} gauge groups $SO ( \widetilde {\cal V}
)$
and ${\cal G}$.

\subsec{Algebraic Description of Equivariant Cohomology}

There are also algebraic models for the ${\cal G}$-equivariant
cohomology of $\widetilde {\cal E}$.
These are far from unique, a fact which partially accounts for the
abundance of TFT for a given moduli space.
For reasons of brevity we shall direct most of our attention to the
{\it Cartan model}.
As was shown by Kalkman\refs{\kalkman}, this model is closely related
to the BRST model which is often used in physics.
We shall discuss other models only briefly; we refer the interested
reader to \refs{\kalkman,\CMRLH,\MatQui} for a fuller description
of the various algebraic models and their equivalence to one another.

The Cartan model proceeds from the complex
\eqn\CartanComplex{
S^\bullet ( \lieg^\ast ) \otimes \Omega^\bullet ( \widetilde {\cal E} ).}
where $\lieg^\ast$ is the dual to $\lieg$ and $S^\bullet ( \lieg^\ast )$
is the symmetric algebra on $\lieg^\ast$, which is freely generated by
$\{ \phi^i \}_{i = 1, \ldots, \dim\ \lieg }$.
A differential, $d_{\cal C}$, may be defined via its action on
the generators of the complex
\eqn\CartanDiffl{\eqalign{
d_{\cal C} \phi^i ~=&~ 0
\qquad\qquad\qquad\qquad\qquad\qquad\qquad\quad\forall \phi^i \in S^2 (
\lieg^\ast )\cr
d_{\cal C} \omega ~=&~ (1 \otimes d - \phi^i \otimes i ( T_i ) ) \omega
\qquad\qquad\qquad\quad\forall\omega\in\Omega^\bullet ( \widetilde {\cal C}
)\cr}}
In analogy to the geometric Lie derivative, one may define an algebraic Lie
derivative on
$S ( \lieg^\ast ) \otimes \Omega^\bullet ( \widetilde {\cal E} )$ to be
$L_i \eqdef [ 1 \otimes i ( T_i ), d_{\cal C} ]_+$.
Note that $d_{\cal C}^2 = - \phi^i \otimes {\cal L} ( T_i )
= \phi^i L_i \otimes 1$; so that $d_{\cal C}$ is nilpotent only on the
subcomplex of {\it equivariant differential forms}, defined as
\eqn\EquivarDF{
\Omega_{\cal G} ( \widetilde {\cal E} )
{}~\eqdef~ \left ( S^\bullet ( \lieg^\ast ) \otimes
\Omega^\bullet ( \widetilde {\cal E} ) \right )^{\cal G}}
where the superscript $( \cdot )^{\cal G}$ denotes the ${\cal G}$-invariant
subcomplex.
That is $\eta \in S ( \lieg^\ast ) \otimes \Omega^\bullet ( \widetilde {\cal E}
)$
is an equivariant differential form iff $\eta \in \bigcap_{i=1}^{\dim\ {\cal
G}} \ker\ L_i$.
This subcomplex corresponds to the basic subcomplex\refs{\kalkman,\CMRLH}.
This motivates the following
\medskip\noindent
{\bf Definition 6.2.1}:
The algebraic definition of the ${\cal G}$-equivariant cohomology of
$\widetilde {\cal E}$ is
\eqn\ECAlgDefi{
H_{{\cal G}, {\rm alg}}^\bullet ( \widetilde {\cal E} )
{}~\eqdef~ H^\bullet \left ( \Omega_{\cal G} ( \widetilde {\cal E} ), d_{\cal
C} \right )}
\medskip\noindent
{\bf Theorem 6.2.2}:
For ${\cal G}$ finite dimensional and compact:
\eqn\ECEquiv{
H_{{\cal G}, {\rm top}}^\bullet ( \widetilde {\cal E} )
{}~\eqsim~ H_{{\cal G}, {\rm alg}}^\bullet ( \widetilde {\cal E} )}
\medskip\noindent
{\it Proof}: Please see \refs{\MatQui,\kalkman,\CMRLH}.
\medskip\noindent
The relationship between the de Rham and Cartan models can be made very
concrete.
Given a connection, $\nabla$, and its curvature, $F_\nabla$, we may
define the Chern-Weil homomorphism\foot{The subscript ${\cal C}$
indicates that this homomorphism acts on the {\it Cartan} complex.
We will in section 6.3.2 define an analogous homomorphism for the
Weil model.}, $w_{{\cal C}, \nabla}$:
\eqn\CWHom{\eqalign{
w_{{\cal C}, \nabla}\colon
\Omega_{\cal G}^\bullet ( \widetilde {\cal E} ) ~\longrightarrow&~
\Omega^\bullet ( \widetilde {\cal E} )_{\lieg-{\rm basic}}\cr
w_{{\cal C}, \nabla}\colon {\cal P} ( \phi )
{}~\longmapsto&~ \left ( {\cal P} ( F_\nabla ) \right )^{\rm hor}\cr}}
The superscript $( \cdots )^{\rm hor}$ indicates projection onto the
horizontal subcomplex via $\nabla$.

\subsec{The Mathai-Quillen Representative of the Thom Class}
\par\noindent
The Thom class is central to the construction of TFT actions\refs{\AJ}.
In this subsection we shall outline constructions of representatives of this
class in the context of various models of equivariant cohomology.

Let $\widetilde {\cal E}$ be an orientable vector bundle
$$\matrix{
\widetilde {\cal E} & \mapleftu{} & \widetilde {\cal V}\cr
\mapdown{\widetilde \pi} & & \cr
\widetilde {\cal C} & & \cr}
$$
We consider $H_{\rm vrd}^\bullet ( \widetilde {\cal E} )$, the cohomology
of forms that are rapidly decreasing in the vertical direction\foot{This is
equivalent to $H_{\rm cv}^\bullet ( \widetilde {\cal E} )$,
the cohomology of forms that are compactly supported along the (vertical)
fiber direction\refs{\MatQui}.}.
On such forms integration along the fiber is well-defined:
\eqn\Pushdownpit{
\widetilde \pi_\ast\colon
\Omega^{\bullet + \rank\ \widetilde {\cal E}}_{\rm vrd}
( \widetilde {\cal E} )\ \to\ \Omega^\bullet ( \widetilde {\cal C} )}
In fact, the cohomologies of these two complexes are isomorphic.
\medskip\noindent
{\bf Theorem 6.3.1}: [{\it Thom Isomorphism}] Integration along the fiber
defines an isomorphism
\eqn\PushdownIso{
\widetilde \pi_\ast\colon
H^{\bullet + \rank\ \widetilde {\cal E}}_{\rm vrd}
( \widetilde {\cal E} )\quad \eqsim\quad H^\bullet ( \widetilde {\cal C} )}
\medskip\noindent
{\it Proof}: Please see Bott and Tu\refs{\BottTu}.
\medskip\noindent
The {\it Thom class}, of $\widetilde {\cal E}$ is defined as
\eqn\ThomClass{
\left [ \Phi ( \widetilde {\cal E} ) \right ]\ \eqdef\
( \widetilde \pi_\ast )^{-1} ( 1 )\quad\in\quad
H^{\rank\ \widetilde {\cal E}}_{\rm vrd} ( \widetilde {\cal E} )}
A {\it Thom form}, $\Phi_\nabla ( \widetilde {\cal E} )$, i.e. a particular
representative of the Thom class, will in general depend on a
connection, $\nabla$, on $\widetilde {\cal E}$.
In terms of such a $\Phi_\nabla ( \widetilde {\cal E} )$, the Thom isomorphism
is explicitly given by:
$$\eqalign{
{\cal T}\colon H^\bullet ( \widetilde {\cal C} )
{}~\longrightarrow&~ H^{\bullet + \rank\ \widetilde {\cal E}}_{\rm vrd}
( \widetilde {\cal E} )\cr
{\cal T} ( \omega )
{}~\longmapsto&~ \widetilde \pi^\ast ( \omega ) \wedge
\Phi_\nabla ( \widetilde {\cal E} )\cr}
$$

{}From remark 6.1.9 we know that $\widetilde {\cal E}$ is associated to a
principal $SO ( \widetilde {\cal V} )$ bundle, $\widetilde {\cal F}$.
A Thom form may therefore be constructed in the context of an algebraic
model of $SO ( \widetilde {\cal V} )$-equivariant cohomology.
An element\foot{The subscript ${\cal C}-SO ( \widetilde {\cal V} )$
indicates that we are working within the Cartan model for
$SO ( \widetilde {\cal V} )$.} $U_{{\cal C}-SO ( \widetilde {\cal V} )}$
is called a {\it universal Thom form}, if it is related via the Chern-Weil
homomorphism to a Thom form.
The relevant complexes fit together as indicated in the diagram below:
$$\matrix{
\left ( S ( so ( \widetilde {\cal V} )^\ast ) \otimes
\Omega^\bullet ( \widetilde {\cal V} ) \right )^{SO ( \widetilde {\cal V} )} &
\maprightu{w_{{\cal C}, \nabla}} &
\Omega^\bullet \left ( \widetilde {\cal V} \times E\ SO ( \widetilde {\cal V})
\right )_{so ( \widetilde {\cal V} )-{\rm basic}}\cr
 & \mapse{\bar w_{{\cal C}, \nabla}} &
\mapup{\widetilde \pi_{SO ( \widetilde {\cal V} )}^\ast}\cr
& & \Omega^\bullet \left ( \widetilde {\cal V}
\times_{SO ( \widetilde {\cal V} )} E\ SO ( \widetilde {\cal V} ) \right )\cr}
$$
\medskip
\centerline{\bf Diagram 6.3.2}
\medskip\noindent
and the various forms are related as follows
$$\matrix{
U_{{\cal C}-SO ( \widetilde {\cal V} )} &
\maprightu{w_{{\cal C}, \nabla}} &
w_{{\cal C}, \nabla} ( U_{{\cal C}-SO ( \widetilde {\cal V} )} )
= \widetilde\pi_{SO ( \widetilde {\cal V} )}^\ast
( \Phi_\nabla ( \widetilde {\cal E} ))\cr
  & \mapse{\bar w_{{\cal C}, \nabla}} &
\mapup{\widetilde\pi_{SO ( \widetilde {\cal V} )}^\ast}\cr
 & & \Phi_\nabla ( \widetilde {\cal E} )
= \bar w_{{\cal C}, \nabla}
\left ( U_{{\cal C}-SO ( \widetilde {\cal V} )} \right )\cr}
$$
\medskip
\centerline{\bf Diagram 6.3.3}
\medskip
Mathai and Quillen\refs{\MatQui} constructed an explicit representative of
the universal Thom class.
Let $x^i$ be orthonormal coordinates for $\widetilde {\cal V}$ and
introduce anticommuting orthonormal coordinates, $\rho_i$, for
$\Pi \widetilde {\cal V}^\ast$.
Here $\Pi$ is the parity change functor\refs{\Man} and indicates that we
are to regard the coordinates of $\Pi \widetilde {\cal V}^\ast$ as being
anti-commuting.
Then a representative of the universal Thom form may be written as
\eqn\CartanRep{
U_{{\cal C}-SO ( \widetilde {\cal V} )}
{}~=~ \left ( {1 \over {4 \pi t}} \right )^{{1 \over 2} \rank\ \widetilde {\cal
E}}
\int_{\Pi \widetilde {\cal V}^\ast} d \rho~ \exp\ \left \{
-{1 \over {4 t}} ( x, x )_{\widetilde {\scriptscriptstyle \cal V}}
+ i \langle \rho, d x \rangle
+ t ( \rho, \phi \rho )_{\scriptscriptstyle \widetilde {\cal V}^\ast} \right
\}}
where $( \cdot, \cdot )_X$ denotes the inner product on $X$, while
$\langle \cdot, \cdot \rangle$ denotes the dual pairing.
Note that we actually have a one parameter family of  representatives
that depend on $t \in \IR$.
\medskip\noindent
{\bf Remarks}:
\itemitem{6.3.4.}
We are often interested in constructing the Thom class of
$\widetilde {\cal E} \times_{\cal G} E {\cal G}
\to \widetilde {\cal C} \times_{\cal G} E {\cal G}$.
{}From Remark 6.1.9 we know that in this case we need to consider
$SO ( \widetilde {\cal V} ) \times {\cal G}$-equivariant cohomology.
In the following we shall construct a number of other universal Thom forms;
we shall leave the total local symmetry group inspecific, to allow for the
possibility of non-trivial ${\cal G}$.
\itemitem{6.3.5.}
Mathai-Quillen representatives of Thom classes play a central role in the
construction of topological field theories\refs{\AJ}.
The argument of the exponential in the Mathai-Quillen representative
is interpreted as the action of a TFT.
For example, the term $\langle \rho, d x\rangle$ is the kinetic term
for the ghost/anti-ghost system.
\itemitem{6.3.6.}
$\ker\ \nabla s$ are the ghost zero modes, while $\coker\ \nabla s$
are the anti-ghost zero modes.
When $\coker\ \nabla s \not = 0$, the Grassmann integral over the
anti-ghosts in the Mathai-Quillen representative of the Thom class
brings down powers of the curvature from the argument of the exponential.
This fact will be made precise in the localization theorems.
\medskip\noindent
The main use of Thom classes stems from the following
\medskip\noindent
{\bf Proposition 6.3.7}:
The pullback of the Thom class, $\Phi_\nabla ( \widetilde {\cal E} )$,
by {\it any} section,
$\widetilde s\colon \widetilde {\cal C} \to \widetilde {\cal E}$,
is the Euler class of $\widetilde {\cal E}$.
\medskip\noindent
{\it Proof}: Please see \refs{\BottTu,\CMRLH}.
\medskip\noindent
We conclude this subsection by briefly indicating a few other representatives
of the Thom class.
Please see \refs{\CMRLH} for more details.

\subsubsec{Another Representative of the Universal Thom Class in the Cartan
Model}

If we introduce further commuting coordinates\foot{Notation:
Conventions and the paucity of alphabets force us to use $\pi$
for both the Lagrange multiplier fields and the projections.
Projections between (in general) infinite dimensional spaces will
have a tilde; projections between (in general) finite dimensional spaces
have a bar.
Lagrange multiplier fields have neither accent.}, $\pi$ for
$\widetilde {\cal V}^\ast$, and extend the Cartan differential
to $S^\bullet ( \lieg^\ast )
\otimes \Omega^\bullet ( \widetilde {\cal V} )
\otimes \Omega^\bullet ( \Pi \widetilde {\cal V}^\ast )$, via
\eqn\QCDef{\eqalign{
Q_{\cal C} ~=&~ 1 \otimes d \otimes 1 + 1 \otimes 1 \otimes d
- \phi^i \otimes i ( T_i ) \otimes 1 - \phi^i \otimes 1 \otimes i ( T_i )\cr
Q_{\cal C} \pmatrix{\rho \cr \pi \cr}
{}~=&~ \pmatrix{ 0 & 1\cr -\phi_i \otimes {\cal L} ( T_i ) & 0 \cr}
\pmatrix{\rho \cr \pi \cr}\cr}}
then we may compactly write
\eqn\QCExactRep{
U_{{\cal C}-{\cal G}}
{}~=~ \left ( {1 \over {2 \pi}} \right )^{\rank\ {\cal E}}
\int_{\widetilde {\cal V}^\ast \times \Pi \widetilde {\cal V}^\ast}
d \pi~ d \rho~ \exp - Q_{\cal C} \left (
- i \left \langle \rho, x \right \rangle
- t ( \rho, \pi )_{\widetilde {\scriptstyle {\cal V}}^\ast} \right )}
The significant feature of this formulation is that the argument of the
exponential (the TFT action) is $Q_{\cal C}$-exact.

\subsubsec{A Representative of the Universal Thom Class in the Weil Model}

The Weil model of the ${\cal G}$-equivariant cohomology of
$\widetilde {\cal E}$ starts from the complex:
$$
{\cal W} ( \lieg )
{}~=~ \Lambda ( \lieg^\ast ) \otimes S ( \lieg^\ast )
$$
where $\Lambda ( \lieg^\ast )$ is the exterior algebra of $\lieg^\ast$
which is freely generated by $\{ \theta^i \}_{i=1,\ldots,\dim\ {\cal G}}$.
The differential of the Weil complex need not concern us here.
A fuller discussion may be found in \refs{\CMRLH,\kalkman,\MatQui}.

The universal Thom form within the Weil model is by:
\eqn\WeilRep{
U_{{\cal W}-{\cal G}}
{}~=~ {1 \over {\pi^{{1 \over 2} \rank\ \widetilde {\cal E}}}}\quad
e^{- ( x, x )_{{\scriptscriptstyle \widetilde {\cal V}}}}\quad
\int_{\Pi \widetilde {\cal V}^\ast} d \rho~
\exp \left \{ {1 \over 4} ( \rho, \phi \rho )_{{\scriptscriptstyle \widetilde
{\cal V}^\ast}}
+ i \langle \nabla x, \rho \rangle \right \}}
where $\nabla x = d x + \theta \cdot x$.

In the context of the Weil model, the Chern-Weil homomorphsim,
$w_{{\cal W}, \nabla}$, simply makes the replacement
$\pmatrix{ \theta\cr \phi\cr} \to \pmatrix{ A\cr F_A\cr}$, where
$A$ is a connection on $\widetilde {\cal E}$ and $F_A$ is its curvature.
The absence of horizontal projection is often convenient.
On the other hand the Chern-Weil homomorphism explicitly introduces a
connection which in many situations is non-local.
In these cases the Weil model is unsuitable for the construction of TFT
actions.

\subsubsec{A Representative of the Universal Thom Class in Hybrid Cartan
and Weil Models}

To construct a universal Thom form on
$\widetilde {\cal E} \times E {\cal G} \to
\widetilde {\cal E} \times_{\cal G} E {\cal G}$, we know from Remark
6.1.9 that it is useful to work within
$SO ( \widetilde {\cal V} ) \times {\cal G}$-equivariant cohomology.
It is useful to use different algebraic models for the equivariant
cohomologies of these groups.
Since the $SO ( \widetilde {\cal V} )$ connection is local, it is convenient
to work in the context of the Weil model, thereby explicitly introducing the
connection, but obviating the horizontal projection.
On the other hand, the ${\cal G}$-connection is generally non-local, so that
for purposes of constructing a TFT action, it is essential that we work
in the context of the Cartan model for ${\cal G}$.
The diagram depicting the interrelation of the various complexes is the
natural generalization of Diagram 6.3.2.
\overfullrule=0pt
\font\tenrmres=cmr10 scaled700
\font\sevenrmres=cmr7 scaled700
\font\fivermres=cmr5 scaled700
\font\tenires=cmmi10 scaled700
\font\sevenires=cmmi7 scaled700
\font\fiveires=cmmi5 scaled700
\font\tensyres=cmsy10 scaled700
\font\sevensyres=cmsy7 scaled700
\font\fivesyres=cmsy5 scaled700
\font\tenitres=cmti10 scaled700

\font\tenbfres=cmbx10 scaled700

\textfont0=\tenrmres \scriptfont0=\sevenrmres \scriptscriptfont0=\fivermres
\def\rm{\fam0 \tenrmres}
\textfont1=\tenires \scriptfont1=\sevenires \scriptscriptfont1=\fiveires
 
\textfont2=\tensyres \scriptfont2=\sevensyres \scriptscriptfont2=\fivesyres
\def\cal{\fam2}
\newfam\itfamres \def\it{\fam\itfamres\tenitres} \textfont\itfamres=\tenitres
\newfam\bffamres \def\bf{\fam\bffamres\tenbfres} \textfont\bffamres=\tenbfres
\medskip\noindent
$$\matrix{
\left ( S ( \underline{g}^\ast ) \otimes
\left ( {\cal W} ( so ( \tilde {\cal V} )) \otimes \Omega ( \tilde {\cal V} )
\right )_{so ( \tilde {\cal V} )-{\rm basic}} \right )^{\cal G} &
\maprightu{w_{{\cal W}, \nabla_{SO ( \tilde {\cal V})}}} &
\left ( S ( \underline{g}^\ast ) \otimes
\Omega ( \tilde {\cal F} \times \tilde {\cal V} )_{so ( \tilde {\cal V} )-{\rm
basic}}
\right )^{\cal G} & & \cr
 & \mapse{\bar w_{{\cal W}, \nabla_{SO ( \tilde {\cal V} )}}} &
\mapup{\tilde \pi_{SO ( \tilde {\cal V} )}^\ast} & & \cr
 & & \left ( S ( \underline{g}^\ast ) \otimes \Omega ( \tilde {\cal E} ) \right
)^{\cal G} &
\maprightu{w_{{\cal C}, \nabla_{\cal G}}} &
\Omega ( \tilde {\cal E} \times E {\cal G} )_{\underline{g}-{\rm basic}}\cr
 & & & \mapse{\bar w_{{\cal C}, \nabla_{\cal G}}} &
\mapup{\tilde \pi_{\cal G}^\ast} \cr
 & & & & \Omega ( \tilde {\cal E} \times_{\cal G} E {\cal G} )\cr}
$$
\medskip\noindent
\font\tenrmnorm=cmr10
\font\sevenrmnorm=cmr7
\font\fivermnorm=cmr5
\font\teninorm=cmmi10
\font\seveninorm=cmmi7
\font\fiveinorm=cmmi5
\font\tensynorm=cmsy10
\font\sevensynorm=cmsy7
\font\fivesynorm=cmsy5
\font\tenitnorm=cmti10

\font\tenbfnorm=cmbx10

\textfont0=\tenrmnorm \scriptfont0=\sevenrmnorm \scriptscriptfont0=\fivermnorm
\def\rm{\fam0 \tenrmnorm}
\textfont1=\teninorm \scriptfont1=\seveninorm \scriptscriptfont1=\fiveinorm
 
\textfont2=\tensynorm \scriptfont2=\sevensynorm \scriptscriptfont2=\fivesynorm
\def\cal{\fam2}
\newfam\itfamnorm \def\it{\fam\itfamnorm\tenitnorm}
\textfont\itfamnorm=\tenitnorm
\newfam\bffamnorm \def\bf{\fam\bffamnorm\tenbfnorm}
\textfont\bffamnorm=\tenbfnorm
\medskip
\centerline{\bf Diagram 6.3.8}
\medskip
The universal Thom form,
$U_{{\cal W}-SO ( \tilde {\cal V} ),{\cal C}-{\cal G}}$,
takes its values in the complex $(S ( \lieg^\ast ) \otimes
( {\cal W} ( so ( \widetilde {\cal V} ))
\otimes \Omega ( \widetilde {\cal V} )
)_{so ( \tilde {\cal V} )-{\rm basic}} )^{\cal G}$.
Applying the Chern-Weil homomorphism we obtain
$$
w_{{\cal W}, \nabla_{SO ( \tilde {\cal V})}} \left (
U_{{\cal W}-SO ( \tilde {\cal V} ),{\cal C}-{\cal G}} \right )
= \tilde \pi_{SO ( \widetilde {\cal V} )}^\ast \left (
\Upsilon_{{\cal C}-{\cal G}} \right )
$$
where $\Upsilon_{{\cal C}-{\cal G}} \in \Omega_{\cal G} ( {\cal E} )$
will play an important role in the general localization theorem
(Proposition 6.5.4).
Altogether, we have
\textfont0=\tenrmres \scriptfont0=\sevenrmres \scriptscriptfont0=\fivermres
\def\rm{\fam0 \tenrmres}
\textfont1=\tenires \scriptfont1=\sevenires \scriptscriptfont1=\fiveires
 
\textfont2=\tensyres \scriptfont2=\sevensyres \scriptscriptfont2=\fivesyres
\def\cal{\fam2}
\def\it{\fam\itfamres\tenitres} \textfont\itfamres=\tenitres
\def\bf{\fam\bffamres\tenbfres} \textfont\bffamres=\tenbfres
\medskip\noindent
$$\matrix{
U_{{\cal W}-SO ( \tilde {\cal V} ),{\cal C}-{\cal G}} &
\maprightu{w_{{\cal W}, \nabla_{SO ( \tilde {\cal V})}}} &
w_{{\cal W}, \nabla_{SO ( \tilde {\cal V})}} \left (
U_{{\cal W}-SO ( \tilde {\cal V} ),{\cal C}-{\cal G}} \right )
= \tilde \pi_{SO ( \tilde {\cal V} )}^\ast \left (
\Upsilon_{{\cal C}-{\cal G}} \right )
 & & \cr
 & \mapse{\bar w_{{\cal W}, \nabla_{SO ( \tilde {\cal V} )}}} &
\mapup{\tilde \pi_{SO ( \tilde {\cal V} )}^\ast} & & \cr
 & & \Upsilon_{{\cal C}-{\cal G}} &
\maprightu{w_{{\cal C}, \nabla_{\cal G}}} &
w_{{\cal C}, \nabla_{\cal G}} \left (
\Upsilon_{{\cal C}-{\cal G}} \right )\cr
 & & & \mapse{\bar w_{{\cal C}, \nabla_{\cal G}}} &
\mapup{\tilde \pi_{\cal G}^\ast} \cr
 & & & & \Phi_\nabla ( \tilde {\cal E} \times_{\cal G} E {\cal G} )\cr}
$$
\medskip\noindent
\textfont0=\tenrmnorm \scriptfont0=\sevenrmnorm \scriptscriptfont0=\fivermnorm
\def\rm{\fam0 \tenrmnorm}
\textfont1=\teninorm \scriptfont1=\seveninorm \scriptscriptfont1=\fiveinorm
 
\textfont2=\tensynorm \scriptfont2=\sevensynorm \scriptscriptfont2=\fivesynorm
\def\cal{\fam2}
\def\it{\fam\itfamnorm\tenitnorm} \textfont\itfamnorm=\tenitnorm
\def\bf{\fam\bffamnorm\tenbfnorm} \textfont\bffamnorm=\tenbfnorm
\medskip
\centerline{\bf Diagram 6.3.9}
\medskip
Explicitly
\eqn\UpsilonDef{
\Upsilon_{{\cal C}-{\cal G}}
{}~=~ \int_{\tilde {\cal V}^\ast \times \Pi \tilde {\cal V}^\ast}
d \pi\ d \rho\ \exp\ -Q_{\cal C} \Psi_{\rm Loc}}
where $Q_{\cal C}$ is the Cartan differential for ${\cal G}$-equivariant
cohomology of $\widetilde {\cal E}$, analogous to \QCDef; and where
\eqn\PsiLocDef{
\Psi_{\rm Loc} ~=~
- i \left \langle \rho, x \right \rangle
+ t ( \rho, \Gamma_{SO ( \widetilde {\cal V} )} \cdot \rho )_{\widetilde {\cal
V}^\ast}
-t ( \rho, \pi )_{\widetilde {\cal V}^\ast}}
\subsec{The Localization Formula for trivial ${\cal G}$}

We shall now apply the construction of Thom classes to sketch a localization
formula for the simpler case when ${\cal G}$ is trivial.
Consider an orientable vector bundle
$$\matrix{
\widetilde {\cal E} & \mapleftu{} & \widetilde {\cal V}\cr
\mapdown{\widetilde \pi} & & \cr
\widetilde {\cal C} & & \cr}
$$
Let $\widetilde s\colon \widetilde {\cal C} \to \widetilde {\cal E}$ be a
section.
In terms of a local trivialization, we may write this as
$$
\widetilde s ~=~ ( {\rm id}, s)\qquad{\rm where}\qquad
s\colon \widetilde {\cal C} ~\longrightarrow~ \widetilde {\cal V}
$$
The subspace of interest is characterized by
\eqn\ModDef{
\widetilde {\cal M}
{}~=~ \{ \varphi \in \widetilde {\cal C}\ \vert\ s ( \varphi ) = 0 \}.}
For every $\varphi \in \widetilde {\cal C}$, the differential of $s$
is a map
\eqn\Diff{
d s \vert_\varphi \colon T_\varphi \widetilde {\cal C}
{}~\to~ T_{s ( \varphi )} \widetilde {\cal V}.}
Actually, since $\widetilde {\cal V}$ is a linear space,
$T \widetilde {\cal V}\ \eqsim\ \widetilde {\cal V}$ and we may view
$d s \vert_\varphi$ as a linear operator:
$$
d s \vert_\varphi \colon T_\varphi \widetilde {\cal C} ~\to~ \widetilde {\cal
V}
$$
\par\noindent
Clearly
$$
\ker\ d s\vert_\varphi ~=~ \ker\ \nabla s \vert_\varphi
{}~\subset~ T_\varphi \widetilde {\cal M}\qquad\qquad
\forall\varphi \in \widetilde {\cal M}
$$
Moreover, if $d s \vert_\varphi$ is injective, then\refs{\CMRLH}
\eqn\KerNablaS{
\ker\ \nabla s \vert_\varphi ~\eqsim~ T_\varphi \widetilde {\cal M}
\qquad\qquad\qquad\forall \varphi \in \widetilde {\cal M}.}
It is also clear that
$$
{\rm Im}\ d s \vert_\varphi\ =\ {\rm Im}\ \nabla s \vert_\varphi
\qquad\qquad\qquad
\forall \varphi\in \widetilde {\cal M}
$$
Now consider the exact sequence of bundles over $\widetilde {\cal M}$:
\eqn\ExactBundleSeq{
0 ~\to~ {\rm Im}\ \nabla s ~\to~ \widetilde {\cal E}
{}~\to~ \coker\ \nabla s ~\to~ 0}
\medskip\noindent
{\bf Proposition 6.4.1}:
Let $\widetilde s = ( {\rm id}, s)$ be a section of the orientable vector
bundle $\widetilde {\cal E}$ as above.
Let $i$ denote the inclusion,
$i\colon \widetilde {\cal M} \to \widetilde {\cal C}$.
If $P \subset \widetilde {\cal M}$ is the Poincar\'e dual to
$e ( \coker\ \nabla s \to \widetilde {\cal M} )$,
then $i ( P ) \subset \widetilde {\cal C}$ is Poincar\'e dual to
$e ( \widetilde {\cal E} \to \widetilde {\cal C} )
= \widetilde s^\ast \Phi_\nabla ( \widetilde {\cal E} )$ in
$\widetilde {\cal C}$.
\medskip\noindent
{\it Proof}:
For a physical proof, please see \refs{\Wiag,\Winm}.
\medskip\noindent
It follows from Propositions 6.3.7 and 6.4.1 that
\medskip\noindent
{\bf Proposition 6.4.2}:
For $\widetilde {\cal O} \in H^{\index\ \nabla s} ( \widetilde {\cal C} )$, we
have
\eqn\LocThmi{
\int_{\widetilde {\cal C}}
\widetilde s^\ast \Phi_\nabla ( \widetilde {\cal E} )
\wedge {\cal O}
{}~=~
\int_{\cal M} e ( \coker\ \nabla s \to {\cal M} )
\wedge i^\ast {\cal O}}
\medskip\noindent
where we have used the fact that for trivial ${\cal G}$,
$\widetilde {\cal M} = {\cal M}$.
\medskip\noindent
{\it Proof}:
For a discussion, please see section 11.10.3 of \refs{\CMRLH}.
\medskip\noindent
Proposition 6.4.2 is the basis for the construction of TFTs without
local symmetries.
There are a number of applications of this localization theorem to super
quantum mechanics and topological sigma models.
For a survey of such applications as well as a more extensive list of
references, please see \refs{\CMRLH}.

\subsec{The Localization Formula for Non-trivial ${\cal G}$}

In order to give a description of $YM_2$ as a topological string theory,
we need to consider a more general localization theorem with ${\cal G}$
non-trivial.
Let $\widetilde {\cal E}$ be an orientable ${\cal G}$-equivariant
vector bundle
$$\matrix{
\widetilde {\cal E} & \mapleftu{} & \widetilde {\cal V}\cr
\mapdown{\widetilde \pi} & & \cr
\widetilde{\cal C} & & \cr}
$$
and $\widetilde s\colon \widetilde {\cal C} \to \widetilde {\cal E}$
a ${\cal G}$-equivariant section.
Then $\widetilde s$ induces a section, $\bar s$:
$$\matrix{
\widetilde {\cal E}  \times E {\cal G} & \mapleftu{\widetilde s} &
\widetilde {\cal C}  \times E {\cal G}\cr
\mapdown{\widetilde \pi_{\cal G}} & & \mapdown{\bar \pi_{\cal G}}\cr
\widetilde {\cal E}  \times_{\cal G} E {\cal G} & \mapleftu{\bar s} &
\widetilde {\cal C}  \times_{\cal G} E {\cal G}\cr}
$$
These maps induce pullback maps between the corresponding de Rham
complexes:
$$\matrix{
\Omega^\bullet ( \widetilde {\cal E}  \times E {\cal G} ) &
\maprightu{\widetilde s^\ast} &
\Omega^\bullet  ( \widetilde {\cal C}  \times E {\cal G} )\cr
\mapup{\widetilde \pi_{\cal G}^\ast} & & \mapup{\bar \pi_{\cal G}^\ast}\cr
\Omega^\bullet ( \widetilde {\cal E}  \times_{\cal G} E {\cal G} ) &
\maprightu{\bar s^\ast} &
\Omega ( \widetilde {\cal C}  \times_{\cal G} E {\cal G} )\cr}
$$
Though $\widetilde {\cal E}  \times E {\cal G}
\to \widetilde {\cal E}  \times_{\cal G} E {\cal G}$ is a principal
${\cal G}$-bundle, we may define an analogue of the Thom isomorphism
for vector bundles\refs{\CMRLH}
\eqn\ThomGDef{\eqalign{
{\cal T}_{\cal G}\colon
H^\bullet ( {\cal E} )
{}~\to&~ H^{\bullet + \dim\ {\cal G}} ( \widetilde {\cal E} )\cr
{\cal T}_{\cal G} ( \omega )
{}~\mapsto&~ \widetilde \pi_{\cal G}^\ast ( \omega )
\wedge \Phi_{\cal G} ( \widetilde {\cal E} )\cr}}
where $\Phi_{\cal G} ( \widetilde {\cal E} )
\in \Omega_{\cal G} ( \widetilde {\cal E} )$ is partially characterized
by the fact that for all $x \in \widetilde {\cal E} \times E { \cal G}$,
$R_x^\ast \Phi_{\cal G} ( \widetilde {\cal E} )$ is the normalized
Haar measure of ${\cal G}$.
Please see \refs{\CMRLH} for a fuller discussion.

Now $\widetilde \pi_{\cal G}^\ast ( \omega ) \in
\Omega ( \widetilde {\cal E} \times E {\cal G} )_{\underline{g}-{\rm basic}}$,
so that by Theorem 6.1.7 it is related to $\varpi \in \Omega_{\cal G} (
\widetilde {\cal E} )$
via the Chern-Weil homomorphism:
\eqn\CDTrivial{
w_{{\cal C}, \nabla_{\cal G}} \left ( \varpi \right )
{}~=~ \widetilde \pi_{\cal G}^\ast ( \omega ).}
We may then define a map
\eqn\SGDef{\eqalign{
{\cal S}_{\cal G}\colon \Omega^\bullet_{\cal G} ( \widetilde {\cal E} )
{}~\to&~ \Omega^\bullet ( \widetilde {\cal E} \times E {\cal G} )\cr
{\cal S}_{\cal G} ( \varpi )
{}~\mapsto&~ w_{{\cal C}, \nabla_{\cal G}} ( \varpi )
\wedge \Phi_{\cal G}\cr}}
The virtue of this map is that it may be readily interpreted in the context
of a TFT.

Introduce $\lambda_a$ and $\eta_a$ as generators of $S ( \lieg )$ and
$\Lambda ( \lieg )$, respectively and extend the action of the Cartan
differential on these generators as follows:
\eqn\ExtendQC{
Q_{\cal C} \pmatrix{\lambda\cr \eta\cr}
{}~=~ \pmatrix{ 0 & 1\cr
- \phi^i \otimes {\cal L} ( T_i ) & 0\cr} \pmatrix{\lambda\cr \eta\cr}}
Then we have the following
\medskip\noindent
{\bf Proposition 6.5.1}:
For all $\varpi \in \Omega_{\cal G}^\bullet ( \widetilde {\cal E} )$,
\eqn\SGDef{
{\cal S}_{\cal G} ( \varpi )
{}~=~ \int d \phi\ \left \{
\varpi \wedge \left ( {1 \over {2 \pi i}} \right )^{\dim\ {\cal G}}
\int_{\underline{g} \times \Pi \underline{g}} d \lambda~ d \eta~
\exp\ - Q_{\cal C} \Psi_{\rm Proj} \right \}}
where
\eqn\PsiProjDef{
\Psi_{\rm Proj} = -i ( \lambda, C^\dagger )_{\underline{g}}}
\medskip\noindent
{\it Proof}:
Please see \refs{\CMRLH}.
\medskip\noindent
{\bf Remarks}:
\itemitem{6.5.2}
We have used the metric on $\widetilde {\cal E}$ to define the adjoint,
$C_x^\dagger$ for all $x \in \widetilde {\cal E} \times E {\cal G}$:
$$\eqalign{
C_x\colon \lieg ~\longrightarrow&~ T_x \widetilde {\cal E}\cr
C^\dagger_x\colon T_x \widetilde {\cal E}
{}~\longrightarrow&~ \lieg\cr}
$$
We may view $C^\dagger$ as a $\lieg$-valued 1-form and so
$( \lambda, C^\dagger )_{\underline{g}} \in T^\ast_x \widetilde {\cal E}$.
\itemitem{6.5.3}
This procedure is distinct from Faddeev-Popov gauge fixing.
Note that no section of $\widetilde {\cal E} \times E {\cal G}
\to \widetilde {\cal E} \times_{\cal G} E {\cal G}$ enters into our
discussion.
For a more careful comparison, please see \refs{\CMRLH}.
\medskip\noindent
Again it is useful to depict the interrelation of the various complexes and
maps diagramatically:
\medskip\noindent
$$\matrix{
\left ( S ( g^\ast ) \otimes \Omega ( \tilde {\cal E} ) \right )^{\cal G} &
\maprightu{{\cal S}_{\cal G}} &
\Omega ( \tilde {\cal E} \times E {\cal G} ) &
\maprightu{\tilde s^\ast}&
\Omega ( \tilde {\cal C} \times E {\cal G} ) \cr
 & \mapse{\bar w_{\nabla_{\cal G}}} &
\mapup{{\cal T}_{\cal G}} & & \mapdown{( \bar \pi_{\cal G} )_\ast}\cr
 & & \Omega ( \tilde {\cal E} \times_{\cal G} E {\cal G} ) &
\maprightd{\bar s^\ast} &
\Omega ( \tilde {\cal C} \times_{\cal G} E {\cal G} ) \cr}
$$
\medskip
\centerline{\bf Diagram 6.5.4}
\medskip\noindent
The object
$\Upsilon_{\nabla_{SO ( \tilde {\cal V} )}, {\cal C}-{\cal G}}
\in \Omega_{\cal G} ( \widetilde {\cal E} )$
which we constructed in \UpsilonDef\ is related to the equivariant
Thom class,
$\Phi_\nabla ( \widetilde {\cal E} \times_{\cal G} E {\cal G} )$,
via the following diagram
\medskip\noindent
$$\matrix{
\Upsilon_{\nabla_{SO ( \tilde {\cal V} )}, {\cal C}-{\cal G}}
& \maprightu{{\cal S}_{\cal G}} &
{\cal S}_{\cal G} \left ( \Upsilon_{\nabla_{SO ( \tilde {\cal V} )}, {\cal
C}-{\cal G}} \right ) &
\maprightu{\tilde s^\ast} &
\tilde s^\ast {\cal S}_{\cal G} \left (
\Upsilon_{\nabla_{SO ( \tilde {\cal V} )}, {\cal C}-{\cal G}}  \right )\cr
 & &
= {\cal T}_{\cal G} \left (
\Phi ( \tilde {\cal E} \times_{\cal G} E {\cal G} ) \right )\cr
 & \mapse{\bar w_{{\cal C}, \nabla_{\cal G}}} &
\mapup{{\cal T}_{\cal G}} & & \mapdown{( \bar \pi_{\cal G} )_\ast}\cr
 & &
\Phi ( \tilde {\cal E} \times_{\cal G} E {\cal G} ) &
\maprightd{\bar s^\ast} &
\bar s^\ast \Phi ( \tilde {\cal E} \times_{\cal G} E {\cal G} )\cr
 & &
= \bar w_{{\cal C}, \nabla_{\cal G}} \left (
\Upsilon_{\nabla_{SO ( \tilde {\cal V} )}, {\cal C}-{\cal G}} \right )
& &
= ( \bar \pi_{\cal G} )_\ast \widetilde s^\ast {\cal S}_{\cal G} \left (
\Upsilon_{\nabla_{SO ( \tilde {\cal V} )}, {\cal C}-{\cal G}} \right )\cr}
$$
\medskip\noindent
\textfont0=\tenrmnorm \scriptfont0=\sevenrmnorm \scriptscriptfont0=\fivermnorm
\def\rm{\fam0 \tenrmnorm}
\textfont1=\teninorm \scriptfont1=\seveninorm \scriptscriptfont1=\fiveinorm
 
\textfont2=\tensynorm \scriptfont2=\sevensynorm \scriptscriptfont2=\fivesynorm
\def\cal{\fam2}
\def\it{\fam\itfamnorm\tenitnorm} \textfont\itfamnorm=\tenitnorm
\def\bf{\fam\bffamnorm\tenbfnorm} \textfont\bffamnorm=\tenbfnorm
\medskip
\centerline{\bf Diagram 6.5.5}
\medskip\noindent
{}From this it is apparent that
\eqn\ProjFermion{
\bar s^\ast \Phi ( \widetilde {\cal E} \times_{\cal G} E {\cal G} )
{}~=~ \widetilde s^\ast {\cal S}_{\cal G} \left (
\Upsilon_{\nabla_{SO ( \tilde {\cal V} )}, {\cal C}-{\cal G}} \right )}
where $\Upsilon_{\nabla_{SO ( \tilde {\cal V} )}, {\cal C}-{\cal G}}$
is given by \UpsilonDef.
\medskip\noindent
We shall assume that ${\cal G}$ acts freely on $\widetilde {\cal E}$
and $\widetilde {\cal C}$:
$$\eqalign{
{\cal E}\ =&\ \widetilde {\cal E} / {\cal G}\cr
{\cal C}\ =&\ \widetilde {\cal C} / {\cal G}\cr}
$$
in order that ${\cal M}$ be a manifold\foot{Hurwitz space is, in fact, not
smooth but possesses orbifold singularities.
In this case, we actually compute orbifold Euler characters.
For more singular spaces, we do not know a general prescription.}.
Then we know from Proposition 6.4.2 that for
${\cal O} \in H^{\index \nabla \bar s} ( {\cal C} )$,
\eqn\Eqi{
\int_{\cal C}
\bar s^\ast \Phi ( \widetilde {\cal E} \times_{\cal G} E {\cal G} )
\wedge {\cal O}
{}~=~ \int_{\cal M}
e ( \coker\ \nabla \bar s \to {\cal M} ) \wedge i^\ast {\cal O}}
and from \ProjFermion
\eqn\Eqii{
\int_{\cal C}
\bar s^\ast \Phi ( \widetilde {\cal E} \times_{\cal G} E {\cal G} )
\wedge {\cal O}
{}~=~ \int_{\widetilde {\cal C}} \widetilde s^\ast\
{\cal S}_{\cal G} \left ( \Upsilon_{{\cal C}-{\cal G}} \right )
\wedge \widetilde \pi_{\cal G}^\ast {\cal O}}
Combining \Eqi\ and \Eqii\ we arrive at the following
\medskip\noindent
{\bf Proposition 6.5.6}:
For ${\cal O} \in H^{\index \nabla \bar s} ( {\cal C} )$,
\eqn\GenLocThm{
\int_{\widetilde {\cal C}} {\cal S}_{\cal G} \left (
\Upsilon_{\nabla_{SO ( \tilde {\cal V} )}, {\cal C}-{\cal G}} \right )
\wedge \widetilde \pi_{\cal G}^\ast {\cal O}
{}~=~ \int_{\cal M} e ( \coker\ \nabla \bar s \to {\cal M} )
\wedge i^\ast {\cal O}}
\medskip\noindent
{\bf Remarks}:
\itemitem{6.5.7.}
If we introduce a supermanifold $\widehat {\cal C}$, whose
odd coordinates are generated from the fibers of
$T^\ast \widetilde {\cal C}$, then
$$
{\cal C}^\infty ( \widehat {\cal C} )
{}~\eqsim~ \Omega^\bullet ( \widetilde {\cal C} ).
$$
On the other hand, $\widehat{\cal C}$ has a natural measure
$$
\widehat \mu
{}~=~ d \varphi^1 \wedge \cdots \wedge d \varphi^n
d \psi^1 \wedge \cdots \wedge d \psi^n
$$
where $( \varphi_i, \psi_i )$ are local coordinates on
$\widehat {\cal C}$.
If $\widehat \omega \in
{\cal C}^\infty ( \widehat {\cal C} )$ corresponds to the differential
form $\omega \in \Omega^\bullet ( \widetilde {\cal C} )$, then
$$
\int_{\widetilde {\cal C}} \omega
{}~=~ \int_{\widehat {\cal C}} \widehat \mu \widehat \omega
$$
so that we can rewrite the integral over $\widetilde {\cal C}$ in
superspace form.
\itemitem{6.5.8.}
The vector bundle $\coker\ \nabla \bar s \to {\cal M}$, though crucial
in the general localization formula, is difficult to work with directly.
$\nabla s\colon T_\varphi \widetilde {\cal C}
\to ( T_{\scriptscriptstyle s ( \varphi )} \widetilde {\cal E} )^{\rm vert}$
is a simpler operator.
Since $s$ is ${\cal G}$-equivariant, $\ker\ \nabla s$ and $\coker\ \nabla s$
are in general infinite dimensional.
However the operator
\eqn\BigODef{
\IO \eqdef \pmatrix{\nabla s\cr C^\dagger\cr}\colon
T_\varphi \widetilde {\cal C} ~\longrightarrow~
( T_{\scriptscriptstyle s ( \varphi )} \widetilde {\cal E} )^{\rm vert}
\oplus \lieg}
defines equivariant vector bundles of finite rank, $\ker\ \IO$ and
$\coker\ \IO$, over $\widetilde {\cal M}$.
These descend to vector bundles over ${\cal M}$.
The operator $\IO$ is of direct importance to TFT as it appears as
the fermionic kinetic term of the complete lagrangian.
\medskip\noindent
Finally using remarks 6.5.7 and 6.5.7 we may rewrite \GenLocThm\
in a way that makes the TFT action
more apparent:
\eqn\TFTrewrite{\eqalign{
&\int_{\cal M} e ( \coker\ \nabla \bar s \to {\cal M} )
\wedge i^\ast {\cal O}\cr
&\quad~=~ \int_{\cal M} e ( (\coker\ \IO \to \widetilde {\cal M})
/{\cal G} ) \wedge i^\ast {\cal O}\cr
&\quad\sim~
\int_{\underline{g}^\ast} [ d \phi ]\
\int_{\widetilde {\cal C} \times \Pi \widetilde {\cal C}}
[ d \varphi ]\ [ d \psi ]\
\int_{\widetilde {\cal V}^\ast \times \Pi \widetilde {\cal V}^\ast}
[ d \pi ]\ [ d \rho ]\
\int_{\underline{g} \times \Pi \underline{g}} [ d \lambda ]\ [ d \eta ]\
\exp\ - Q_{\cal C} \left ( \Psi_{\rm Loc} + \Psi_{\rm Proj} \right )\cr}}
where we have absorbed the normalizations into the measures
$[ d \cdots ]$.
The TFT action may be identified with
$$
I_{\rm Top} ~=~ Q_{\cal C} ( \Psi_{\rm Loc} + \Psi_{\rm Proj} )
$$
where $\Psi_{\rm Loc}$ is given by \PsiLocDef\ and $\Psi_{\rm Proj}$
is given by \PsiProjDef.

\newsec{Topological String Theory and the Chiral Theory}

\subsec{Standard Topological String Theory}

The basic configuration space is given by\foot{
The localization to ${\rm Met}_{-1} ( \Sigma_W ) \subset {\rm Met} ( \Sigma_W
)$
is standard (For a review and more extensive references, please see
\refs{\CMRLH}.)
For the localization Lagrangian we introduce a scalar antighost $\rho$ and its
Lagrange multiplier $\pi$.
Then the gauge fermion for localizing to ${\rm Met} ( \Sigma )_k$ is:
\eqn\WeylLocGF{
\Psi_{\rm Weyl\ Loc} = \int d^2 z \sqrt{h}~ \rho~ ( R +1 )}

Using the following two relations:
\eqn\Useful{\eqalign{
Q_{\cal C} \Gamma^\alpha_{\beta\gamma}
=& \half h^{\alpha\delta} \left (
D_\beta Q h_{\gamma\delta} + D_\gamma Q h_{\beta\delta} - D_\delta Q
h_{\beta\gamma}
\right )\cr
Q{\cal C} R
=& - \half D_\alpha D^\alpha ( h^{\gamma\beta}~ Q h_{\beta\gamma} )
+ D^\alpha D^\beta~ Q h_{\alpha\beta}
- \half R~ h^{\alpha\beta}~ Q h_{\alpha\beta}
\cr}}
we may write this action as
\eqn\WeylLocAct{
I_{\rm Weyl\ Loc}=
\int d^2 z \sqrt{h}~ \left \{ \pi~ ( R + 1 ) - \rho~ L^{\alpha\beta}
\psi_{\alpha\beta}
\right \}}
where
\eqn\LOpDef{
L^{\alpha\beta} = D^\alpha D^\beta - \half h^{\alpha\beta} D^2 + \half
h^{\alpha\beta}}
}.
\eqn\WeylFixed{
\widetilde {\cal C}
{}~=~ \{ ( f, h )~ \vert~ f \in {\cal C}^\infty ( \Sigma_W, \Sigma_T )~
{\rm and}~ h \in {\rm Met}_{-1} ( \Sigma_W ) \}}
where ${\rm Met}_{-1} ( \Sigma_W )$ is the space of metrics on $\Sigma_W$
with constant Ricci scalar curvature\foot{We assume for simplicity that
the genus of the world sheet is greater than one.} $-1$
Hurwitz space may be described by the Gromov equation for (pseudo-)
holomorphic maps:
\eqn\HurSpDescr{\eqalign{
{\cal M} ~=&~ \widetilde {\cal M} / {\rm Diff}^+ ( \Sigma_W )\cr
\widetilde {\cal M} ~=&~ \{ ( f, h ) \in \widetilde {\cal C}~ \vert~
d f + J~ d f~ \epsilon [ h ] = 0 \}\cr}}
At $( f, h ) \in \widetilde {\cal M}$, the tangent space is described by
(See Appendix B for a derivation.)
\eqn\THursSp{
T_{\scriptscriptstyle ( f, h )} \widetilde {\cal M}
{}~=~ \{ ( \delta f, \delta h ) \in
T_{\scriptscriptstyle ( f, h )} \widetilde {\cal C}~ \vert~
D ( \delta f ) + J~ D ( \delta f )~ \epsilon[ h ] + J~ d f~ k [ \delta h ]
= 0 \}}
where $D$ is the pulled-back connection
$(D_\alpha \delta f )^\mu = \partial_\alpha \delta f^\mu
+ \Gamma^\mu_{\kappa\lambda} \partial_\alpha f^\kappa
\delta f^\lambda$; and $k [ \delta h ]$ is the variation of the complex
structure.
\THursSp\ suggests that we introduce an operator
\eqn\BoldD{\eqalign{
\ID_{\scriptscriptstyle ( f, h )}\colon
T_{\scriptscriptstyle ( f, h )} \widetilde {\cal C}
{}~\longrightarrow&~ \widetilde {\cal V}_{\scriptscriptstyle ( f, h )}\cr
\ID_{\scriptscriptstyle ( f, h )} ( \delta f, \delta h )
{}~\longmapsto&~ D ( \delta f ) + J~ D ( \delta f )~ \epsilon [ h ]
+ J~ df~ k [ \delta h ]\cr}}
where $\widetilde {\cal V}_{\scriptscriptstyle ( f, h )}$ will be defined
shortly.

To construct a topological string theory action, we regard the Gromov
equation, \HurSpDescr, as a ${\rm Diff}^+ ( \Sigma_W )$-equivariant
section
\eqn\GomovEq{\eqalign{
s\colon \widetilde {\cal C}~ \longrightarrow&~ \widetilde {\cal E}\cr
s ( f, h ) ~\longmapsto&~ d f + J~ df~ \epsilon [ h ]\cr}}
where $\widetilde {\cal E}$ is a ${\rm Diff} ( \Sigma_W )$-equivariant
vector bundle whose fiber above $( f, h ) \in \widetilde {\cal C}$
is given by
\eqn\VDef{
\widetilde {\cal V}_{\scriptscriptstyle ( f, h )}
\colon= \Gamma [ T^\ast \Sigma_W \otimes f^\ast ( T \Sigma_T ) ]^+}
The superscript $( \cdots )^+$ indicates that the sections must
satisfy the self-duality constraint:
\eqn\SDConstraint{
\rho \in \Gamma [ T^\ast \Sigma_W \otimes f^\ast ( T \Sigma_T ) ]^+
{}~\Longleftrightarrow~ \rho - J~ \rho~ \epsilon[ h ] = 0}

$\widetilde {\cal E}$ admits an $SO (\widetilde {\cal V} )$-connection,
$\nabla_{SO ( \widetilde {\cal V} )}$, characterized by \KerNablaS
$$
\ker\ \nabla_{SO ( \widetilde {\cal V} )} s
\vert_{\scriptscriptstyle ( f, h )}
{}~=~ T_{\scriptscriptstyle ( f, h )} \widetilde {\cal M}
\qquad\qquad\forall
( f, h ) \in \widetilde {\cal M}
$$
Let $s_\alpha{}^\mu [ f, h ]$ be a local section of
$\widetilde {\cal E} \to \widetilde {\cal C}$, and define
\eqn\SOVConnect{\eqalign{
( \nabla_{SO ( \widetilde {\cal V} )} s )_\alpha{}^\mu
{}~=&~ \int d^2 \sigma \sqrt{h} \left (
{{\delta s_\alpha{}^\mu} \over {\delta f^\kappa ( \sigma )}}
\delta f^\kappa ( \sigma ) +
{{\delta s_\alpha{}^\mu} \over {\delta h_{\beta\gamma} ( \sigma )}}
\delta h_{\beta \gamma} ( \sigma ) \right.\cr
&\qquad\qquad\qquad\qquad\qquad\qquad
- \Gamma^\mu_{\kappa\lambda} [ f ( \sigma ), h ( \sigma ) ]
s_\alpha{}^\kappa \delta f^\lambda ( \sigma ) \biggr )\cr}}
Then for $s [ f, h] = d f + J~ df~ \epsilon$, one may
readily check that
\eqn\RightConnect{
\nabla_{SO ( \widetilde {\cal V} )} s ~=~ \ID ( \delta f, \delta h )}
Having determined $\nabla_{SO ( \widetilde {\cal V} )}$, we may construct
$\Upsilon_{\nabla_{SO ( \widetilde {\cal V} )}, {\cal C}-{\cal G}}$:
$$
\Upsilon_{{\cal C}-{\cal G}}
{}~=~ \int_{\tilde {\cal V}^\ast \times \Pi \tilde {\cal V}^\ast}
d \pi\ d \rho\ \exp\ -I_{\rm Top\ \sigma}
$$
where from \UpsilonDef\ and \PsiLocDef:
$$
I_{\rm Top\ \sigma}
{}~=~ Q_{\cal C}
\left [
i \left \langle \rho, s ( f, h ) \right \rangle
- t ( \rho, \Gamma_{SO ( \widetilde {\cal V} )} \cdot \rho )_{\widetilde {\cal
V}^\ast}
+t ( \rho, \pi )_{\widetilde {\cal V}^\ast} \right ]
$$

$\widetilde {\cal C}$ is a ${\rm Diff} ( \Sigma_W )$-manifold, so that
for all $( f, h ) \in \widetilde {\cal C}$, there is a canonical isomorphism
between ${\rm diff} ( \Sigma_W )$ and
$( T \widetilde {\cal C} )^{\rm vert}$ given by
\eqn\CanonIso{\eqalign{
C_{\scriptscriptstyle ( f, h )}\colon {\rm diff} ( \Sigma_W )
{}~\longrightarrow&~ ( T \widetilde {\cal C} )^{\rm vert}\cr
C_{\scriptscriptstyle ( f, h )} ( \gamma ) ~\longmapsto&~
\pmatrix{ {\cal L}_\gamma f\cr {\cal L}_\gamma h\cr}\cr}}
Hence from \PsiProjDef\ the projection fermion is given by
$$
\Psi_{\rm Proj} ~=~ - i ( \lambda, C^\dagger )_{\underline{g}}
$$

$\Psi_{\rm Weyl\ Loc} + \Psi_{\rm Top\ \sigma} + \Psi_{\rm Proj}$,
and therefore the action we have produced thus far, is
${\rm Diff} ( \Sigma_W )$ invariant.
Therefore, in order to compute anything using the standard methods of
local quantum field theory, we still need to fix this symmetry\refs{\BauSin}.
This may be done by adding the (gauge-fixing) action
\eqn\DiffFix{
\Psi_{\rm GF} = \int d^2 z \sqrt{h}~
b^{\alpha \beta} ( h_{\alpha \beta} - h_{\alpha \beta}^{{\scriptscriptstyle
(0)}} )}
where the action of $Q_{\cal C}$ extends to the symmetric tensor fields,
$b^{\alpha\beta}$ and $d^{\alpha\beta}$ as
$Q_{\cal C} b^{\alpha\beta} = d^{\alpha\beta}$.
Altogether the action of standard topological string theory is given
by
\eqn\StandTopString{
I_{\rm TS} ~=~ I_{\rm Weyl\ Loc} + I_{\rm Top\ \sigma}
+ I_{\rm GF} + I_{\rm Proj}}

Our objective is to find a string theory whose connected partition function
is
\eqn\StringThy{\eqalign{
Z_{\rm string}
{}~\sim&~ \chi_{\rm orb} ( {\cal M} )\cr
{}~=&~ \int_{\cal M} e ( T {\cal M} \to {\cal M} )\cr}}
%
If we pursue the construction outlined in section 6, it is apparent that
we indeed obtain a theory that localizes to Hurwitz space.
{}From \TFTrewrite\ the measure is given by $e( \coker\ \IO/ {\cal G})$, where
\SOVConnect\ and \CanonIso\ together define
$\IO_{\scriptscriptstyle ( f, h )}$.
Unfortunately, however\refs{\CMRLH},
\eqn\KerCokerO{\eqalign{
\ker\ \IO_{\scriptscriptstyle ( f, h )}/ {\cal G}
{}~\eqsim&~ T_{\scriptscriptstyle ( f, h )} {\cal M}\cr
\coker\ \IO_{\scriptscriptstyle ( f, h )}
{}~\eqsim&~ \{ 0 \} \cr}}
so that the standard topological string theory clearly does not produce
the desired measure, $e ( T {\cal M} \to {\cal M} )$.
It is clear that we have to modify this theory somewhat.
The following gives a clue about what this modification ought to
entail.

Since $\widetilde {\cal V}$ is endowed with a metric, we may
define the adjoint of $\IO_{\scriptscriptstyle ( f, h )}$, which we may
view as an operator
$$
\IO_{\scriptscriptstyle ( f, h )}^\dagger\colon
T_{\scriptscriptstyle s ( f, h )} \widetilde {\cal V} \oplus \lieg
{}~\longrightarrow~ T_{\scriptscriptstyle ( f, h)} \widetilde {\cal C}
$$
{}From \KerCokerO\ it follows that
\eqn\CokerKerO{\eqalign{
\ker\ \IO^\dagger
{}~\eqsim&~ \{ 0 \} \cr
\coker\ \IO^\dagger/ {\cal G}
{}~\eqsim&~ T_{\scriptscriptstyle ( f, h )} {\cal M}\cr}}
Clearly we want to produce a TFT wherein the fermion kinetic term is
\eqn\DesiredFermKin{
\IO^{\rm Total} ~=~ \IO \oplus \IO^\dagger}
In this case, $\ker\ \IO^{\rm Total} = \coker\ \IO^{\rm Total}
= T {\cal M}$, so that such a theory would produce the correct measure.

\subsec{``Cofields''}

In order to obtain \DesiredFermKin\ as the fermion kinetic operator,
we must extend the field space relative to that of the standard topological
string theory.
The  new fields are completely determined by two requirements
\item{1.}
$\IO^\dagger$ maps ghosts to antighosts,
\item{2.}
$Q_{\cal C}$ extends to act on the new fields as the Cartan differential
of ${\rm Diff} ( \Sigma_W )$-equivariant cohomology.
\par\noindent
To describe the additional fields, it is easiest to begin with the new
set of ghosts, $\widehat \IG$; we shall sometimes refer to these
as the ``Co-Ghosts".
These take values in the domain of $\IO^\dagger$, so that
\eqn\CoGhosts{
\widehat \IG
{}~\in~ \Gamma( T \Sigma_W \otimes f^\ast ( T^\ast \Sigma_T ))^+
\oplus \Gamma( T \Sigma_W )}
and their index structure is given by
$$
\widehat \IG ~=~ \pmatrix{\widehat \chi_\mu{}^\alpha\cr
                              \widehat \psi^\alpha\cr}
$$
As usual, the superscript $( \cdots )^+$ indicates that the
$\widehat \chi_\mu{}^\alpha$ satisfy a self-duality constraint:
$$
\widehat \chi_w{}^z ~=~ 0
$$

The Co-Ghosts represent differential forms on an enlarged field space,
$\widetilde {\cal D}$.
This enlarged field space may itself be viewed as the {\it total} space
of a vector bundle, $\widetilde {\cal D} \to \widetilde {\cal C}$,
where the fibre at $( f, h ) \in \widetilde {\cal C}$ is given by
$$
\widetilde {\cal D}_{\scriptscriptstyle ( f, h)}
{}~=~ \Gamma \left ( T \Sigma_W \otimes f^\ast ( T^\ast \Sigma_T ) \right )^+
\oplus \Gamma ( T \Sigma_W )
$$
We refer to the additional fields as ``Co-Fields"; like the Co-Ghosts,
their index structure is given by
\eqn\CoFields{
\widehat \IF
{}~=~ \pmatrix{\widehat f_\mu{}^\alpha \cr
                 \widehat h^\alpha\cr}}
where $\widehat f_w{}^z=0$.

In turn $\widetilde {\cal D}$ forms the base space of a
${\rm Diff} ( \Sigma_W )$-equivariant vector bundle,
$$\matrix{
\widetilde {\cal E} \oplus \widetilde {\cal E}_{\rm cf} & \longleftarrow &
\widetilde {\cal V} \oplus \widetilde {\cal V}_{\rm cf}\cr
\mapdown{\widetilde \pi} & & \cr
\widetilde {\cal D} & & \cr}
$$
Now consider the following section
\eqn\CoSection{\eqalign{
{\bf S} \colon \widetilde {\cal D} ~\longrightarrow&~
\widetilde {\cal V} \oplus \widetilde {\cal V}_{\rm cf}\cr
{\bf S} \colon ( \IF, \widetilde \IF)
{}~\longmapsto&~ ( df + J~ df~ \epsilon [ h ],~
\IO^\dagger \widehat \IF)\cr}}
The zero set of this section is still Hurwitz space
\eqn\ZeroSetEnl{
\{ ( \IF, \widehat \IF) \in \widetilde {\cal D}~ \vert~
s ( \IF, \widehat \IF ) = 0 \}
{}~=~ {\cal M} \times \{ 0 \}}
since $\ker\ \IO^\dagger = \{ 0 \}$.
Moreover, when restricted to ${\cal M} \times \{ 0 \}$, the operator
appearing in the total fermion kinetic term is given by
$\IO^{\rm Total} = \IO \oplus \IO^\dagger$.
Our choice of section dictates that the antighost bundle be dual to
$\widetilde {\cal V} \oplus \widetilde {\cal V}_{\rm cf}$ where
$\widetilde {\cal V}$ is defined in \VDef, while the range of
$\IO^\dagger$ defines
\eqn\CoFieldDualBundle{
\widetilde {\cal V}_{\rm cf} = \Gamma( f^\ast (T  \Sigma_T ))
\oplus \Gamma( \Sym ( T \Sigma_W^{\otimes 2} )).}
or
\eqn\coantighst{
\widetilde \IA
{}~=~ \pmatrix{\widehat \rho^\mu\cr
                 \widehat \eta_{\alpha \beta} \cr}.}
$Q_{\cal C}$ extends to the Co-Fields as the Cartan differential for
${\rm Diff} ( \Sigma_W )$-equivariant cohomology.
$$
Q_{\cal C} \pmatrix{\widehat \IF\cr \widehat \IG\cr}
{}~=~ \pmatrix{0 & 1\cr -{\cal L}_\gamma & 0\cr}
\pmatrix{\widehat \IF\cr \widehat \IG\cr} \qquad
Q_{\cal C} \pmatrix{\widehat \IA\cr \widehat \IPi\cr}
{}~=~ \pmatrix{0 & 1\cr -{\cal L}_\gamma & 0\cr}
\pmatrix{\widehat \IA\cr \widehat \IPi\cr}
$$
The addition of the cofields does not  change the $Q_{\cal C}$-cohomology,
so we expect  to have the same observables as in topological string theory.
(Please see section 7.3).

The Lagrangian for the ${\rm YM}_2$ string will be a sum of a Lagrangian for
the topological
string theory $\Sigma_W \to \Sigma_T$ plus a Lagrangian for localizing to
$\widehat \IF=0$:
\eqn\LagrangianYMT{
I_{YM_2 } = I_{\rm TS} + I_{\rm cf}}
Following section 6, we write down the gauge fermion for the co-fields:
\eqn\GaugeFermionCF{
\Psi_{\rm cf} ~=~ \left\langle
\widehat \IA,  \IO^\dagger \widehat \IF \right\rangle
- t \left ( \widehat \IA, \widehat \IPi \right )}
where $t \in \IR$ and
\eqn\ODaggerDef{
\IO^\dagger
{}~=~\pmatrix{- ( \delta_\mu{}^\nu \nabla_\gamma + J_\mu{}^\nu \nabla_\beta
\epsilon^\beta{}_\gamma ) &
- G_{\mu\nu} \partial_\gamma f^\nu\cr
- \delta^\alpha{}_\gamma J^\nu{}_\mu \partial_\delta f^\mu
\epsilon^{\delta\beta} &
{1\over2} ( \delta_\gamma{}^\alpha D^\beta + \delta_\gamma{}^\beta D^\alpha )
\cr}}

To derive the localization theorem for this theory, we must analyze the
exact bundle sequence analogous to \ExactBundleSeq.
For the choice of section, ${\bf S}$, and connection,
$\nabla_{SO ( \widetilde {\cal V} )}$, we have
We find
$$
0~ \longrightarrow~ {\rm Im}\ ( \IO \oplus \IO^\dagger )
{}~\longrightarrow~ \widetilde {\cal V} \oplus
\widetilde {\cal V}_{\rm cf}
{}~\longrightarrow~ \coker\ ( \IO \oplus \IO^\dagger)
\rightarrow 0
$$
as a sequence of bundles over ${\cal M} \times \{ 0 \}$.
Then by the general principles we have explained in the previous section,
we see, by combining \KerCokerO\ and \CokerKerO, with \TFTrewrite,
that the path integral computes the Euler character of the cokernel
bundle, $T {\cal M}$, which is the problem we set out to solve.

\subsec{Observables in the theory}

The most obvious observables in topological string theories are made
from gravitational descendents of the primaries of the corresponding
topological sigma model.
The observables in topological string theories are of two types:
(a) homology observables and (b) homotopy observables\nref\WTopPhase{E.~
Witten, ``On the Structure of the Topological Phase of Two-Dimensional
Gravity", Nucl. Phys. {\bf B340} (1990) 281.}\nref\DVV{E. Verlinde
and H. Verlinde , Nucl. Phys. {\bf B348} (1991) 457;
R. Dijkgraaf, E. Verlinde and H. Verlinde Nucl. Phys. {\bf B348}
(1991) 435; {\bf B352} (1991) 59; in String Theory and quantum
Gravity, Proc. Trieste Spring School, April 1990 (World Scientific,
Singapore, 1991).}\nref\horavats{ P.Horava, ``Two dimensional
string theory and the topological torus,'' Nucl. Phys. {\bf B386},
383-404,1992.}\refs{\donaldson,\WTopPhase,\DVV, \horavats.}.

\noindent
$\underline{\rm Homology\ Observables:}$

These observables are built from  cohomology classes of the target space.
For a target space  a Riemann surface of genus $G$, the cohomology
classes are described by:
\eqn\Cohom{\eqalign{
\{ 1 \} ~\in&~ H^0 ( \Sigma_T )\cr
\{ \xi^{A}, \xi^{\bar A}  \}
{}~\in&~ H^1 ( \Sigma_T )\qquad A =
1,\ldots,G\cr
\{ \omega \} ~\in&~ H^2 ( \Sigma_T )\cr}}
where $\omega$ is the K\"ahler class.
So the homology observables of the topological string theory are given by:
\eqn\HomolObserv{\eqalign{
&\sigma_n ( 1 )\cr
&\sigma_n ( \xi^{A} _i \chi^i ),\sigma_n ( \xi^{\bar A} _i \chi^i
)\qquad
A=1,\ldots,G\cr
&\sigma_n ( \omega_{ij} \chi^i \chi^j )\cr}}
where $\sigma_n ( \cdots )$ represents the gravitational dressing of
the operator.
In essence\refs{\BauSinii}
$$
\sigma_n ( {\cal O} )
{}~=~ ( \epsilon_{\alpha\beta} \partial^\alpha \gamma^\beta )^n {\cal
O}.
$$

\noindent
$\underline{\rm Homotopy\ Observables:}$

There is a ring of homotopy observables with $2G$ generators:
\eqn\HomotObserv{
{\cal O}_{\vec k}
{}~=~ \exp \left\{ 2 \pi i \Biggl( \sum_{A=1}^{G}
k_{A} \cdot
\int_{w_0}^{f ( z) }\xi^{A} +
\bar k_{\bar A} \cdot
\int_{\bar w_0}^{\bar f ( z) }\xi^{\bar A}\Biggr)
\right \}}
where ${\vec k} = ( k_{{\scriptscriptstyle [ 1]}}, \ldots,
k_{{\scriptscriptstyle [ G]}} )$ are
vectors in the dual to the period lattice, $\Lambda$.

We expect that these operators will form a ring related to the group ring
of the fundamental group of the target manifold.
(Since the matter is not a topological conformal field theory we expect it
will involve a nontrivial deformation of that ring.)

Via the descent equations we may obtain $1$-form versions of the
operators \HomotObserv, which generate an algebra of symmetries which
we naturally expect to be related to $w_\infty$.
The operators relevant to the topological string theory are, of course, the
gravitational dressings of the above.
Indeed, 2D string theories are famous for having $w_{\infty}$-type
symmetries in the target space theory.
This should be true in our case and should explain the area-preserving
diffeomorphism invariance of $YM_2$ from the string perspective.
We have not carried this out in detail.

\newsec{Turning on the area}
\par
\subsec{Area Polynomials in $YM_2$}

The same basic reasoning we have used in the
$A=0$ case can be applied to the $A>0$ case.
We begin with the $1/N$ expansion of
the chiral \ymt\ partition function.
Manipulations identical to those leading to
 \Zchiii give
\eqn\ZchiAi{\eqalign{
Z^+&(A,G,N)\cr
&=\sum_{n,\ell>0}^\infty N^{n ( 2 - 2G)-\ell}
e^{-\half A(n-{n^2\over N^2})}{(-A)^\ell\over \ell !}
\sum_{s_i,t_i \in S_n}  {1\over n!} \delta(\Omega_n^{2-2G}
T_{2,n}^\ell \prod_1^G[s_i,t_i])\cr
{}~=&~ \sum_{n=0}^\infty\sum_{\ell=0}^\infty e^{-\half A(n-{n^2\over
N^2})}
{(-A)^\ell \over \ell !}\cr
&\sum_{p_1,\ldots,p_k \in T_2 \subset S_n}
\sum_{L^\prime=0}^\infty \sum_{v_1,\ldots,v_{L^ \prime }\in
S_n}^\prime
\bigl({1 \over N}\bigr)^{n ( 2 G - 2 ) + \ell + \sum_{j=1}^{L^\prime}
(
K_{v_j} - n )}\cr
&
\sum_{s_1,t_1,\ldots,s_G,t_G\in S_n}
{{d( 2 - 2G,L^\prime  )}\over{n!}}
\delta( p_1 \cdots p_k v_1 \cdots v_{L^\prime}
\prod_{i=1}^G[s_i,t_i]). \cr}}
where $T_{2,n}\in \IC[S_n]$ is the sum of transpositions.
Recall that to establish a correspondence between homomorphisms
from $F_{G,L}$ to $S_n$ and branched coverings over a set $S$ we
make a choice
of generators of $\pi_1( \ST -S,y_0 ) $. This choice leads to an
association of
a set of $k$ points
with the permutations $p_1 \cdots p_k$. After collecting powers of
$N$ we get
\eqn\Zchiiv{\eqalign{
Z^+(A,G,N)
&=\sum_{n=0}^{\infty} \sum_{B=0}^{\infty} e^{-nA/2}
e^{{n^2\over 2N^2} A}
\biggl({1\over N}\biggr)^{2h-2} \sum_{L=0}^{B}
P_{n,B,L}(A) \cr}}
where $P_{n,B,L}(A)$ is a polynomial defined by:
\eqn\dfply{
P_{n,B,L}(A)=
\sum_{k=0}^{B}  {{(-A)^k}\over {k!}} \sum_{L=k}^{B}
{\cal \chi}({\CC_{L-k}( \ST )})
\sum_{\Psi (n,B,G,L,k)}
       {1\over {\vert C(\psi)\vert} } .}
where $\Psi(n,B,G,L,k)$
is the set of homomorphisms $F_{L,G} \rightarrow S_n$ which, via
Theorem 3.1, correspond to branched coverings over some set  $S$ of
branch points, with the property that  over a  {\it fixed subset of $k$
points } the branching is simple.

 For example, if $\psi$ has only
simple branch points, i.e., corresponds to a map in
the simple Hurwitz space then the formula becomes:
\eqn\splbp{
\psi \in  \Psi( n, B, G, L=B ) ~\Longrightarrow~
P_\psi(A)= \sum_{k=0}^{B}
{{(-A)^k}\over {k!}} {\cal \chi}({\CC_{B-k}(\ST)}).}

In order to obtain the full $1/N$ expansion we must also
expand the factor
\eqn\tubhan{
e^{{n^2\over 2N^2} A}
}
in \Zchiiv .  This has been interpreted in  \Min\ and
in \GrTa\ in terms of contributions of ``collapsed tubes and
handles.''

\subsec{Area Polynomials from Perturbations}

In the topological field theory
there is a natural mechanism by which the area can be
included: perturbation by BRST  closed but nonexact operators.
This is how, e.g.,  one explores the more physical phases
of 2d gravity, studied in the double scaling limit of matrix models,
in the
framework of topological 2D gravity.
In the present case we  will study the perturbation of  the action by
a BRST
invariant
operator $\half \int \CA^{(2)}$.
Here $\CA^{(2)}$ fits into the area operator descent multiplet:
\eqn\areamult{\eqalign{
\CA^{(0)}&=\sigma _0\bigl(\omega_{ij}(f(x)) \chi^i \chi^j\bigr)\cr
\CA^{(1)}&=\sigma _0\bigl(dx^\alpha \omega_{ij}(f(x))\p_\alpha f^i
\chi^j\bigr)\cr
\CA^{(2)}&=\sigma_0\bigl(dx^\alpha\wedge dx^\beta
\omega_{ij}(f(x))\p_\alpha
f^i
\p_\beta f^j\bigr)\cr}
}
Here $\omega$ is the  K\"ahler two-form from the target space.
The form degree $0$, ghost number $2$ member of this multiplet
has a geometric interpretation as a $2$-form on $\cF$.
Thus  insertions of ${\cal A}^{{\scriptscriptstyle (0)}}$ compute
intersection numbers on $\cF$.
The deformed action is
\eqn\defdact{
I_0 ~\longrightarrow~ I_0 + \half \int {\cal A}^{{\scriptscriptstyle
(2)}}}
Naively, the contribution of $\half \int \CA^{(2)}$ in
a path integral over maps $f$ of index $n$ is
$e^{-\half n A}$. This accounts nicely for the
genus-independent exponential
factors in \Zchiiv, but fails to explain the polynomial
of $A$ in \Zchiiv.

We can understand some features of the polynomial
in $A$ in \Zchiiv\ by considering more carefully the
``conformal perturbation series'' in question:
\eqn\cps{Z^+(A,G, N)=\biggl\langle e^{-\half \int \CA^{(2)} }
\biggr\rangle_{A=0}=
\sum_{\ell=0}^\infty {(-1)^\ell\over \ell !}\Biggl\langle
\biggl(\half \int_\Sw f^* \omega\biggr)^\ell \Biggr\rangle_{A=0}
}
The measure $\langle ~\cdots \rangle$ implicitly contains
further operators, such as the four fermion terms in the
curvature, which arise from the \pseudo sigma model.
It is important to understand that the expression \cps\
is ill-defined.
Evaluation of the terms in the series involves integration
over operators inserted at coincident points. As in all
theories of gravity, merely  identifying the operators
as in \areamult\ does not fully specify their correlators,
because we must choose contact terms, i.e., we must
carefully specify the terms in the correlators of $\CA^{(2)}$
and $ \langle \tbA, \bf R [ \bG , \bG ]~ \tbA \rangle $ which have
delta-function support
on two or more points.

In the following subsections we will show how a consideration
 of contact terms can account for the
area polynomials  \splbp\ which arise from the contributions of
{\it simple Hurwitz space.}
We will not try to account for the other types of
coverings in the sum  over $\Psi$ in \Zchiiv.
Similarly we do not try to account for the
the terms arising from expanding \tubhan. We firmly
believe that these more complicated polynomials can
also be explained by looking at more complicated
contact terms from the higher codimension boundaries
of Hurwitz space and the space of Maps$\times$ Metrics.

\subsec{Measure on the space of simple covers}

Let $\cF^{{\scriptscriptstyle (1)}}$ be the simple Hurwitz
space of   maps with
$B$ simple branch points.
Denote these simple branch points by $P_I$ with
corresponding ramification
points of index 2 at $R_I$: these are the unique ramification
points above $P_I$.
We can choose a basis
$\{ \bG_I \}_{I=1,\ldots,2B}$ for $T \cF$,
such that $\bG_{2I-1}$ and $\bG_{2I}$ have support only at
the $I$-th ramification point.
The analogue  in ordinary string theory is a choice of
Beltrami differentials which have support only at punctures.
This is a well-defined choice away from the boundary of
moduli space.
\par
Now consider the curvature insertions in these local
coordinates:
\eqn\CurvIns{
\int \cD [ \tbA ] \exp \biggl[ -\quarter \tbA^I {\cal R}_{IJ} \tbA^J
\biggr]
{}~=~ {{(-1)^B}\over{2^B}} {\rm Pfaff} ( {\cal R}_{IJ} )}
where $B$ is even and
the matrix, ${\cal R}_{IJ}$, takes the following form
in an oriented orthonormal basis
\eqn\CurvDiag{
{\cal R}_{IJ}
{}~=~ \pmatrix{\matrix{0 & {\cal R}_{12}\cr -{\cal R}_{12} & 0\cr} &
&
\cr
                           & \ddots & \cr
                            &              & \matrix{0  & {\cal
R}_{2B-1~2B}\cr -{\cal R}_{2B-1~2B} & 0 \cr}\cr}}
so that

\eqn\PfExp{
{\rm Pfaff} ( {\cal R}_{IJ} )
{}~=~ \prod_{I=1}^B {\cal R}_{2I-1~2I} [ \bG^{2I-1}, \bG^{2I}] ( R_I
)}
and the full measure for the \top\ string theory is
\eqn\FullMeas{\eqalign{
&{{(-1)^B}\over{(2\pi)^B}}
\int_{  \cF^{(1)}} \cD [ \bF, \bG ]~  \prod_{I=1}^B {\cal R}_{2I-1~
2I}
[  \bG^{2I-1},
\bG^{2I}] ( R_I )~
\exp  \biggl[ -\half \int_\Sw f^\ast \omega \biggr] \cr
&={1\over{(2\pi)^B}} \sum_{k=0}^\infty {{(-1)^k}\over{k!}}
\int_{   \cF^{(1)} } \cD [ \bF, \bG ]~ \prod_{I=1}^B {\cal R}_{2I-1~
2I}
[\bG^{2I-1}, \bG^{2I}]
\left ( \half \int_\Sw f^\ast \omega \right )^k\cr
&=~ {1\over{(2\pi)^B}} \sum_{k=0}^\infty {{(-1)^k}\over{2^k k!}}
\langle\!\langle
\cA^{{\scriptscriptstyle (2)}}  \cdots
\cA^{{\scriptscriptstyle (2)}}
\rangle\!\rangle_{\cF (B,k) }\cr}}

In the last line we have introduced a space
${\cal F }(B,k)$, which is the product space
\eqn\CBK{
 \cF(B,k)=   \cF^{(1)}\times (\Sw)^k \qquad .}
The integral over this space, $\langle\!\langle \cdots
\rangle\!\rangle_{{\cal F}(B,k) }$
is formally defined by \FullMeas.
In order to define the integrated correlators we have to
describe possible delta-function supported contributions
at places where area operators collide with curvature
insertions, and where area operators collide with themselves.
Collisions of curvature operators $R_i=R_j$ belong to
higher codimension boundaries outside the space of
simple covers and contribute to other terms in
 \Zchiiv. In the next three sections we analyze the other
collisions.

\subsec{Plumbing Fixtures}

Suppose  a simple ramification point $R$ and a marked unramified
point $S$
 collide on the worldsheet.
The corresponding images of these points $P$
and $Q$ collide on the target space.
Let $U_1 \subset \ST$ be a disk containing $P$ and $Q$.
Let the disk $V_1 \subset \Sw$ be the preimage of $U_1$.
$V_1$ contains $R$ and $S$.
It is a double covering of the disk $U_1 \subset \ST$.
The disk $U_1$ glues into the annulus $U_2$ on $\ST$.
$V_2$ is the double-covering preimage of $U_2$.
The figure below describes this collision

\ifig\fhhpiv{Collision of Area and Curvature Operators}
{\epsfxsize3.0in\epsfbox{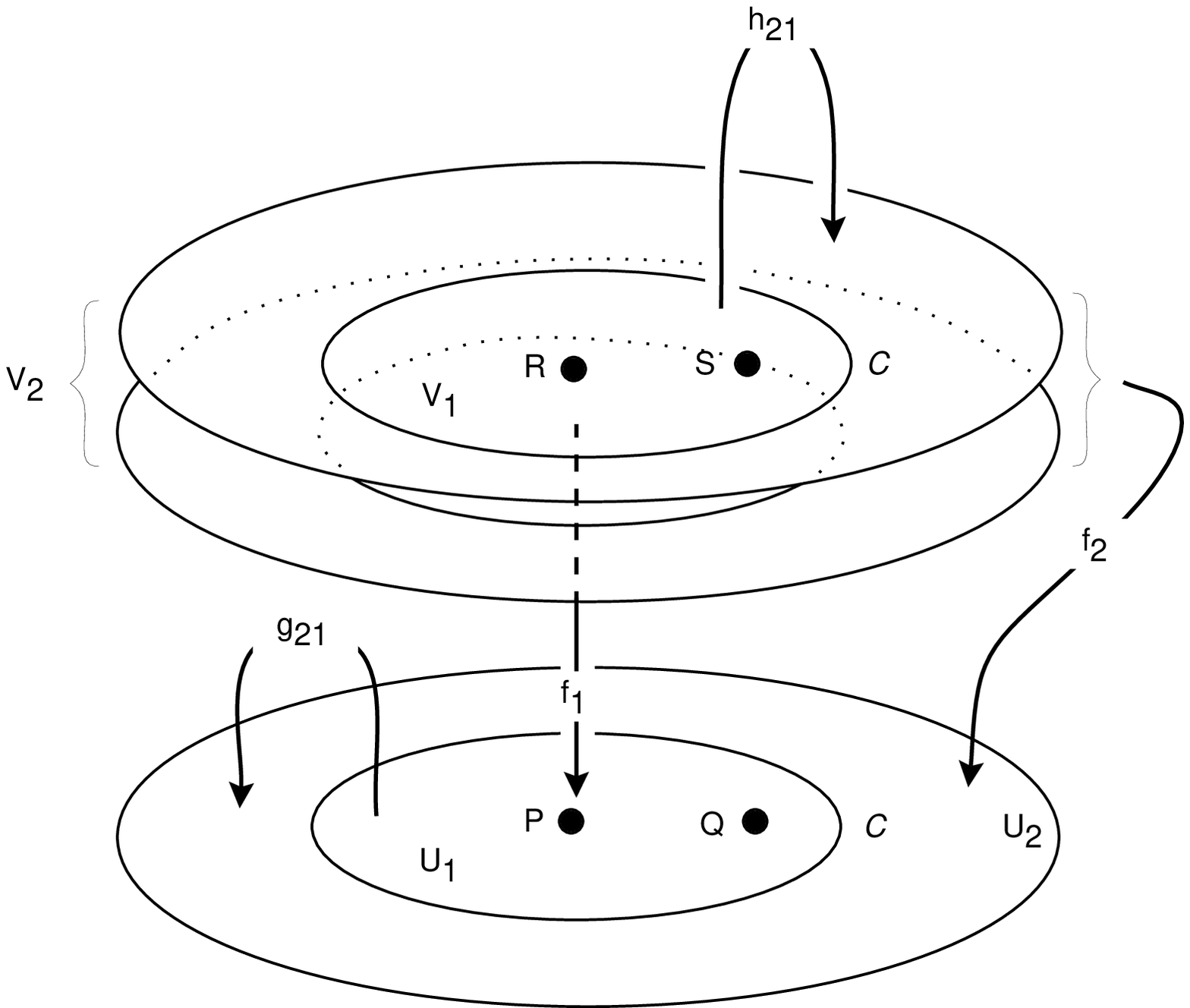}}

In the plumbing fixture description of the degeneration,
the points are kept fixed in their local coordinates.
The transition functions depend on the modulus which
describe the relative position of $R$ and $S$.

To describe the plumbing fixture more explicitly, we shall need to
specify transition maps between coordinate patches both on the
worldsheet as well as on the target.
\eqn\CACD{\matrix{
V_1                    & \mapright{h_{21}} & V_2\cr
\mapdown{f_1} &                                          &
\mapdown{f_2}\cr
U_1                     & \mapright{g_{21}}                     &
U_2\cr}}
We introduce local coordinates $w_{1,2}$ on $U_{1,2}$
and $z_1, z_2$ on $V_1, V_2$, respectively.
Then the transition functions of the \holo\ family of degenerations
are given by
\eqn\CATF{\eqalign{
g_{21} ( w_1 ) ~=&~ {q \over w_1}\cr
h_{21} (z_1 )  ~=&~ \eta {q^{1/2} \over {z_1}}, \cr}}
where $\eta=\pm$.
Note that the transition function $g_{21}$ does not change the
complex structure of the target Riemann surface without
marked points.
Similarly the function $f:\Sw \to \ST$ is locally defined by
\eqn\CALF{\eqalign{
f_1 ( z_1 ) ~=&~ z_1^2\cr
f_2 ( z_2 ) ~=&~ z_2^2\cr}}
Th modular deformation which leads to this degeneration
corresponds to a diffeomorphism generated by a quasi conformal
vector field with support on $V_2$ and a discontinuity along $\cC$.
This quasi conformal vector field is given by%
\eqn\ACqcvf{
V_2 ~=~ z_2 {{\delta q}\over q} {\partial\over{\partial z_2}}}
Integrating against the stress tensor, we obtain
\eqn\StressInt{
{{\delta q}\over q} L_0}
We exponentiate this to obtain the plumbing fixture:
\eqn\PlumbFix{\eqalign{
\int& d^2 \tau d^2 \hat\sigma
\exp 2\pi i ( \tau L_0 + \hat \sigma l_0 )
\exp -2\pi i ( \bar\tau \bar L_0 + \bar{\hat \sigma}\bar
l_0 )\cr
{}~&=~ \int {{d^2 q}\over{\vert q \vert^2}}
q^{L_0} \bar q^{\bar L_0} l_0 \bar l_0\cr}}
where $l_0$ and $\bar l_0$ are modes of the \top\ superstress tensor,
$t$, which
is the BRST partner to the ordinary stress energy tensor, $T$.
\eqn\sstt{
\{ Q,  t  \} ~=~ T}
\par
\subsec{Area-Curvature Contact Terms}

To begin, let us note that simple considerations of
quantum numbers and scaling constrain the contact term
to be of the form:
\eqn\simcont{ \int    {\bf R}_{2I-1,2I}  [ \bG , \bG ]  {\cal
A}^{{\scriptscriptstyle (2)}}
=c {\cal A}^{{\scriptscriptstyle (0)}}}
for some constant $c$.
The integral is over an infinitesimal disc surrounding the curvature
insertion,
which is nonzero because of a delta function contribution from the
collision of
two operators.
Since $R$  is a two form on $\cF$, it has ghost number two. $A^{(2)}$
is a two
form on the
target pulled back to the worldsheet. After doing the integral, it is
clear
that the result should have ghost number two and should be a zero
form on the
worldsheet. Moreover, the LHS scales like the area of the
target space.  The unique operator in section 6.5 which satisfies all
these
criteria  is $A^{(0)}$.

We now describe how the above contact term can be
directly derived using ideas along the lines of those used
in the case of pure topological gravity  \DVV. The derivation is
only heuristic.
The contact term we wish to compute here is the collision of an
integrated area operator and a curvature insertion
\eqn\ACContactTerm{\eqalign{
{{1}\over{2\pi}} \int_{\vert q \vert \le \epsilon}& {{d^2 q} \over
{\vert q
\vert^2}}~
q^{L_0} \bar q^{\bar L_0} l_0 \bar l_0~
\left \{
{\cal A}^{{\scriptscriptstyle (0)}} ( 1 ) {\cal R}_{IJ} [ \bG, \bG ]
( 0 ) \right\}\cr
{}~=&~ {{1}\over{2\pi}} \int_{\vert q \vert \le \epsilon}
d^2 q~ \partial_q \partial_{\bar q} \left\{
q^{L_0} \bar q^{\bar L_0}
\left [ {\cal A}^{{\scriptscriptstyle (0)}} ( 1 ) \Phi_{IJ} ( 0 )
\right ] \right \}\cr}}
where ${\cal A}^{{\scriptscriptstyle (0)}}$ is BRST invariant,
$0$-form
descendant  of ${\cal A}^{{\scriptscriptstyle (2)}}$.
$\epsilon << 1$ is some small positive number.
We leave the zero mode projections $b_0$ implicit.
In writing the second line we have used a contour-deformation
argument for the BRST current, the equation
 \sstt, and the fact that
\eqn\curv{\eqalign{
{\cal R}_{IJ} [ \bG_K, \bG_L ]
{}~=~ \delta_K \delta_L \Phi_{IJ}.\cr}}
However, as noted at the end of section 4.4 in \hvb, since $T\CF$ is
a
holomorphic Hermitian vector bundle we may identify the matrix
$\Phi_{IJ}=\log h_{IJ}$ where $h_{IJ}$ is the matrix of inner
products of the fields projected onto the zero-mode sector.

We evaluate the total derivative by first expanding:
\eqn\CancelProp{\eqalign{
{{1}\over{2 \pi}} \int_{\vert q \vert \le \epsilon}& d^2 q~
\partial_q
\partial_{\bar q}
\left \{ q^{L_0} \bar q^{\bar L_0}
\left  [ {\cal A}^{{\scriptscriptstyle (0)}} \Phi_{IJ}  \right ]
\right \}\cr
{}~=&~ {{1}\over{2 \pi}} \int_{\vert q \vert \le \epsilon}
d^2 q~ \partial_q \partial_{\bar q} \left \{  (
1 + \log q~ L_0 + \cdots )
( 1 + \log \bar q \bar L_0 + \cdots ) \left [ {\cal
A}^{{\scriptscriptstyle (0)}} \Phi_{IJ}  \right ]
\right \}\cr}}
At this point we do not integrate by parts (this is part of
our {\it choice} of contact term). Instead, we argue that
if  $L_0+\bar L_0$ does not annihilate the term in
brackets we will pick up a $\delta$-function when $q \to 0$.
Therefore, we  must evaluate
\eqn\scalingi{
(L_0 +\bar L_0)\left [ {\cal A}^{{\scriptscriptstyle (0)}} \Phi_{IJ}
\right ]
}
This may be done - at least heuristically - by remarking that
$(L_0 +\bar L_0)$ is the generator of scaling transformations.
{}From \natmetric\ we see that under  a Weyl  transformation
$h_{\alpha \beta}\to \lambda^2 h_{\alpha \beta}$,$\CA^{(0)}$
the matrix $h_{IJ}$ scales as $h_{IJ}\to \lambda^2 h_{IJ}$
so that
$\Phi_{IJ}\to \Phi_{IJ}  + 2 \log \lambda$ leading to
\foot{
The extra minus sign arises because we have computed
the change in $\Phi_{IJ}$ under an {\it active} transformation
on $h_{\alpha \beta}$ whereas the eigenvalue of $L_0+\bar L_0$
measures the response of an operator under a {\it passive}
transformation $z\to \lambda z$.}
\eqn\scalingiio{
(L_0 +\bar L_0)
\left [ {\cal A}^{{\scriptscriptstyle (0)}} \Phi_{IJ} \right ]
{}~=~ - 2 {\cal A}^{{\scriptscriptstyle (0)}}}
Note also that
$(L_0 + \bar L_0)^n
\left [ {\cal A}^{{\scriptscriptstyle (0)}} \Phi_{IJ} \right ] = 0$
for all $n > 1$, so that \CancelProp\ becomes
\eqn\FCancelProp{
\longrightarrow~  \int_{\vert q \vert \le \epsilon}
\delta^{{\scriptscriptstyle (2)}} ( q )
{\cal A}^{{\scriptscriptstyle (0)}} ( 1 )}
which corresponds to an area operator inserted at a ramification
point.
In conclusion, we have derived \simcont\ with $c=-2$.

\subsec{Area-Area Contact Terms}

The argument \ACContactTerm\ to \FCancelProp\ can
be repeated for the collision of two
integrated area operators.
The plumbing fixture in this case is the one familiar from bosonic
string theory
\eqn\AAContactTerm{\eqalign{
{{1}\over{2\pi}} \int_{\vert q \vert \le \epsilon}&
{{d^2 q} \over {\vert q \vert^2}}~
q^{L_0} \bar q^{\bar L_0} l_0 \bar l_0~
\left \{
{\cal A}^{{\scriptscriptstyle (0)}} ( 1 ) {\cal
A}^{{\scriptscriptstyle (0)}} ( 0 ) \right\}\cr
{}~=&~ {{1}\over{2\pi}} \int_{\vert q \vert \le \epsilon} d^2 q~
\partial_q
\partial_{\bar q} \left\{
q^{L_0} \bar q^{\bar L_0}
\left [ {\cal A}^{{\scriptscriptstyle (0)}} ( 1 ) k ( 0 ) \right ]
\right \}\cr}}
where $k$ is the K\"ahler potential,
\eqn\Kpot{
\omega
{}~=~ \partial \bar\partial k}
The calculation now proceeds as in the previous
subsection.
The analog of \scalingiio\ is
\eqn\scalingiit{
(L_0 +\bar L_0)\left [ {\cal A}^{{\scriptscriptstyle (0)}} k \right ]
{}~=0}
since both $\CA^{(0)}$ and $k$ are invariant under
worldsheet scale transformations.

\noindent
{\bf Remark}:

The absence of ${\cal A} \cdot {\cal A}$ contact terms may at
first seem a bit counter-intuitive. Indeed,
the collisions of analogous operators in conformal field theory
are well known to play an important role
\ref\dkutasov{D. Kutasov, ``Geometry on the space of
conformal field theories and contact terms,''
Phys. Lett. {\bf 220B}(1989)153}.
For example, when changing the
``compactification data'' of a product of Gaussian
models by conformal perturbation theory  it is exactly
these contact terms which account for the dependence
of the conformal weights and operator product coefficients
on the compactification data
 (see e.g. \ref\GregF{G.~ Moore,
``Finite in All Directions", hep-th/9305139.}.)
In the present case
we have made implicit choices in our evaluation of contact
terms. Our choices are related to the preservation of
BRST invariance of the theory.

\noindent
{\bf Conjecture 8.1}. Within the family of contact terms
preserving the BRST Ward identities the area polynomials
will remain unchanged and will be given by \dfply\ above.

\subsec{Recursion Relations and Calculation of an Area Polynomial}

We now combine the above results on contact terms
to derive recursion relations for the integrated
area correlators. We attempt to remove the
area operators $\CA^{(2)}$ successively. Each operator
contributes a bulk term and a contact term. If
$r$ such operators have collided with curvature
operators producing a contact term of type
\simcont\ then the remaining correlator is
integrated over a space $\CF(B,\ell;r)=\CF^{(1)}\times (\Sw)^\ell$
where $B-r$ copies of $\CA^{(0)}$ are inserted at
simple ramification points, $r$ copies of the
curvature operator are inserted at the remaining
simple ramification points, and  $\ell$
area operators $\CA^{(2)}$ are integrated over the
worldsheet $\Sw$. If we try to remove an
area operator $\CA^{(2)}$ we obtain the
recursion relation:
\eqn\RecursePrimitive{\eqalign{
\langle\!\langle
&\overbrace{
{\cal A}^{{\scriptscriptstyle (0)}}\cdots
{\cal A}^{{\scriptscriptstyle (0)}}
}^{B-r}
\overbrace{
{\cal A}^{{\scriptscriptstyle (2)}}
 \cdots
{\cal A}^{{\scriptscriptstyle (2)}}
}^k
\rangle\!\rangle_{\cF(B,k; r  )}\cr
&~=~ n A \langle\!\langle
\overbrace{
{\cal A}^{{\scriptscriptstyle (0)}}\cdots
{\cal A}^{{\scriptscriptstyle (0)}}
}^{B-r}
\overbrace{
{\cal A}^{{\scriptscriptstyle (2)}} \cdots
{\cal A}^{{\scriptscriptstyle (2)}}
}^{k-1}
\rangle\!\rangle_{\cF(B,k-1;r )}\cr
&\qquad -2r  \langle\!\langle
\overbrace{
{\cal A}^{{\scriptscriptstyle (0)}}
\cdots
{\cal A}^{{\scriptscriptstyle (0)}}
}^{B-r+1}
\overbrace{
{\cal A}^{{\scriptscriptstyle (2)}} \cdots
{\cal A}^{{\scriptscriptstyle (2)}}
}^{k-1}
\rangle\!\rangle_{\cF(B,k-1 ;r-1)}\cr}}
The first term represents the bulk contribution.
In the second
term there is one extra insertion of
 ${\cal A}^{{\scriptscriptstyle (0)}}$ which
has replaced a curvature operator
at a ramification point, and there is one  fewer
${\cal A}^{{\scriptscriptstyle (2)}}$
operator.  The coefficient $r$ in the second term comes from the fact
that for each area integral there are $r$ collisions with  curvature
insertions
at ramification points. The factor of $-2$ comes from the
normalization of the contact term.
 Iterating this recursion relation,
we are led to the following
\eqn\Recurse{\eqalign{
&\langle\!\langle
{\cal A}^{{\scriptscriptstyle (2)}} \cdots
{\cal A}^{{\scriptscriptstyle (2)}}
\rangle\!\rangle_{\cF(B,k)}\cr
{}~
&=~\sum_{l=0}^k \pmatrix{k \cr l\cr}  { 2^l B! (-1)^l\over {(B-l)!} }
( n A
)^{k-l}
\langle\!\langle  {\cal A}^{{\scriptscriptstyle (0)}}(R_1) \cdots
{\cal A}^{{\scriptscriptstyle (0)}}(R_l)  \rangle\!\rangle_{{\cal
F}({B,0}
;B-l) } .\cr} }
 When  $l>B$ it is clear that the correlation function on the right
vanishes,
by ghost
number counting.
 So that altogether
\eqn\FinalMeas{\eqalign{
&{1\over{(2\pi)^B}}
\int_{\cF^{(1)}} \cD [ \bF, \bG ]~ \prod_{I=1}^B {\bf R}_{2I-1~ 2I}
[  \bG^{2I-1}, \bG^{2I}] ( Q_I )~
\exp -\half \int_\Sw f^\ast \omega\cr
&=~\sum_{k=0}^\infty { B!\over {k! (B-k)!} }
\sum_{l=0}^{{\rm min} [ k , B]}
\pmatrix{k \cr l\cr} ( - \half n A )^{k-l}
\langle\!\langle  {\cal A}^{{\scriptscriptstyle (0)}}(R_1) \cdots
{\cal A}^{{\scriptscriptstyle (0)}}(R_l)  \rangle\!\rangle_{{\cal
F}({B,0};
B-l) }
 \cr}}

Substituting in the RHS of \FinalMeas\ we obtain
\eqn\simpoly{
 e^{- {1\over2} n A} \sum_{k=0}^B {{B!} \over{k!(B-k)!}}
\langle\!\langle  {\cal A}^{{\scriptscriptstyle (0)}}(R_1) \cdots
{\cal A}^{{\scriptscriptstyle (0)}}(R_k)  \rangle\!\rangle_{{\cal
F}(B,0;
r=B-k ) }
}
So we are left with the integral
\eqn\intF { \int_{  \cF^{(1)}} \cD [ \bF, \bG ]~ \prod_{I=1}^{B-k}
{\bf
R}_{2I-1~ 2I}
[  \bG^{2I-1}, \bG^{2I}] ( R_I )~
{\cal A}^{{\scriptscriptstyle (0)}}(R_{B-k+1})
\cdots
{\cal A}^{{\scriptscriptstyle (0)}}(R_B) }
Now we use again the fact that we are only interested in
the contribution of
simple Hurwitz space. This space
is a bundle over $ C_{0,B} /S_B$ with discrete fiber
the set
$\Psi( n,B,G,L=B)$.
Further the measure on Hurwitz space inherited from the
path integral divides out by
diffeomorphisms. Therefore the correlator in
\simpoly\ is:
\eqn\rewrit{\eqalign{
\sum_{\psi\in \Psi(n,B,G,L=B)} {1\over \vert C(\psi)\vert} \times
{1\over {B!}} \int_{C_{0,B}}
&
\bigl\langle
\prod_{I=1}^{B-k} {\bf
R}_{2I-1~ 2I}
[  \bG^{2I-1}, \bG^{2I}] ( R_I )~ \cr
&
{\cal A}^{{\scriptscriptstyle (0)}}(R_{B-k+1})
\cdots
{\cal A}^{{\scriptscriptstyle (0)}}(R_B)
\bigr\rangle\cr}
}
The correlation function has singularities when any
two ramification points $R_I$ collide. In isolating the
contributions of simple Hurwitz space we must ignore
the singularities from the collisions of $R_I$, $I\leq B-k$
with $R_J$, $J\geq B-k+1$. Thus we replace  \rewrit\
by the expression:
\eqn\rewritii{\eqalign{
\sum_{\psi\in \Psi(n,B,G,L=B)}
&
{1\over \vert C(\psi)\vert} \times
{1\over {B!}} \times \cr
\int_{C_{0,B-k}\times (\ST)^k}
\bigl\langle
\prod_{I=1}^{B-k} {\bf
R}_{2I-1~ 2I}
[  \bG^{2I-1},
&
 \bG^{2I}] ( R_I )~
\bigr\rangle
\wedge \omega(P_{B-k+1}) \wedge \cdots \wedge \omega(P_{B})
\cr}
}
where $P_J\in \ST$ are the images of the
simple ramification points $R_J$.
We can do the integrals over the  the wedge product of Kahler classes
separately to get
 $A^k$ (the area of $\ST^{\times k}$). The remaining
integral over the $B-k$ curvature insertions is the same correlator
appearing in the partition function. Thus we have:
\eqn\vlofo{\eqalign{
&\langle\!\langle  {\cal A}^{{\scriptscriptstyle (0)}}(R_1) \cdots
{\cal A}^{{\scriptscriptstyle (0)}}(R_k)  \rangle\!\rangle_{{\cal
F}({B,0
}; r=B-k)}\cr
&= {{(-A)^k}\over B!}   {\cal \chi} ( C_{B-k} ( \ST ))
\sum_{\psi\in \Psi(n,B,G,L=B)} {1\over \vert C(\psi)\vert}\cr
}}
Substituting in \simpoly, and comparing with \splbp, we see that the
contribution of simple Hurwitz space to the Path integral perturbed
as in
\defdact\ agrees with the  conjecture that the perturbation in
\defdact\ is
equivalent to 2D Yang Mills at finite area.

\noindent
{\bf Remark}. Our discussion has only focused on the
 contact terms needed to reproduce the area
polynomial of simple Hurwitz space. Higher contact terms
will be affected by the presence of gravitational descendents
of the area operator in the action. Thus we should consider
perturbations generalizing \defdact\ like:
\eqn\genpurr{
I_0\longrightarrow I_0 +\sum_{n\geq 0} \tau_n \int
\sigma_n(\CA^{(2)})
}
In the original \ymt\ theory there is a similar class
of deformations of the theory obtained by adding higher
Casimirs to the heat kernel Boltzman weight:
\eqn\ymgenpurr{
C_2(R)\longrightarrow \sum_{n\geq 2} t_n C_n(R)
}
It is natural to conjecture that these classes of deformed theories
are in fact equivalent. Experience from 2D gravity
\ref\mss{
G. Moore, N. Seiberg, and M. Staudacher,
``From Loops to States in $2D$ Quantum Gravity,''
 Nucl. Phys. {\bf B362}(1991)665}
leads us to expect that the change of variables
$\{t_n\}\to \{\tau_n\}$ can involve complicated nonlinear
terms. Indeed nothing in the present discussion precludes the
possibility that the pure $C_2(R)$ \ymt\ theory is equivalent
to a perturbation of type \genpurr\ with $\tau_n\not=0$
for $n>0$.

\newsec{Wilson Loops}

The techniques of \GrTa\  extend to Wilson
loop expectation values. In general the answer
is expressed in terms of rather intricate
gluing rules  \GrTa.  In this section we will restrict attention
to the simplified case of the chiral theory. The string
interpretation of
these quantities is given by macroscopic loop amplitudes
(familiar from gravity) with certain Dirichlet boundary
data on the boundary of the worldsheet.

\subsec{Observables}

The natural observables in gauge theory are the
Wilson loops. Let $R$ be a finite-dimensional
representation of $SU(N)$ and let $\G$ be a
piecewise-differentiable oriented
curve $\G: S^1\to \ST$. Such curves generically have
at most double points as self-intersections and we will
assume this to be the case.
We define:
\eqn\willoop{\eqalign{
W(R,\Gamma) &\equiv \tr_R(U_\G)\cr
U_\G& =Pexp\oint_\G A\cr}
}
we will often denote the image $\G\subset \ST$
by the same symbol.

As pointed out in \GrTa\  a more natural basis of
observables for the $1/N$ expansion are the
{\it loop functions}:
\eqn\shuri{
\Upsilon(\vec k_\G,\G)  \equiv \prod_{j=1}^\infty (\tr
U_\G^j)^{k_\G^j}
}
The vector $\kG=(k_\G^1,k_\G^2,\dots)$ determines a
conjugacy class (via cycle decomposition) in $S_{m_\G}$
where $m_\G=\sum j k_\G^j$. By Frobenius reciprocity
we have
\eqn\shurii{
\Upsilon(\vec k_\G,\G)  =
\sum_{R\in Y_{m_\G}} \chi_R(p_\G) W(R,\G)
}
where $p_\G$ is any element in the conjugacy class
$\kG$.

\subsec{Exact Answer: Nonintersecting Loops}

Suppose that we have a collection $\{ \G \}$
of {\it nonintersecting} curves in $\ST$.   Let
\eqn\ccomp{
\ST-\amalg \G = \amalg_c \Sc
}
be the decomposition into disjoint connected components.
Each component has $G_c$ handles and $b_c$ boundaries.
Since $\ST$ and $\G$ are each oriented, each curve $\G$
can be deformed into two curves $\G^\pm$ as in

\vskipabit
\ifig\fhhpiv{Using the orientation of the surface and of the
Wilson line we can define two infinitesimal deformations of
the Wilson line $\Gamma^\pm$.}
{\epsfxsize2.0in\epsfbox{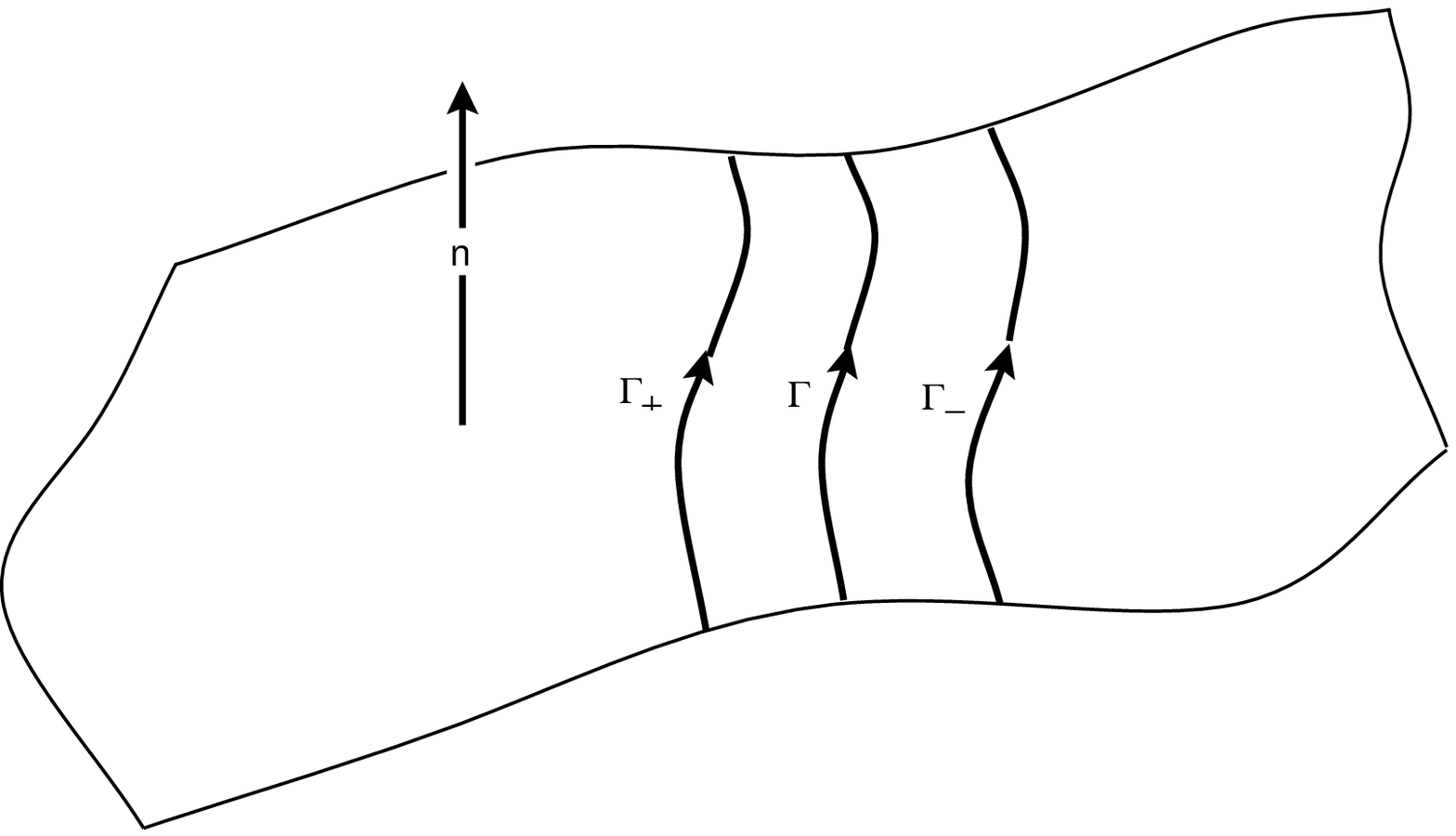}}

We let $c_\G^\pm$ denote the label of the
component $\Sc$ which contains
$\G^\pm$.
The exact answer for correlation functions of Wilson loops is
easily obtained from standard cutting and gluing techniques.
One finds:
\eqn\wilavi{\eqalign{
\biggl\langle \prod_\G W(R_\G,\G)\biggr\rangle&=
\sum_{R(c)} \prod_c \bigl(\dim R(c)\bigr)^{\chi(\Sc)}
e^{-\half A_c c_2(R(c))/N}
\prod_\G N_{R(\cgm),R_\G}^{R(\cgp)}\cr}
}
where we sum over unitary irreps $R(c)$ for each
component $c$,
 $N_{R_1,R_2}^{R_3}$ are the ``fusion numbers''
defined by the  decomposition of a tensor product into irreducible
representations
\eqn\fusruls{
R_1\otimes R_2=\oplus_{R_3} N_{R_1,R_2}^{R_3} R_3 \qquad ,}
and $A_c$ denote the areas of the components $\Sc$.
Note that $\Sc$ are open manifolds. When we speak of
the Euler character we glue back in the $b_c$
boundary circles.

\subsec{Chiral Expansion: Nonintersecting Loops}

The chiral expansion of Wilson loop averages
 may be obtained directly
from \wilavi\ without recourse to the gluing
rules of \GrTa.  To begin one derives a formula for the
fusion numbers in terms of a sum over the symmetric
group. This may be done by expressing them as integrals
of characters, passing to the  loop function basis, and then
expressing
in answer in terms of the symmetric group. The result,  in
a form useful for us, is:
\eqn\fsnmb{\eqalign{
\sum_{R \in Y_{m_\G}} \chi_R (p_\G) &
 N_{R(\cgm),R }^{R(\cgp)}=
\delta_{n(\cgp), n(\cgm)+m_\G} {d_{R(\cgp)} \over (n(\cgp)!)}
{d_{R(\cgm)} \over (n(\cgm)!)} \cr
\sum_{u_\G^-\in S_{n(\cgm)} }
\sum_{u_\G^+, x_\G\in S_{n(\cgp)} }&
{\chi_{R(\cgp)}(u_\G^+) \over d_{R(\cgp)} }
{\chi_{R(\cgm)}(u_\G^-) \over d_{R(\cgm)} }
\delta_{n(\cgp)} \bigl[u_\G^+,x_\G (u_\G^-\cdot p_\G)
x_\G^{-1}\bigr]\cr}
}
where $d_R$ is the dimension of the representation of the
symmetric group, and, in the last factor $(u_\G^-\cdot p_\G)$
is the image  under the natural embedding of symmetric
groups
\eqn\emsym{
\iota_\G^\pm:
S_{n(\cgm)} \times S_{m_\G}  ~ \longrightarrow ~ S_{n(\cgp)}
}
which takes the first permutation to a permutation of the first
$n(c_\G^-)$
entries
and the
second permutation to a permutation of the last $m_\G$ entries.

As usual we obtain the chiral sum by making the
replacement in \dfchsm.
%
%
Using the standard
set of identities from \GrTa\  together with \fsnmb,
the chiral expansion of \wilavi\ becomes:
\eqn\wilavii{\eqalign{
\biggl\langle {\prod_{\G} { \vert \kG \vert  \over {m_\G ! } }
\Upsilon(\kG,\G)} \biggr\rangle&=
\sum_{n(c)\geq 0}
\sum_{\ell(c)\geq 0}
\sum_{L(c)\geq 0} \cr
\sum_{v_1(c),\dots v_{L(c)}(c)\in S_{n(c)} }
\sum_{s_1(c),\dots t_{G_c}(c)\in S_{n(c)} }&
\sum_{u_\G^\pm \in S_{n(c_\G^\pm)} }
\sum_{x_\G \in S_{n(c_\G^+)} }
\sum_{p_\G \in S_{m_\G}}
\cr
\prod_c  \Biggl[
e^{- \half A_c(n(c)-{n(c)^2\over N^2})}
{\bigl(-A_c\bigr)^{\ell(c)} \over \ell(c)!}
 \biggl( {1\over N} &
\biggr)^{
n(c)\chi(\Sc)-\sum_i (n(c)-k_{v_i(c)}) }
\chi(\CC_{L(c)} (\Sc))  \Biggr]\cr
\prod_c \Biggl[{1\over n(c)!} \delta_{n(c)}\biggl(
\prod_{\G:\cgm=c} u_\G^-
\prod_{\G:\cgp=c} u_\G^+ &
\prod_{1}^{L(c) } v_i(c) T_{2,n(c)}^{\ell(c)}
\prod_1^{G_c} [s_i(c),t_i(c)] \biggr)\Biggr]\cr
\prod_\G  {1\over {m_\G !}} \biggl[ \delta_{n(\cgp),n(\cgm)+m_\G} &
\delta_{n(\cgp)}
\bigl(u_\G^+, x_\G (u_\G^-\cdot p_\G)x_\G^{-1}\bigr)\biggr]\cr}
}
The normalization factors in front of $\Upsilon$ are chosen for later
convenience;
$\vert \kG \vert$ is the order of the conjugacy class determined by
$p_\G$.
Despite its extremely
cumbersome appearance, this expression
has an elegant geometrical content as we shall
see.
The fourth line defines the
coverings $\Sw^c$ of components $\Sc$.
The last line describes how these covering spaces
$\Sw^c$ are glued together.

\subsec{Chiral Expansion: Intersecting Loops}

If the loops $\amalg\G$ have intersections (including
self-intersections) then the exact answer for \ymt\ is
much more complicated than \wilavi\ and involves
summing over $6j$ symbols at the intersection vertices
of the loops \witten. Nevertheless  the chiral $1/N$ expansion
for intersecting Wilson loops
has a relatively
simple set of rules which have been  worked out in
\GrTa. The only modification of \wilavii\ is the
replacement:
\eqn\parcomp{
\Sc \rightarrow \tilde \Sc
}
where $\Sc$ is constructed from the open manifold
$\Sc$ by gluing in open intervals and vertices along
$\p\Sc$. The rules for constructing $\tilde \Sc$
are as follows.  Consider $\amalg\G$ as a graph. It has  open
edges $E_j$ and vertices $v_j$. Using the orientation
we can define deformations $E_j^+$ and $v_j^{++}$.
The edge $E_j^+$ is the deformation of the edge in
the direction of $\G^+$, the vertex is obtained by deforming
into the $+$ region for each of the two intersecting
curves. We glue the edges $E_j$ to the boundary of
the component
containing $E_j^+$ and we glue the vertices $v_j$ to
the boundary of the component containing $v_j^{++}$.
We may define the Euler character of $\tilde \Sc$ to be
\eqn\ectsc{
\chi(\tilde \Sc)=\chi(\Sc) + \sum_{E_j\in \tilde \Sc} (-1)
+ \sum_{v_j\in \tilde \Sc} (+1)
}
This is {\it not} a homotopy  invariant, but it is a  homeomorphism
invariant. In the previous case of nonintersecting Wilson
loops the modification $\Sc\to \tilde\Sc$ makes no
difference since $\chi(S^1)=0$. Gross and Taylor's
rule says that the only change we must make in \wilavii\ is
the change $\Sc\to \tilde\Sc$ of \parcomp!

%
%

\subsec{String interpretation}

The string interpretation of the chiral
nonintersecting Wilson loop
averages in the $\Upsilon$ basis is stated very simply.
The vectors $\kG$ may be thought of as specifying
the homotopy class of a map from a disjoint
union of circles to $\G$: We have $k_\G^j$ $j$-fold
coverings of the circle by the circle.
\foot{Since $j\geq 1$ the homotopy class has an orientation
preserving representative.}
 The
only change that is needed in the path integral of sections 6,7
is that we have a {\it macroscopic loop amplitude}:
The worldsheet $\Sw$ has a boundary. Data
specifying the Wilson loops is encoded in  the boundary
conditions on $f:\Sw \to \ST$. These boundary
conditions state that
$f:\p \Sw \to \amalg \G$ is in the homotopy class
$\{ \kG \}$.  Boundary conditions on the metric are standard
\ref\Alvarez{O.~ Alvarez, ``Theory of strings with boundaries: fluctuations,
topology and quantum geometry," Nucl. Phys {\bf 216} (1983) 125.},
and follow from the requirements that
1) The loop $\G $ is unparametrized, and
2) $P^\dagger$ is the adjoint of $P$.
Let $n$ denote a normal vector and $t$ a tangent vector to $\p \Sw$.
We take
$g(n,t)= 0$ on $\p \Sw$. Correspondingly, vector fields $\xi$
generating
diffeomorphisms satisfy
$n.\xi=0$ and $n^at^b\nabla_{(a} \xi_{b)} = 0 $.
Boundary conditions for other fields follows from
BRST invariance and invariance of the action.
This string interpretation will be justified in the
next section.

\noindent
{\bf Conjecture 9.1}. The string interpretation for the
case of chiral intersecting Wilson loop amplitudes
is obtained by the boundary condition
that $f:\p \Sw \to \amalg \G$
is in the homotopy class
\eqn\hmtpycl{
\p \Sw {\buildrel \{ \kG\}\over \longrightarrow} \amalg_{\G} S^1
{\buildrel \amalg \G\over \longrightarrow} \amalg \G
}
The first arrow describes a covering of circles by
circles. The second is the homotopy class of the curves
defining the Wilson loops.

\subsec{Hurwitz spaces for surfaces with boundary }

We now give an argument for the claim of the
previous subsection.

\noindent
{\bf Definition 9.1}. A {\it boundary-preserving branched
covering} is a map
\eqn\bpbc{
f:\bigl(\Sw,\p\Sw\bigr)\rightarrow \bigl(\ST,\p\ST\bigr)
}
such that

1. $f:\p\Sw\rightarrow\p\ST$ is a covering map.

2. $f:\Sw - \p\Sw \rightarrow \ST- \p\ST$ is a branched covering.

Equivalence and automorphism of such maps are defined in
the obvious way. Note that the boundary components
$\p\Sw$ are unlabelled so
$\phi:\bigl(\Sw,\p\Sw\bigr)\rightarrow \bigl(\Sw,\p\Sw\bigr)$
can permute the boundaries.

By (1) $f$ determines a class $\kG$ for each component
$\G$
of $\p\ST$. Let us assume that $\ST- \p\ST$ is connected.
Then, by  (2) $f$ determines a branch locus
$S(f)\subset \ST- \p\ST$, an index $n$,
 and an equivalence class of a homomorphism
$\psi_f: \pi_1(\ST-S(f), y_0) \to S_n$.  We have the direct
analog of the Riemann existence theorem Theorem 3.1:

\noindent
{\bf Proposition 9.1}. Let $\ST$ be a connected,
closed surface with
boundary. Let $S\subset \ST- \p\ST$ be a finite set,
 and let $n$ be a positive integer.
There is a one-one
correspondence between equivalence classes of boundary-preserving
branched covers \bpbc\ with branch locus $S$ and
equivalence classes of homomorphisms
$\psi: \pi_1(\ST-S(f), y_0) \to S_n$.

{\it Proof}. The proof proceeds as before. We choose a
representative for $\psi$ and a basis of generators for
$\pi_1$. Then we glue together $n$ copies  $\ST$ according
to the data given by the homomorphism. $\spadesuit$

The maps that we will need for nonintersecting Wilson loops
are considerably more complicated than boundary-preserving
covers: we must allow for the possibility that the inverse
images of the loops $\G$ contain loops which lie in the
interior of $\Sw$.  This leads us to introduce:

\noindent
{\bf Definition 9.2}. Suppose $\Sw$ is a closed
oriented surface with boundary
and $\{\G\}$ is a collection of nonintersecting
oriented closed curves in
$\ST$. By a {\it covering map}  $f:\Sw\to \ST$ { \it with boundaries
over}
$\{\G\}$ we mean a continuous
orientation-preserving map $f$ such that

1. $f:\p\Sw\rightarrow \amalg \G$ is a covering map.

2. $f^{-1}(\amalg \G)$ is a disjoint union of circles.

3. $f:\Sw - f^{-1}[\amalg \G] \rightarrow \ST- \amalg \G$
is a branched covering.

Equivalence and automorphism are defined as before.

We now describe these maps in some detail.
As before,
by (1) $f$ determines a homotopy class $\{\kG\}$
of $\p\Sw\to\amalg \G$.
By (3),
we have a covering
\eqn\compcov{
f^c: \Sw^c\to \Sc
}
where $\amalg_c \Sw^c=\Sw-f^{-1}(\amalg \G)$.
The number of sheets of a covering will be
different for different components of $\Sc$. An elementary
example is:

\vskipabit
\ifig\fhhpiv{Different components of the target space
can be covered by different number of sheets.}
{\epsfxsize3.5in\epsfbox{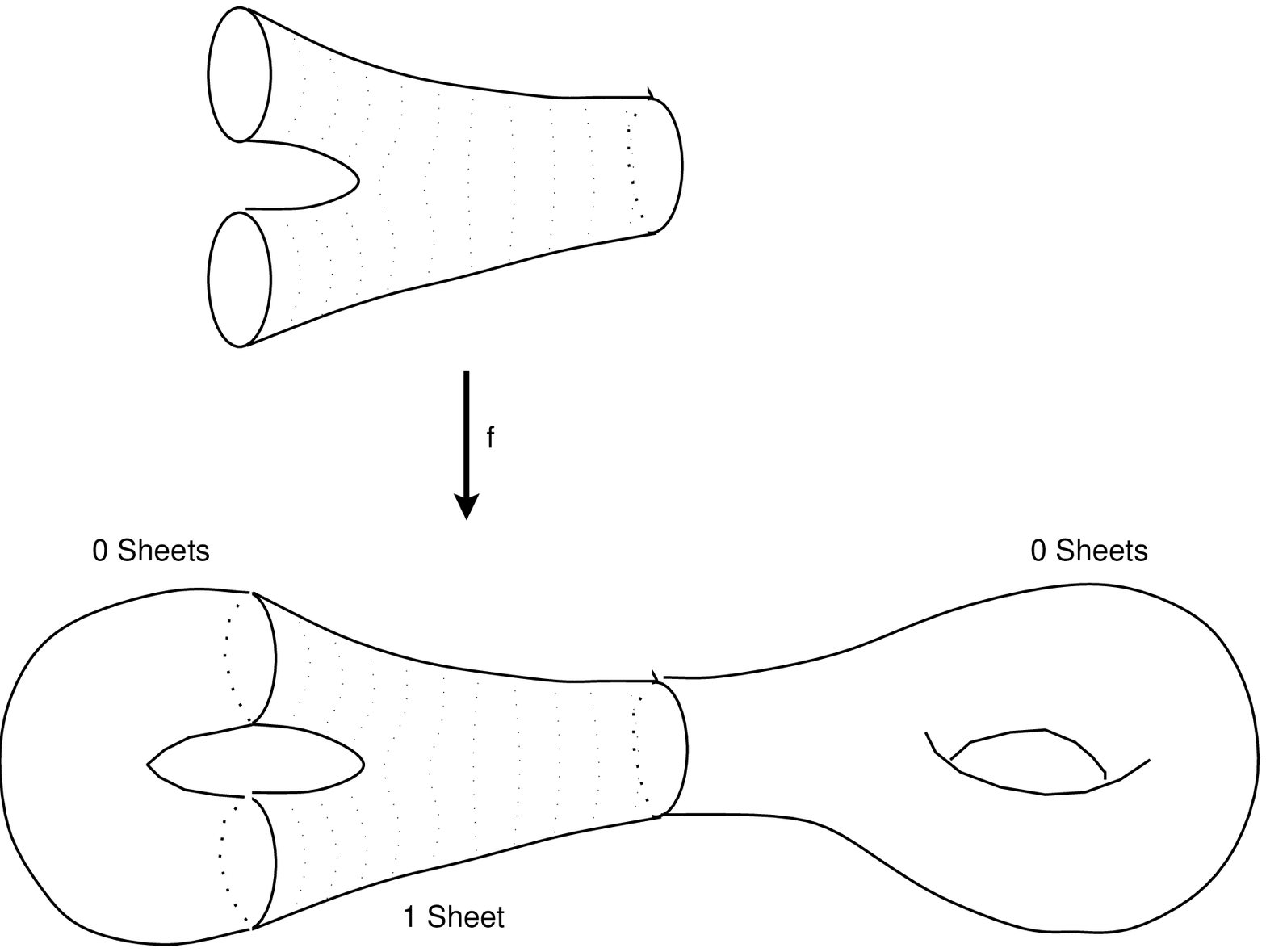}}

 In general, the   coverings $f^c$ are boundary-preserving
branched coverings, the boundaries of $\Sw^c$
covering the boundaries of $\Sc$.
 From each covering $f^c$ we obtain
a branch locus  $S_c\subset \Sc$, index $n(c)$, and equivalence
classes of homomorphisms:
$\psi_c: \pi_1(\Sc- S_c, y_{0,c})\rightarrow S_{n(c)} $.
  By (2) the inverse
image under $f$ of any loop $\G$ may be divided into
{\it interior} loops and {\it boundary} loops, that latter
living in  $\p \Sw$.
The different surfaces $\Sw^c$
must be smoothly glued together along the interior loops
of $f^{-1}(\G)$.
This requirement results in the gluing conditions $(9.5)$ to $(9.20)$
below.

First, above a loop $\G$ there are
 $\sum_j k_\G^j$ components with $m_\G$ sheets
belonging to $\p\Sw$. (Recall that
$m_\G=\sum j k_\G^j$.)
The remaining components lie in the interior of
$\Sw$. Using the orientation we see that if we perturb
the curves $f^{-1}(\G)$
 in the plus direction we get an $n(\cgp)$-sheeted
covering of $\G^+$. On the other hand, since a perturbation
of the boundary curves of $\Sw$ in
the minus direction takes us off the surface $\Sw$
 we can only perturb the interior
curves of $f^{-1}(\G)$ in the minus direction. Thus we get
an $n(\cgm)$-sheeted covering of $\G^-$.  The situation
may be summarized in the following figure:

\vskipabit
\ifig\fhhpiv{Locally the covering map looks like this.
Above $\Gamma^-$ there are only interior curves.
Above $\Gamma^+$ there are interior and boundary
curves. }
{\epsfxsize5.0in\epsfbox{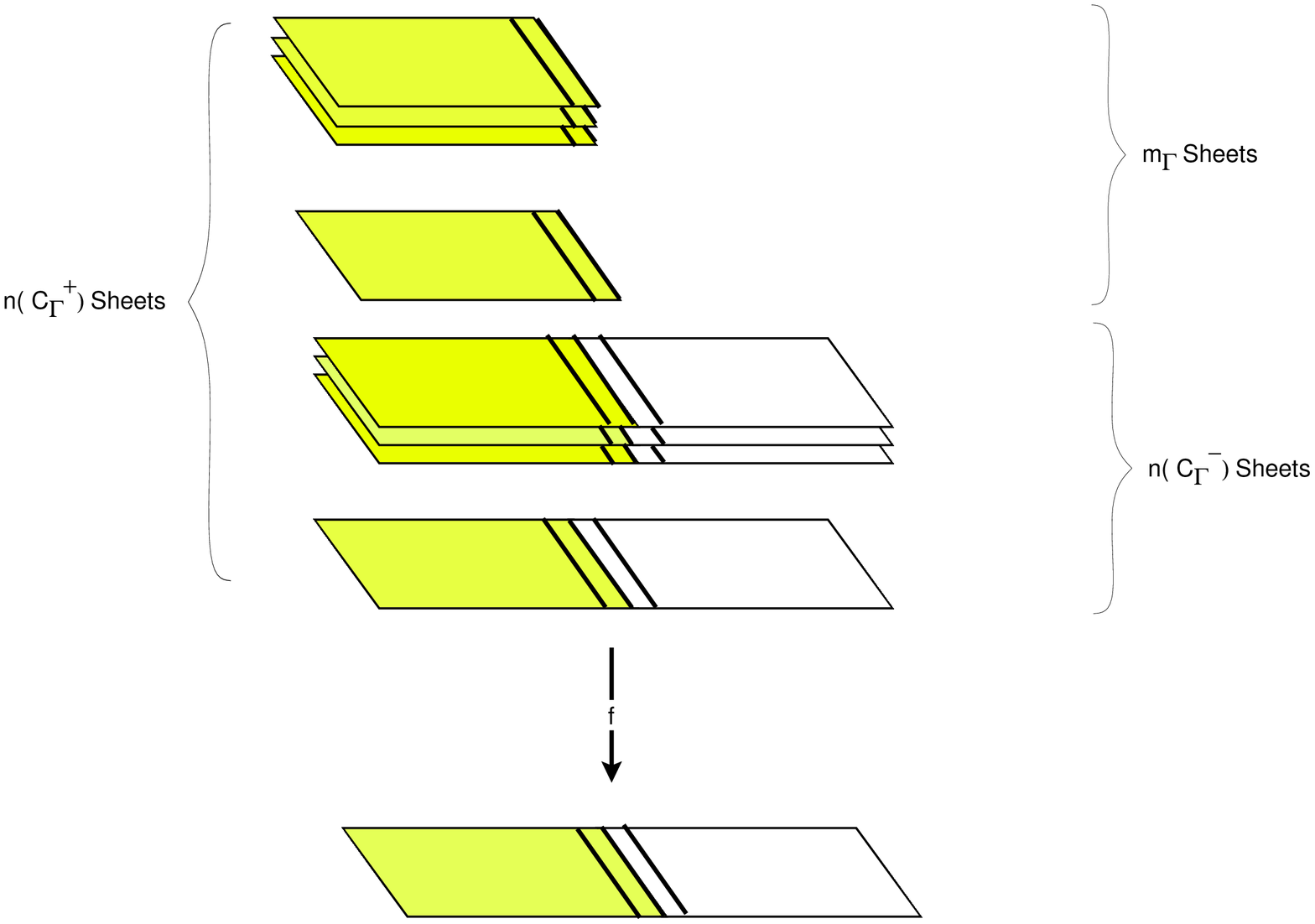}}

{}From the figure the first gluing condition
\eqn\match{
n(\cgp)= n(\cgm) + m_\G
}
becomes evident.

We obtain the second gluing condition
by starting from Proposition 9.1. As in Fig. 1 of sec. 3.1
 we can choose
generators
\eqn\comgens{
\alpha_i(c),\beta_i(c),\sigma_i(c),\gamma_i^+,\gamma_i^-
}
of
\def\PiC {\pi_1(\Sc- S_c, y_{0,c})}
$\PiC$, such that $\alpha_i(c),\beta_i(c)$, run around
handles, $\sigma_i(c)$ become trivial
if we fill in branch points,
$\gamma_i^\pm$ become trivial if we fill in $\G^\pm$,
we have the relation:
\eqn\comrels{
\prod \gamma_i^+ \prod \gamma_i^-
\prod \sigma_i(c) \prod [\alpha_i(c),\beta_i(c)]
=1
}
and the covering $\Sw^c\to \Sc$ can be constructed by
glueing together copies of $\Sc$ using the homomorphism
$\psi_c$ as in the LHS or RHS of  Figure 12.

\vskipabit
\ifig\fhhpiv{Figure showing construction of branched
covers with boundary using the data $\CD$. }
{\epsfxsize4.5in\epsfbox{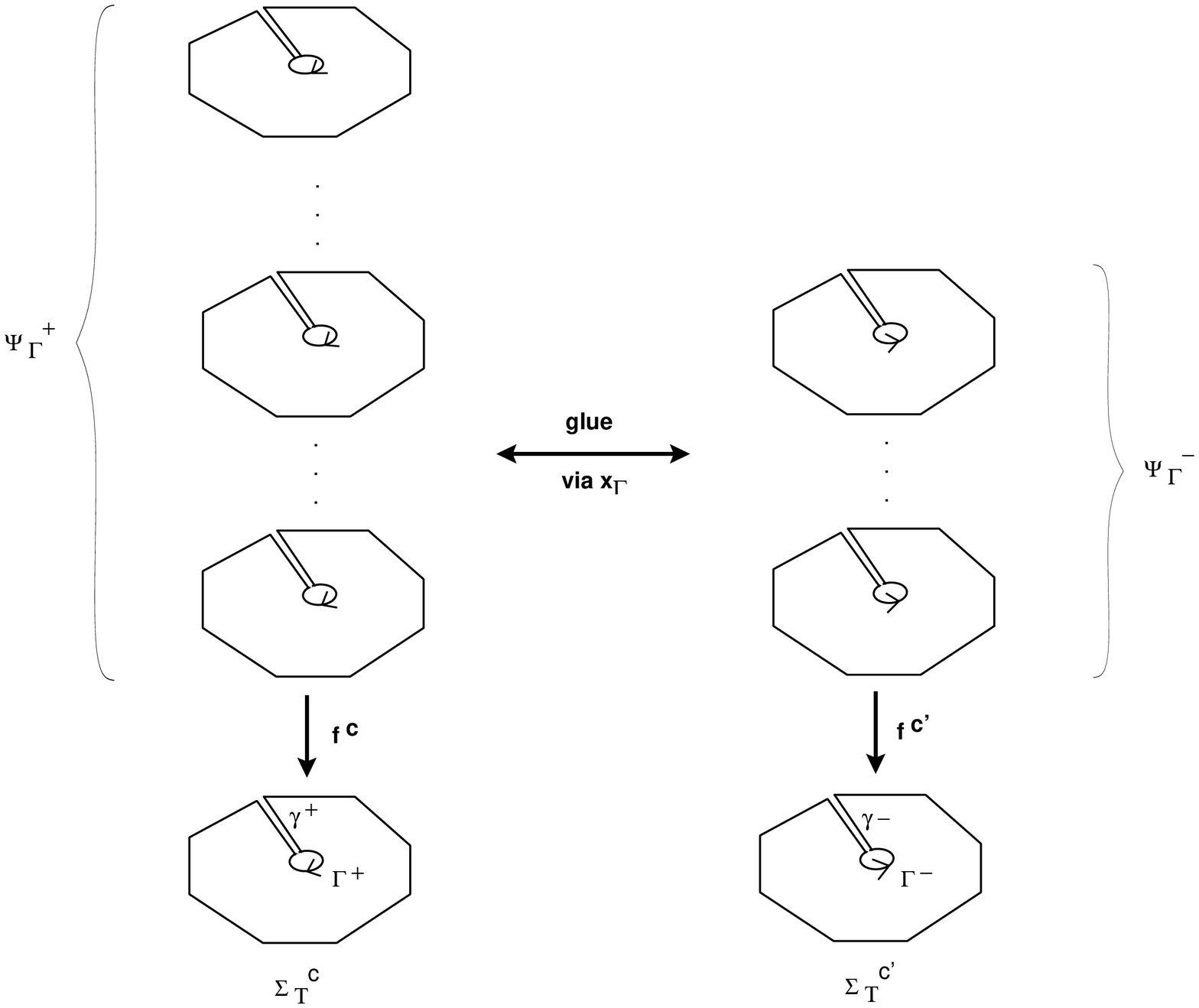}}

Consider now two components
$\Sc$ and $\ST^{c'}$ which must be glued along
a loop $\G$, as in Figure 12.
 Suppose that $\cgp=c$, $\cgm=c'$,
and that $\gamma^\pm$ surrounds $\G^\pm$.
The homotopy type of the covering of the interior circles
above $\G^-$ is given by the conjugacy class of
\eqn\psgm{
\psi_\G^-=\psi_{\cgm}(\gamma^-)
}
while the homotopy type of the covering of the
interior circles above $\G^+$ is given by the
conjugacy class of
\eqn\psgp{
\psi_\G^+ =\psi_{\cgp}(\gamma^+)
}
Because the interior curves must be smoothly
glued together there must exist a re-labelling
$x_\G\in S_{n(\cgp)}$ of
the sheets above $\ST^{\cgp}$ such that
\eqn\matchii{
\psi_\G^+ =x_\G \bigl[
\iota^\pm_\G(\psi_\G^- p_\G) \bigr] x_\G^{-1}
}
where $\iota_\G^\pm$ is the embedding \emsym\ and
 $p_\G$ is any element in
the conjugacy class $\kG$.
Finally two equivalent coverings $\tilde f=f\circ \phi $ will lead to
the same data but with $\psi_c,\tilde \psi_c$ differing
by an inner automorphism of $S_{n(c)}$.

In summary,  to any map satisfying Definition 9.2
we can unambiguously associate an equivalence class
of  {\it covering data} $\CD$. This data is  composed of

\noindent
D1. Branch loci $S_c\subset \Sc$, basepoints $y_{0,c}$,
indices $n(c)$, boundary data $\kG$.

\noindent
D2. Generators \comgens\  of $\PiC$

\noindent
 D3. Homomorphisms $\psi_c:\PiC\to S_{n(c)}$.
These are obtained after  choosing  a labelling of the inverse images
of
$y_{0,c}$.

\noindent
D4. Elements $p_{\Gamma} \in S_{m_{\Gamma}}$. These are obtained
after choosing
basepoints $y_\G \in \G $ and choosing a labelling of the points
lying on the
boundaries
of $\Sw$ in $f^{-1}(y_\G)$.

\noindent
D5. Elements $x_\G\in S_{n(\cgp)}$

These data are required to satisfy the conditions:

\noindent
C1. $n(\cgp)= n(\cgm)+m_\G$

\noindent
C2. $\forall \G~  \exists p_\G\in S_{m_\G}$ such
that $[p_\G]=\kG$ and
$\psi_\G^+ =x_\G \iota^\pm_\G(\psi_\G^- p_\G) x_\G^{-1}$

Two sets of data $\CD$ and $\tilde \CD$ satisfying
these conditions will be considered to be equivalent if
the following relations holds:

\noindent
E1.  The bases $\alpha_i(c),\dots $, $\tilde \alpha_i(c),\dots $
differ by $Aut(\PiC)$.

\noindent
E2. $\tilde \alpha_i=\alpha_i,\dots$, and
 $\exists w(c)\in S_{n(c)}$, $w(\Gamma)\in S_{m_{\Gamma}}$ such that
\eqn\edata{\eqalign{
\tilde \psi_c(\cdot) &= w(c) \psi_c(\cdot) w(c)^{-1} \cr
\tilde x_\G  &= w(\cgp) x_\G w(\cgm)^{-1}w(\G)^{-1}   \cr
\tilde p_\G &=  w(\G) p_\G w(\G)^{-1} \cr
} }

We define $ C(  \psi_c,  x_\G, p_\G  )$ to be the  subgroup of
$\prod S_{n_c} \prod S_{m_\G}$ which leaves the set  $\{ \psi_c,
x_\G, p_\G
\} $
fixed under the action defined above.

The above discussion has proved half of:

\noindent
{\bf Proposition 9.2}. There is a one-one correspondence between
equivalence classes of
covering maps with boundaries over $\amalg \G$ with
prescribed data:

a1. $f$ defines an $n(c)$-fold cover of $\Sc$, with branch locus
$S_c$

a2. $\p \Sw\to \amalg \G$ is of homotopy type $\kG$

and
equivalence classes of covering data $\CD$
as defined above.

{\it Proof}.
Given the data $\CD$ we first construct boundary-preserving
branched coverings $f^c:\Sw^c\to \Sc$ in the standard way.
Then, with the labelling of the sheets specified by the
data we glue the loops covering $\G^-$ to the interior
loops covering $\G^+$ using the relabelling $x_\G$.
This gluing is smooth by the conditions defining $\CD$.
Equivalent data induce equivalent coverings. Note in particular
that $\prod_c S_{n(c)} \prod_\G S_{m_{\G}} $ acts transitively on
equivalence
classes of relation $E2$ and that the number of
distinct elements in a class is given by
\eqn\degen{
{\prod n(c)! \prod m_\G ! \over  \vert C(  \psi_c,x_\G, p_\G  )\vert
}
}
It is also clear from this construction that the automorphism
group of the covering is $Aut(f)=  C(  \psi_c, x_\G , p_\G ) $.
$\spadesuit$

\noindent
{\bf Definition 9.3}.
The Hurwitz space
$H(h,b,n(c),S_c,\kG)$ is the space of equivalence classes
of covers $f$ with boundaries
above $\amalg \G$, where $\Sw$ has $h$ handles,
$b$ boundaries, such that $f$ is an $n(c)$-fold covering
of $\Sc$ with branch locus $S_c$,
and restricts to $\p\Sw\to \amalg \G$ of
homotopy type $\kG$. The union of these
spaces over sets of branch points $S_c$ with
$L(c)$ elements defines the Hurwitz space
$H(h,b,n(c),L(c),\kG)$. The compactification of
this space with different $L(c)$ defines the
Hurwitz space $H(h,b,n(c),\kG)$.

A corollary of Proposition 9.2 is  that $H(h,b,n(c),L(c),\kG)$
is a discrete fibration over $\prod_c \CC_{L(c)}(\Sc)$.
Therefore, the orbifold Euler characteristic is defined as usual:
\eqn\orbeulb{
\chi_{\rm orb}\biggl(H(h,b,n(c),L(c),\kG )\biggr)
= \prod_c \chi(\CC_{L(c)}(\Sc) )\sum_{[f]\in
H(h,b,n(c),S_c,\kG)} {1\over \vert Aut(f)\vert}
}
where, on the RHS we may choose any set $S_c$ of
$L(c)$ distinct points in $\Sc$.

Now finally let us compare with the chiral expansion
\wilavii. If we first put $A(c)=0$ then the effects of
homotopically trivial loops $\G$ can be shown to be
trivial, simply contributing overall powers of $1/N$.
Nevertheless, homotopically nontrivial loops
have nontrivial correlators. Combining \wilavii\ with
Proposition 9.2 we see that
 the chiral expansion becomes:
\eqn\wilaviii{
\sum_{n(c),h,b\geq 0} \biggl({1\over N}\biggr)^{2h+b-2}
\chi_{\rm orb}\biggl(H(h,b,n(c),\kG )\biggr)
}
as in the previous sections.
Thus, with an appropriate
choice of contact terms decoupling the $\pm$ sectors
\foot{See section 11 below.}
the macroscopic loop path integral described in
section 9.4 will produce the product of
chiral Wilson loop averages, in complete analogy to the
partition function.

Finally,
we include the effects of the area in
the Wilson loop averages, as computed in
\wilavii. The structure is exactly the same as that
found in sec. 8 and the same discussion shows that -
at least on the simple Hurwitz sub-space
$H(h,b,n(c),L(c)=B(c),\kG)$ the contact terms account for
the area.

\subsec{Divers Remarks on Wilson Loop Averages}

We gather several miscellaneous remarks in this
subsection.

1. First, even in the chiral theory the
other
$A_c$-dependent polynomials in \wilavii\
associated with non-simple Hurwitz spaces remain
to be analyzed.

2. The discussion should be generalized to the chiral case
allowing intersections of the graph $\amalg \G$.
The conjectural string formulation (Conjecture 9.1) will
follow from

\noindent
{\bf Conjecture 9.2}. The chiral intersecting loop
amplitudes may be discussed as in section 9.6 with
by simply modifying condition 2 in definition 9.2
by allowing $f^{-1}(\amalg \G)$ to be a graph and
modifying Proposition 9.2  and Definition 9.3
by allowing the branch locus $S_c$ to {\it intersect}  $\G$.
In order to avoid double-counting we must let
$S_c$ approach from the $+$ side only.  The only effect
on Proposition 9.2 and \orbeulb\ is the replacement:
$\Sc\to \tilde \Sc$.

3. The extension of our
discussion to the coupled theory is very
nontrivial.  The rules of \GrTa\ become
considerably more elaborate. Some of the
issues arising in this extension appear to
be quite relevant to establishing a string
picture of {\it four-dimensional} QCD along
the lines of\ref\Kostov{I.K.~ Kostov, ``Continuum QCD2 in terms of
Discretized Random Surfaces with Local weights,'' Saclay-SPht-93-050,
Jun 1993, hep-th/9306110.}.

4. An interesting open problem is the derivation of
the Migdal-Makeenko loop equations from the
topological string theory point of view. Given our
experience with 2D gravity, we may guess that there
is an analog of $W_\infty$ constraints on the partition
function which is equivalent to the loop equations, and
that these $W_\infty$ constraints
 follow from a contact term analysis.

5. Further investigation of Wilson loop amplitudes also
promises to yield some extremely interesting
insights in mathematics.
Recently, V.I. Arnold has discovered new invariants
of plane curves (immersions)
\ref\arnold{See V. Arnold, `` Remarks on enumeration of plane
curves'';
`` Plane curves, their invariants, Perestroikas and
classifications.''}.
We  remark that if $S_k(x_1,x_2,\dots )$
are elementary symmetric polynomials then
$S_l({\p \over \p A_c})\vert_{A_c=0} \langle \prod \Upsilon\rangle$
are also invariants of immersions (by the area-preserving
diffeomorphism invariance of \ymt.) The relations of
these invariants to the mathematics of covering spaces
may well be very rich.

6. Finally we mention that, following
\ref\strominger{A. Strominger,  ``Loop space solution of
Two-Dimensional QCD''
Phys. Lett. {\bf 101B}, 271(1981)}
\ref\barshan{I. Bars and A. Hanson,
Phys. Rev. {\bf D13},1744(1976);
Nucl. Phys. {\bf B111}, 413(1976)}
\GrTa\ one should be able to incorporate dynamical quarks into
the present framework. One must modify the string theory
by turning it into an open-closed (Dirichlet) string.
%
%

We hope to return to these issues in future work.

\newsec{The  coupled  theory}

We have seen that some aspects of the chiral $YM_2$ theory are
related
to the topological string theory of sec. 6.
In the next two sections we will show how the full ("coupled") theory
also fits into the framework of the theory of a \top\ string theory.
We will restrict attention to reproducing the partition function
\fullgt. The
key observation is that $Z$ in \fullgt\ differs from a product of
chiral
partition functions $Z^+Z^-$ through the contribution of boundaries
of
the space ${\rm Maps} \times {\rm Met(\Sw)}$.

In sections 10.1-10.5 we will use the geometrical picture  of \GrTa\
to  show that the $A=0$  partition function $\fullgt$ computes Euler
characters of spaces of maps from singular surfaces $\Sigma_W$ to
$\Sigma_T$.  In 10.1, 10.2 we describe in detail
 the maps and worldsheets in question.
In 10.3 we develop a combinatoric approach to the space of maps.
In 10.4 we describe the relation of the space of maps to
configuration
spaces. In 10.5 we write the zero area QCD sum in terms of
Euler characters of the spaces of maps.

\subsec {Degenerated  coupled covers}

Let  $\Sw$ be a smooth surface, perhaps with double
points. Topologically this means that there is a set of points
$\{P_1,\dots,P_d\}\subset \Sw$ such that a local
neighborhood $D_i$ of $P_i$ is the one-point union of
disks  $D_i^{(1)},D_i^{(2)}$:
\eqn\disun{
D_i = D_i^{(1)}\amalg D_i^{(2)}/ (P_i^{(1)} \sim P_i^{(2)})
}
The {\it normalization}  of $\Sw$, $N(\Sw)$ is the smooth
surface obtained by replacing $D_i\to D_i^{(1)}\amalg D_i^{(2)}$.
$N(\Sw)$ may be connected or disconnected. A map
$f: \Sw\to \ST$ defines a normalized map $N(f)$ in a natural way.

\noindent
{\bf Definition  10.1} Let $\Sw$ be a surface with double points.
A {\it degenerate branched cover}  $f:\Sw\to \ST$ is a
continuous map such that

1. $N(f):N(\Sw)\to \ST$ is a branched cover, and

2. If $P_i^{(1)}, P_i^{(2)}$ are the normalizations of the
double points $P_i$ then
\eqn\rameq{
 Ram\bigl(N(f),P_i^{(1)}\bigr) = Ram\bigl(N(f),P_i^{(2)}
\bigr)=e_i
}
where $Ram$ is the ramification index.

The cover in the neighborhood of the double point may be
thought of as a degeneration of a family of  $2e_i$-sheeted
covers of  annuli by annuli degenerating to the cover of one
disk by two disks.

One of the very strange aspects of the \ymt\ partition
function is that it appears to involve maps which are
branched covers which are neither holomorphic nor
anti-holomorphic.

\noindent
{\bf Definition 10.2} A  {\it coupled map}  $f:\Sw\to \ST$ of Riemann
surfaces is a
continuous map such that there  are  circles
$S_i$ which separate $\Sw$ into two disjoint surfaces
\eqn\dissurf{
\Sw=\Sw^+ \amalg  \Sw^-/(S_i^+ \sim S_i^-)
}
and such that $f:\Sw^+\to \ST$ is holomorphic while
$f:\Sw^-\to \ST$ is antiholomorphic.

An example of such a map is a mapping of  the complex plane to a
closed
disk given by $f(z)= z^n$ for $\vert z\vert\leq 1$ and $=1/\bar{z}^n$
for
$\vert z\vert\geq 1$. Note that $df$ is discontinous along the unit
circle.

Our main object of interest combines the above two
notions and will be called a degenerated coupled
cover. It is a coupled map, where the circles $S_i$
have been shrunk to points.  More formally, we state

\noindent
{\bf Definition 10.3} A {\it degenerated coupled cover}   (dcc)
$f:\Sw\to \ST$
of
Riemann
surfaces is a map such that if we take the normalization of
the double points $\{Q_1,\dots Q_d\}$ then we have a
disjoint decomposition into smooth surfaces
$N(\Sw)=N^+(\Sw)\amalg N^-(\Sw)$ such that
$f^+:  N^+(\Sw)\to \ST$ is holomorphic,
$f^-:  N^-(\Sw)\to \ST$ is antiholomorphic and
such that
 \eqn\rameqi{
Ram(f^+,Q_i^+)= Ram(f^-,Q_i^- )
}
Two dcc's $f_1$ and $f_2$
 are said to be equivalent if
there is a homeomorphism $\phi:\Sigma_W \rightarrow \Sigma_W$ such
that $f_1 \circ \phi = f_2$. A homeomorphism $\phi$ of $\Sw$ is an
automorphism of the dcc if $f\circ \phi = f$.
 If $\Sigma_W$ has no double points then the map is just a
branched cover, either holomorphic or antiholomorphic.

As with branched covers we may associate several natural
quantities to a dcc. $f^\pm$ define indices $n^\pm(f)$,
branch loci:
\eqn\brloci{
S^\pm(f)=\{ f^\pm(Q)\vert  Ram(f^\pm,Q)>1\}
}
double point locus:
\eqn\tubelocus{
S^T(f)=\{ f(Q)\vert {\rm Q \ is \ a\ double\ point}\}
}
tube number
\eqn\tubenumber{
d(f,P)=Card[\{ Q\vert P=f(Q), {\rm Q \ is \ a\ double\ point}\}]
}
ramification vectors:
\foot{These vectors are considered to be infinite tuples of positive
integers with almost all entries zero.}
\eqn\ramvct{\eqalign{
\vec r^\pm(f,P) &=(r_1^\pm,r_2^\pm, \dots )\cr
r_j^\pm(f,P)&=Card\biggl[ \{ Q\vert  f^\pm(Q)=P, Ram(f^\pm,Q)=j\}
\biggr] \cr}
}
and, finally, homomorphisms :
\eqn\indhomo{
\psi^\pm: \pi_1(\ST-S^\pm(f),y_0)\to S_{n^\pm}
}

In the ordinary case the specification of the branch locus and the
homomorphism $\psi$ essentially specified the  equivalence class of
the
map $f$ as in Theorem 3.1.
This is no longer the case for dcc's because there are
many ways in which the ``double points'' connecting the different
ramification points can be introduced. This leads us to the
combinatoric
discussion of subsection 10.3

\subsec{Degenerating coupled covers}

The answer provided by \ymt\ demands a further
refinement of the above ideas. We must take into
account the way in which a family of covering maps
has degenerated to a dcc.

We will define a local degenerating family of
coupled covers to be specified by the following
data.

\noindent
1. We have a plumbing fixture degenerating to
 the double point of $\Sw$:
\eqn\plumb{
U_q = \{(z_1,z_2)\vert  z_1 z_2=\eta q,  q \le \vert z_1\vert
,\vert z_2\vert <1\}
}
where $0\leq q <1$ and $\eta$ is an $n^{th}$ root of
unity for a positive integer $n$.

\noindent
2. On the plumbing fixture we have a family of
covering maps
\eqn\mdleff{
f^{q,n} (z) ~=~ \cases{z_1^n & for $ q ^{1/2}\le\vert z_1\vert <1$\cr
                         \bar z_2^n & for $ q ^{1/2}\le \vert
z_2\vert <1$\cr}}
Notice that  $n$ different degenerating complex structures on
$\Sw$ determine maps $f^{q,n} $ projecting to the same target space.
Fig. 13 illustrates the model in the case where
$n=2$.

\vskipabit
\ifig\fhhpiv{Local model for a degenerating coupled cover with n=2.
The region
between stripes single-covers the annulus. }
{\epsfxsize3.5in\epsfbox{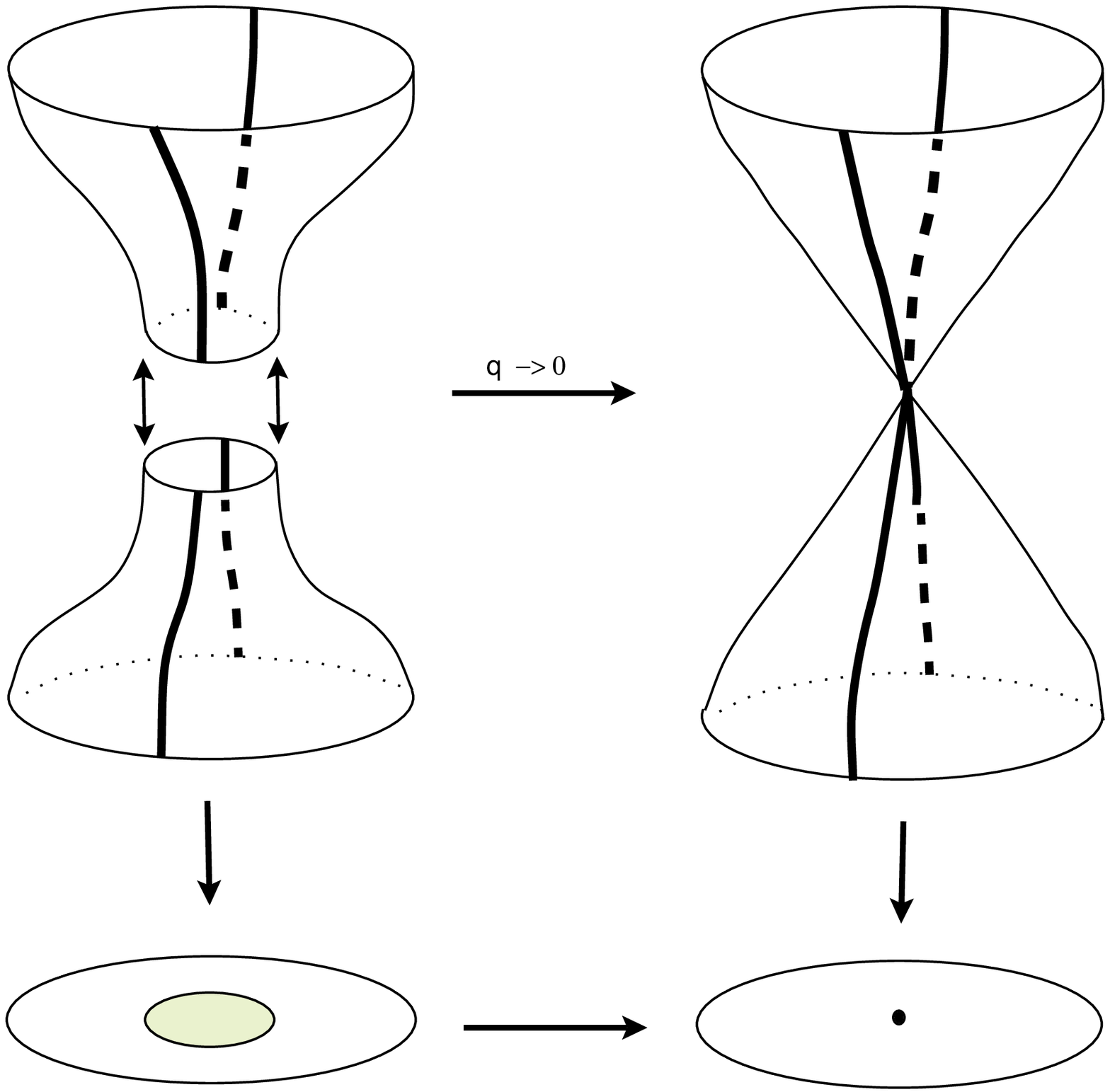}}

We
will want to
distinguish different $f$'s corresponding to these $n$
different degenerations, so we introduce:
\noindent

{\bf Definition 10.4} A  {\it degenerating} coupled cover (Dcc)
is a dcc equipped with a
choice of a locally degenerating family of
coupled covers (\plumb,\mdleff) for each double point.

\noindent
{\bf Remarks.}
1. In a degenerating family
$f^{q,n} $ is not differentiable along $\vert z_i \vert =  q  ^{1/2}$
because the normal derivative is discontinuous. It lives in the space
of
piecewise differentiable maps
rather than in the space of $C^\infty$
maps. The discontinuity is proportional to $q^{n-1 \over 2}$ and goes
to zero
when $q$ goes to
zero, for $n>1$.

2.  Our definition is admittedly somewhat
{\it ad hoc} and could be considerably improved.
The partition function of  $YM_2$
computes the Euler characters of
spaces of `degenerating coupled covers ,' as
opposed to those of dcc's (see Proposition 10.1).  The introduction
of the
discrete choice of degenerating family
accounts for an important combinatoric factor proportional to
the product over all the double points,
of the index of the ramification points
being joined at each double point.

3. Much has been made
on the suppression of ``folds'' and ``fold degrees
of freedom'' in \ymt.  It is perhaps worth noting that at
$q\ne 0 $
the map \mdleff\ has a fold,
 but this fold disappears for $q \rightarrow 0$.
In general, in the formulation of this
paper folds are suppressed because they are
incompatible with holomorphy of the map $f$.

\subsec{Combinatoric description of degenerated coupled covers}

We give now a combinatoric description of dcc's, establishing a 1-1
correspondence between data defined in terms of symmetric groups and
maps defined geometrically in the previous subsection. We will
discuss here
dcc's with parameters $n^\pm,B^{\pm},S= S^+ \cup S^- \cup S^T $
fixed. We
denote $L=\vert S \vert $. When we wish to emphasize the cardinality
of $S$ we
write $S(L)$.

Consider a dcc. Pick a base point $y_0$ on the target
space and label the inverse images $x_1^+$ to $x_{n^+}^+$ on
the holomorphic side and
$x_1^-$ to $x_{n^-}^-$ on the antiholomorphic side.
Inclusion
gives natural maps:
\eqn\inclus{
\pi_1(\Sigma_T - S(L),y_0)
{\buildrel i_* \over \longrightarrow} \pi_1(\Sigma_T - S^\pm(f),y_0)
{\buildrel j_* \over \longrightarrow} \pi_1(\Sigma_T ,y_0)
}
The map $i_*$ naturally defines homomorphisms
$\psi_L^\pm:\pi_1(\Sigma_T - S(L),y_0)\to S_{n^\pm}$
which factor through the homomorphisms $\psi^\pm$
of the previous section.
Now choose a set of generators $\alpha_i,\beta_i,\gamma(P)$,
$P\in S(L)$ of $\pi_1(\Sigma_T - S(L),y_0)$.
Once we have
chosen a set of generators,
each loop
$\alpha_i,\beta_i,\gamma(P)$ defines
a pair of permutations $(s_i,\tilde s_i)$, $(t_i,\tilde t_i)$,
and $(v (P), w (P)) $ in $S_{n^+} \times S_{n^-}$.

\bigskip
The behavior of a dcc at a double
point determines some further data from the following construction.
Let us
choose a representation of $\ST-S$ as in Figure 1 of section 3.1. If
$\gamma(y_0,P)$ is a curve from $y_0$ to $P$ then we may lift this
curve with
$f^\pm$. We denote the endpoint of the lifted curve by
$x_a^\pm.\gamma(y_0,P)
$, where we choose $x_a^\pm$ as the
lift of the initial point.
If $v(P)$    has a cycle $(a_1,\cdots, a_k)$ of length $k$ then
$x_a^+.\gamma(y_0,P)$ will be
ramification point  $Q^+$ over $P$ of index $k$. Thus,
a choice of curves $\gamma(y_0,P)$ allows us to define a pairing of
the cycles
in
$v(P)$ with those in $w(P)$. To be precise, we introduce the
following
definition.
\noindent

{\bf Definition  10.5}
Let $v  \in S_{n^+}$ and $w  \in S_{n^-}$.
Let $Cyc(v )$ be the set of cycles in the cycle decomposition of
$v $, and $Cyc(w )$ be the set of cycles of $w $. A {\it pairing}
of
$(v ,w )$ is a subset $K \subseteq Cyc(v )\times Cyc(w )$
such
that

1. $(\alpha , \beta) \subset K$
only for cycles $\alpha,\beta$ of equal length.

2. Projections  $K \rightarrow Cyc (v )$ and $K \rightarrow
Cyc(w )$
   are injective.

The second condition expresses the fact
that a ramification point in the holomorphic
sector can be connected to at most one ramification point in
the antiholomorphic sector, i.e there are only double points and no
higher
singularities. The cardinality  $\vert K\vert $ is the number of
pairings. Let
$J_{v  w }$ denote the set of all pairings of $(v ,w )$.

{\bf  Example}
 Suppose  $v  = (12)^+ (3)^+ (4)^+$
and $w  = (12)^- (34)^-$. If $ K= \{ \bigl(  (12)^+, (12)^- \bigr)
\}$ , then
one pairing has been made and
the other cycles  have been left unpaired.  This pairing is
illustrated by the
figure below.
This means that in the degenerate
branched cover inducing this map the point $x_1^+.\gamma(y_0,P)$ is a
double
point coinciding with   $x_1^-.\gamma(y_0,P)$.

\vskipabit
\ifig\fhhpiv{Glueing together ramification points according
to the data of a pairing.}
{\epsfxsize2.5in\epsfbox{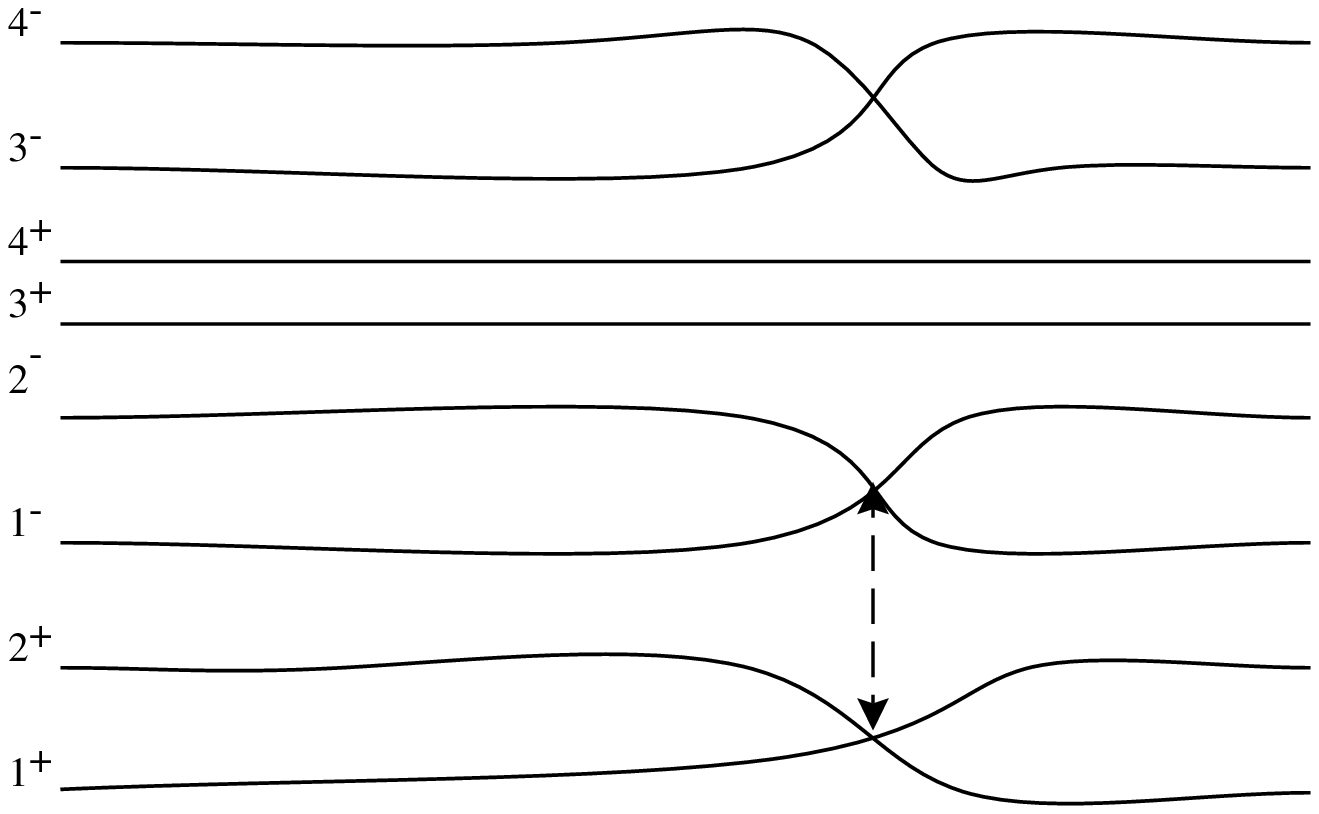}}

Thus, in summary, a dcc, together with a choice of basis of $\pi_1$
and curves
$\gamma(y_0,P), P\in S(L)$ describes elements

(1) $(s_i,t_i)\in S_{n^+}$

(2) $(\tilde s_i,\tilde t_i) \in S_{n^-}$

(3) $\bigl( v (P), w (P)) \in S_{n^+} \times S_{n^-} \forall P\in
S(L)$.

(4) $K_P \in J_{v (P),w (P)} \forall P \in S^T $

This data we call a {\it configuration}.
Moreover, conjugation by $S_{n^+} \times S_{n^-}$ acts on the data
$1-4$,
and defines an equivalence relation. We let CFG stand for the set of
equivalence classes. The subgroup of $S_{n^+} \times S_{n^-}$ which
leaves an
element $e$
of CFG invariant is called $C(e)$.


{\bf Example}  : Suppose L is $1$, $n^+=n^-=4$,  and there is one
tube over the
branch
point $P$. Let  $v(P)=v$ and $w(P)=w$, where v and w are
the same as in the previous example.
A  possible CFG, $e$,   has as representative the {\it configuration}
defined
by
$ K= \{ \bigl ( ( 12)^+, (12)^- \bigr) \} $ together with the $s$'s,
$t$'s and
${\tilde s}$'s,
${\tilde t}$'s. Conjugating by the permutation  $ (12)^- (34)^-$
leaves this pairing
invariant.  Suppose  it also leaves the ${\tilde s}$'s and ${\tilde
t}$'s
invariant. Then  $ \bigl ( 1, (12)^-(34)^- \bigr ) $  is an element
of $C(e)$.
The permutation
$ \bigl ( 1, (14)^-
(23)^- \bigr) $,  on
the other hand,  does not leave the pairing $K$   invariant although
it does
leave
$w $ invariant. So it cannot be an element of $C(e)$.

\bigskip
{\bf Proposition 10.1}
There is a one-one correspondence between elements of CFG and
equivalence classes of degenerated coupled covers (dcc's).

\bigskip
{\bf Proof}

We describe the proof in three steps.

1. Equivalent {\it configurations} come from equivalent dcc's.
Suppose two maps $f_1:\Sigma_1 \rightarrow \Sigma_T$ and $f_2:
\Sigma_2
\rightarrow \Sigma_T$ determine equivalent {\it configurations}
related by a
permutation
$g$ in $S_{n^+} \times S_{n^-}$.
 Delete the set $S(L)$ from $\Sigma_T$
and its inverse images from $\Sigma_1$ and $\Sigma_2$
to give ${\bar \Sigma_1}$ and ${\bar \Sigma_2}$ . Then $f_1$,$f_2$
 restrict to
 unbranched covers of $\Sigma_T - S(L)$ by
${\bar \Sigma_1}$ and ${\bar \Sigma_2}$
respectively. The proof that $f_1^\pm$ and $f_2^\pm$ give equivalent
branched covers follows from Theorem 3.1.
In the inverse image of $y_0$, $\phi$ restricts to the
permutation $g$ which conjugates the {\it configuration} associated
to $f_1$
into
that
associated with $f_2$.
In the case of  ordinary branched covers we just use
continuity to complete the homeomorphism $\phi$ over the
deleted points.
In the  case  of dcc's we have to prove that when
two points of ${\bar \Sigma_1}$ over P get identified their
 images under the
homeomorphism $\phi$ also get identified as $\Sigma_2$ is
reconstructed
from ${\bar \Sigma_2}$. This
follows from the uniqueness
of  lifted paths \Mas\  with  given starting point on the cover,
which
implies that for any $x_i$
\eqn\lift{
(\phi (x_i)).\gamma (y_0,P^{'})=  \phi (x_i.\gamma (y_0,P^{'})) }
where  $P^{'}$ is a point infinitesimally close to P.

2. Equivalent dcc's determine equivalent  {\it configurations}.
 This part again uses the proof in the case of ordinary branched
covers together with uniqueness of lifted paths.

3. The map from CFG to equivalence classes of dcc's is onto.
This  third part follows from the usual
proof in the case of branched covers
together with our choice of allowed pairings in CFG.{$\spadesuit$}

\bigskip

 A corollary of Proposition 10.1  is that the group
$Aut f$ is isomorphic to
$C(e)$, where e $\in $ CFG represents the equivalence class of $f$.

{\bf Remarks}

1. Each  element $e$ of CFG corresponds to one {\it degenerated }
coupled
cover.
 We denote by  $m(e)$, the number of
{\it degenerating} coupled covers associated to it. $m(e)$  is the
product
of the  common cycle lengths over all the pairings.

2. $Aut f $ is a subgroup of $Aut f^+ \times Aut f^-$, and therefore
$C(e)$ is
a subgroup
of $C(\psi^{+}) \times C(\psi^-) $ where $e$ is an equivalence class
of dcc's
corresponding to
a pair of
equivalence classes $\psi^+$, $\psi^-$ of branched covers.  In
general $C(e)$
is a proper subgroup.  As in section  4.2  we  see that $Aut f$ is a
subgroup
of $Aut (\Sw)$,
where  $\Sw $ can now have double points.

\subsec{Coupled Hurwitz Space}

Now let $CHS(h,G)$ be the space of
degenerating coupled covers from a surface of
genus $h$ to a surface of genus $G$. Let $S$ be the union of the
branch and tube loci. Define
$CHS(h,G,L)$ to be the space of Dcc's for which the
set $S$ has  $L$ points.
We isolate the subspace of coupled branched covers
where $n^+$ is the total degree of the map from holomorphic-sector,
$B^+$ is the holomorphic
branching number, $n^-$, $B^-$  are corresponding
quantities for the anti-holomorphic sector; and $D$ is the total
number
of double points. We call this $CHS(n^\pm,B^\pm,L,D)$.
If we specify the locus $S(L)$ then we define the {\it finite set }
$CHS(n^\pm,B^\pm,S(L),D)$. The space $CHS(n^\pm,B^\pm,L,D)$ is
a
bundle over the configuration space of $L$ points on $\Sigma_T$, with
discrete fibre $CHS(n^\pm,B^\pm,S(L),T)$.

Following the
reasoning of section 5 we may write a formula for the
orbifold Euler characteristic of $CHS(n^\pm,B^\pm,L,D)$.
By definition this may be taken to be:
\eqn\eulcha{\eqalign{
\chi&\biggl( CHS(n^\pm, B^\pm, L, D) \biggr)_{\rm orb}\cr
&\equiv~ \chi(\CC_L(\ST))
\sum_{[f]\in CHS(n^\pm,B^\pm,S(L),D)} {1\over \vert Aut(f)\vert}\cr}}

We first show how to count equivalence
classes of degenerating coupled covers  $[f]$
inducing the data \brloci {--} \indhomo\ compatible with
$n^\pm,B^\pm,L,D$.

It will be convenient to introduce the following
quantities. Given
two vectors $\vec r^\pm$, of nonnegative integers with almost all
entries zero,
 we define the polynomial
\eqn\polyn{\eqalign{
\wp(\vec r^+,\vec r^-, x) &=\prod_{j=1}^\infty
\biggl[ 1+\sum_{\ell=1}^\infty
x^\ell
\ell ! {r_j^+ \choose \ell } {r_j^- \choose\ell } j^\ell\biggr]\cr
&= \sum_{t=0}^{\infty} x^t \wp (t,\vec r^+,\vec r^-) .\cr}
}
The binomial coeficient ${r_j^{\pm} \choose \ell }$ is defined to be
zero
for $r_j^{\pm} < \ell$.  If $v$ and $w$ are in symmetric groups
we also define   $p_{v \otimes w }^{(d)}$ to be
equal to  $\wp(d,\vec r^+,\vec r^-) - \delta (v, w )$ where  $r^\pm$
encode the  cycle
decompositions of $v,w$
and $\delta (v , w) = 1$ if  $(v=1,  w=1)$ and  zero otherwise.

{\bf Proposition 10.2}. Suppose we are given $n^\pm >0$, $B^\pm$,
D and the set  $S (L)$.
Then, we have:
\eqn\wttdsum{\eqalign{
 \sum_{e \in CFG}  {{m(e)} \over {\vert C(e)\vert}}
&
=
\sum_{[f]\in CHS(n^\pm,B^\pm,S(L),D)} {1\over {\vert Aut f\vert} }
\cr
=\sum_{   { \scriptstyle {d_1,d_2\cdots d_L} \atop
           \scriptscriptstyle{d_1 + \cdots d_L = D}
          }
      }
&
\sum_{s^\pm_i,t^\pm_i, v _i, w _i}
{1\over {n^+! n^- !}}\qquad\qquad\qquad
\cr
\delta \biggl ( \prod_{i=1}^{L} v _i \otimes w _i
        \prod_{i=1}^{G}
&
[s^+_i\otimes s^-_i, t^+_i\otimes t^-_i] \biggr )
\prod_{i=1}^{L} p_{v _i \otimes w _i}^{(d_i)}
\cr
}}
where the first sum is over elements of CFG  compatible with the
data, the
second  sum
  is over equivalence
classes of  {\it degenerating } coupled covers inducing the specified
data,
and the
$v_i,w_i$  in the third  sum are compatible with total branching
numbers
$B^\pm$.

{\it Proof.}
The first equality follows from Proposition 10.1.
 To derive the second equality we note that
for a given pair of homomorphisms   which determine
  (anti-)holomorphic
maps $f^\pm: N^\pm(\Sw)\to \ST$, there are many ways to introduce
 double points to define a dcc. We sum over all possible
numbers of double points $d_1 \cdots d_L$ compatible with $d_1 + d_2
\cdots d_L
= D$.
Now there are a number of ways of introducing double points  $d_1,
\cdots d_L$
to define
dccs compatible with the  maps $f^\pm$.
 Let $d(f,P)$ denote the tube number above $P$. The case $d(P) = 2 $
is
illustrated
in Figure 15.

\vskipabit
\ifig\fhhpiv{$d(P)$ is the number of tubes above the point
$P$}
{\epsfxsize2.0in\epsfbox{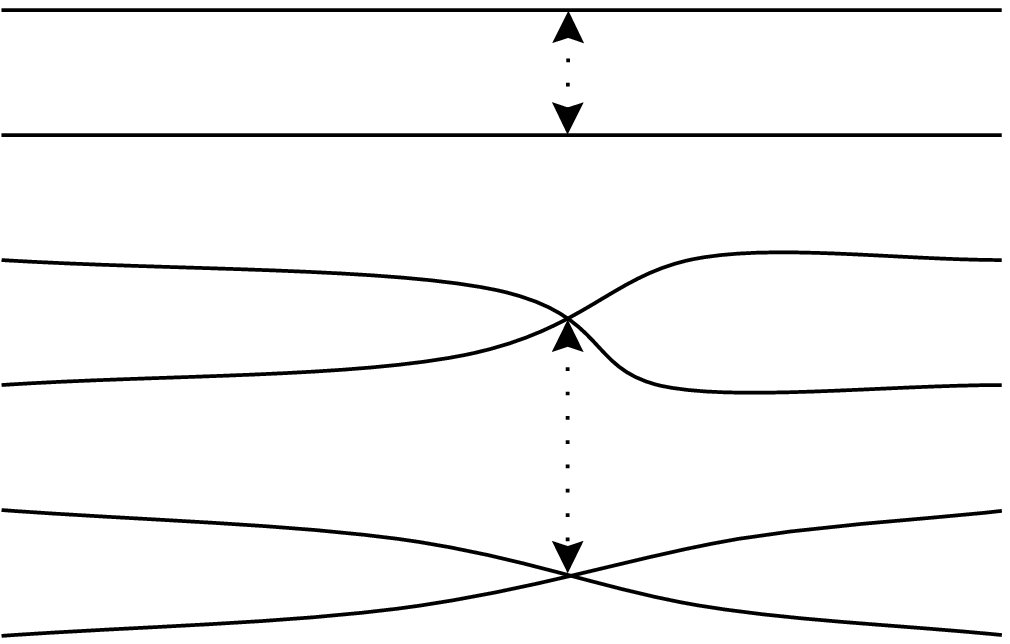}}

Because we count distinct degenerations separately, a double point
joining
ramification points of index  $j$ is counted $j$-times.
Thus there
are exactly $ \prod_{P}  p ^{d(f, P)}(v(P),w(P))$ ways of
introducing double points above points in $S(L)$  compatible with the
specified
homomorphisms. Therefore each pair of homomorphisms in the third line
is weighted by the total number configurations compatible with it,
multiplied by the multiplicity
appropriate for counting {\it degenerating } coupled covers.
$S_{n^+}\times S_{n^-}$ acts on the set of configurations and the
number of
times a given equivalence class occurs is ${{n^+!n^-!}\over{\vert
C(e)\vert}}$.
 This establishes the second equality.
$\spadesuit$

Now,  combining Proposition 10.2 and \eulcha, we may write the Euler
character
as
\eqn\eulchari{\eqalign{
&
\chi\biggl(CHS(n^\pm,B^\pm,L,D)\biggr)_{\rm orb}
\qquad\qquad\qquad
\cr
&
=\cch ( \cC_L(\Sigma_T))
\sum_{   { \scriptstyle {d_1,d_2\cdots d_L} \atop
           \scriptscriptstyle{d_1 + \cdots d_L = D}
          }
      }
\sum_{s^\pm_i,t^\pm_i, v _i, w _i}
{1\over {n^+! n^- !}}
\cr
\delta
&
\biggl ( \prod_{i=1}^{L} v _i \otimes w _i
        \prod_{i=1}^{G}
[s^+_i\otimes s^-_i, t^+_i\otimes t^-_i] \biggr )
\prod_{i=1}^{L} p_{v _i \otimes w _i}^{(d_i)}  .
 \cr}
}

\subsec{The nonchiral \ymt\ sum and Euler characters of CHS}

Having completed our geometrical preliminaries we return to the $1/N$
expansion of $YM_2$.
The full partition function \fullgt  can be written, in the zero area
limit,
as
\eqn \Zfull{\eqalign{
Z(0,G, N) =&\sum_{n^\pm =0}^{\infty}
N^{( n^+ +n^- ) (2-2G)}\cr
&\sum_{s^\pm_1,t^\pm_1,\ldots,s^\pm_G,t^\pm_G\in S_n}
\biggl[ {1\over {n^+! n^-!}}\delta \bigr ( \Omega_{n^+,n^-}^{2-2G}
\prod_{j=1}^G
[ s^+_j, t^+_j ] \otimes [ s^-_j,  t^-_j ] \bigr ) \biggr], \cr
                        } }
The delta function is over the group $S_{n^+} \times S_{n^-}$.
The element $\Omega_{n^+, n^-}$, introduced in \GrTa,  is related to
the
dimension of
$SU(N)$ representations by
 \eqn\coup{ \dim (R{\overline S}) = {{N^{n^++ n^-}}\over {n^+! n^-!}}
\chi_{R {\overline S}} (\Omega_{n^+ n^-}) }
where $R$ has $n^+$ boxes and $S$ has $n^-$ boxes, and
$(R\overline S)$ is the irreducible representation of largest
dimension
in the  tensor product of $R$ with the complex conjugate of $S$.
Explicitly,  $\Omega_{n^+, n^-}$ is an element of the group algebra
of
$S_{n^+}\times S_{n^-}$ given by

\eqn\Omegc{ \Omega_{n^+,n^-}
{}~=~ \sum_{v  \in S_{n^+}, w  \in S_{n^-}}
          (v  \otimes w ) P_{v  ,w } ({1\over N^2})
             \bigl({1\over N}\bigr)^{ (n^+ -K_{v }) + (n^- -K_{w })}}
The polynomials $P_{v,w }({1\over {N^2}})$ are given by $P_{v,w } (
{1\over
{N^2}})
= \wp(\vec r(v ),\vec r(w
),-1/N^2) $
 where
$\vec r$ is the vector of non-negative integers describing the
cycle decomposition of the permutations $v ,w $ and $\wp $ was
defined in
\polyn.

We write
\eqn\Ocp{\eqalign{ \Omega_{n^+, n^-} &= 1 \otimes 1  +
\sum_{v  \in S_{n^+}, w  \in S_{n^-}}
v  \otimes w
\bigl({1\over N}\bigr)^{ (n^+ -K_{v }) + (n^- -K_{w } ) }
p_{v  ,w }({1\over N^2})\cr
&= 1\otimes 1 + \sum_{d } ({-1\over N^2})^{d}
\sum_{v  \in S_{n^+}, w  \in S_{n^-}}
v  \otimes w   \bigl({1\over N}\bigr)^{
           (n^+ - K_{v }) + (n^- -K_{w } ) }
p_{v  ,w }^{(d)}
\cr}}
where  we have pulled out the leading term of
$1\otimes 1$ so that $p_{v , w } = P_{v , w }  - \delta(v,w) $.
In the second line we have collected terms with a given power
of $1/N$.

Using \Ocp\ and expanding the inverse $\Omega$ point
as in section five  leads to the following expression for the
partition function
\eqn\Zfuli{\eqalign{
Z(0,G,N) &= \sum_{h} N^{2-2h}
\sum_{ {\scriptstyle{n^\pm, B^\pm, D}}
    \atop{\scriptscriptstyle{(n^++n^-)(2G-2)+B^+ + B^- +2D=2h-2}}}
(-1)^D\cr
&\sum_{L} \cch ( \cC_L (\Sigma_T))
\sum_{   {\scriptstyle {d_1,d_2\cdots d_L} \atop {
\scriptscriptstyle{d_1 + \cdots + d_L = D}}}}\cr
&\sum_{s^\pm_i,t^\pm_i, v _i,w _i}
{1\over {n^+! n^- !}}
\delta( \prod_{i=1}^{L} v _i \otimes w _i
\prod_{i=1}^{G} [s^+_i \otimes s^-_i, t^+_i\otimes t^-_i] )
\prod_{i=1}^{L} p_{v _i \otimes w _i}^{(d_i)} .\cr}}
Since we have collected together the contributions with fixed
branching
number $B^+$ and $B^-$, the sum
over the permutations in the last line is required to obey the
condition $\sum_{i} (n^{\pm}-K_{v _i})= B^{\pm}$.
Note that the sum on $L$ is actually finite, and can be bounded
above by $B^+ + B^- +D$.
In the above, the sum appearing after the Euler character of the
configuration space of $L$  points in $\Sigma_T$ is a sum over a
discrete set which we have described in the previous subsection
as a sum over equivalence classes of dcc's.
Indeed, using \eulchari\ we finally arrive at

{\bf Proposition 10.3}
The full $A=0$ partition function of $YM_2$ is a generating
functional for the
orbifold Euler characters of coupled Hurwitz spaces:
\eqn\Zfulii{\eqalign{
Z(0,G,N) &= \sum_{h} N^{2-2h}
\sum_{{\scriptstyle{n^\pm,B^\pm, D}} \atop
    { \scriptscriptstyle{(n^++ n^-)(2G-2)+B^++ B^-+2D=2h-2}}}
(-1)^D\cr
&\sum_{L=0}^{B^+ + B^- +D} \cch_{\rm orb} ( CHS(n^\pm,B^\pm,L,D)
).\cr }}
 Note that the Euler character of configuration spaces which appears
involves the configurations of both  {\it branch points and
images of double points } on the target.

\newsec{The Nonchiral Topological String Theory}

The nonchiral analog of the theory of section 7 must localize on
both the space of holomorphic {\it and} antiholomorphic maps.
When we regard the topological string path integral as an
infinite dimensional version of an equivariant Thom class, it becomes
clear that we need a section ${\bf T}$ of some bundle
which localizes on the submanifolds $\widetilde {\cal M}^\pm$
of $\widetilde {\cal C}$ defined by $df \pm J~ df~ \epsilon [ h ]=0$.
It is therefore natural  to choose a section of the form:
\eqn\NonChiralSection{\eqalign{
{\bf T}\colon \widetilde {\cal C} ~\longrightarrow&~
\widetilde {\cal V}_{\rm nc} \oplus
\widetilde {\cal V}_{\rm nc}^{\rm cf}\cr
{\bf T}~ ( f, h ) ~\longmapsto&~
( df + J~ df~ \epsilon[ h ] ) \otimes ( df - J~ df~ \epsilon [ h ] )\cr}}
Following the considerations for the construction of a general TFT, we
have the following fields, ghosts, antighosts, and Lagrange-multipliers:
\vskip0.1truein
\hbox{\hfill
\divide\hsize by 2{
\vbox{
$$\eqalign{
\IF ~=&~ \pmatrix{f^\mu\cr h_{\alpha\beta}\cr}\cr
  A ~=&~ \rho_{\alpha \beta}^{\mu\nu }\cr}
$$}
\hfill
\vbox{
$$\eqalign{
\IG ~=&~ \pmatrix{ \chi^\mu\cr \psi_{\alpha\beta}\cr}\cr
\Pi ~=&~ \pi_{\alpha \beta}^{\mu\nu }. \cr}
$$}
\hfill}}
Only the anti-ghosts and Lagrange multipliers of the sigma model
have changed relative to the chiral theory.
In particular, the appropriate bundle for the antighosts $\rho$
has fiber:
\eqn\FibreVpm{
\widetilde {\cal V}^{\rm nc}_{{\scriptscriptstyle ( f, h)}}
{}~=~ \Gamma \left [ ( T ^\ast \Sigma_W )^{\otimes 2}
\otimes ( f^\ast  ( T \Sigma_T ))^{\otimes 2} \right ]_\pm}
where the subscript $\pm$ indicates that the sections must
satisfy ``self-duality" constraints:
\eqn\GenSDCon{
\rho ~\in~ \widetilde {\cal V}^{\rm nc}_{{\scriptscriptstyle ( f, h)}}
\qquad\Longleftrightarrow\qquad
\cases{
\rho - ( J \otimes {\bf 1} )~ \rho~ ( \epsilon \otimes {\bf 1} ) ~=~ 0 & \cr
\qquad{\rm or}
    & \cr
\rho + ( {\bf 1} \otimes J )~ \rho~ ( {\bf 1} \otimes \epsilon ) ~=~ 0 & \cr}}

The  BRST transformations are the same as above.
The nonchiral theory has an action
\eqn\NonChiralAction{
I_{\rm YM_2  string} ~=~ I_{\rm tg} +
I^{\rm nc}_{\rm t \sigma} +I^{\rm nc}_{\rm cofield}}
The gravity part of the action is the same as before.
The topological sigma model part is
\eqn\TopSigma{
I^{\rm nc}_{t\sigma}
{}~=~ Q_{\cal C} \int d^2 z~ \sqrt{h} \left \{
\rho^{\alpha \beta}_{\mu \nu } \bigl [
i  t^{\mu\nu }_{\alpha \beta}
-\Gamma^\mu_{\lambda\rho} \chi^\lambda \rho^{\rho \nu }_{\alpha \beta}
-\Gamma^\nu_{\lambda\rho} \chi^\lambda \rho^{\mu \rho }_{\alpha \beta}
+ \half \pi_{\alpha \beta}^{\mu\nu } \bigr ]  \right \}}
where the indices on $\rho$ and $\pi$ are raised and lowered with the
metrics on the worldsheet ($h$), and target space ($G$) .

If we expand \TopSigma\ and integrate out the Lagrange multiplier then the
bosonic term becomes (in local conformal coordinates)
\eqn\TopSigmai{
I^{\rm nc}_{\rm t \sigma}
{}~=~ \int h^{z \bar z} G_{w \bar w}^2~
\vert \partial_z f^w \vert^2 \vert \partial_{\bar z} f^w \vert^2
+ \cdots}
thus clearly localizing on both holomorphic and antiholomorphic maps.
Moreover, when we work out the quadratic terms in the fermions
we find that many components of $\rho$ do not enter the Lagrangian.
These components are eliminated by the  constraints \GenSDCon.
In locally conformal coordinates the only non-trivial components of
$\rho \in \widetilde {\cal V}_{{\scriptscriptstyle ( f, h)}}^{\rm nc}$ are
$\rho_{zz}^{\bar w w}, \rho_{z \bar z}^{\bar w \bar w},
\rho_{\bar z z}^{ww}$, and $\rho_{\bar z \bar z}^{w \bar w}$.
(Note that $\rho_{\alpha\beta}^{\mu \nu }$ is not symmetric in interchanging
$\{(\alpha\beta)(\mu\nu)\} \leftrightarrow \{(\beta\alpha)(\nu\mu)\}$.).
The kinetic term for the fermions is given by
\eqn\TopSigmaii{
I^{\rm nc}_{\rm t \sigma}
{}~=~ i \int d^2 z~ \sqrt{h}~
\pmatrix{ \rho & \eta\cr} \IO^{nc}
\pmatrix{\chi\cr \psi\cr} +\cdots}
where $\IO_{\rm nc}$ is a $2 \times 2$ matrix operator with entries:
\eqn\NonChiralO{\eqalign{
\IO^{{\scriptscriptstyle nc}}_{11}
{}~=&~ \CD^+ \otimes [ df - J~ df~ \epsilon ]
+ [ df + J~ df~ \epsilon ] \otimes {\cal D}^-\cr
\IO^{{\scriptscriptstyle nc}}_{12}
{}~=&~ J~ d f~ k \otimes [ df - J~ df~ \epsilon ]
- [ df + J~ df~ \epsilon ] \otimes J~ d f~ k\cr
\IO^{{\scriptscriptstyle nc}}_{21}
{}~=&~ \partial f\cr
\IO^{{\scriptscriptstyle nc}}_{22} ~=&~ P^\dagger\cr}}
where $( P^\dagger \delta h )_\beta
= D^\alpha \delta h_{\alpha\beta}$, ${\cal D}^\pm \chi^\mu
=  D \chi^\mu \pm J ( D \chi^\mu) \epsilon$
and, as usual, $k[ \delta h ]$ is the variation of the complex structure
on $\Sigma_W$ induced from a variation of the metric $\delta h$.
The co-model is introduced using the same principles as before.

\subsec{Singular Geometries}

The full path integral of the topological string theory
involves field configurations $(f,h) \in Map( \Sw, \ST ) \times Met(
\Sw )$,
which are  not necessarily $\CC^\infty$.
An appropriate  completion of this space
will introduce, among other things,
piecewise continuous or even singular maps and
geometries.
Usually in field theory such considerations
of analysis are of interest only in
constructive quantum field theory
\ref\cnstrctv{See, e.g., J. Glimm and A. Jaffe,
{\it Quantum Physics}, Springer 1981}\
but in the present case the contribution of singular
field configurations
 $(f,h)$ in the path integration domain becomes
an issue of great importance because the path integral
can localize on subspaces of singular geometries.
By allowing piecewise differentiable maps and
metrics we can incorporate the coupled Hurwitz space
 in the fiber bundle picture we described for
ordinary Hurwitz space in section 4.4.

In the theory of analytic functions one can show
\foot{This
is sometimes called Goursat's theorem.}
that the weak assumption of {\it differentiability}
of a solution to the Cauchy Riemann
equations  implies the function is $C^\infty$ and
even analytic
\ref\conway{J.H. Conway, {\it Functions of One Complex
Variable}, Springer.}. Thus it might seem that
one gains nothing by replacing the $C^\infty$ assumption
by the assumption of differentiability
when searching for zeroes of $df \pm J df \epsilon$.
This reasoning breaks down if the place where
derivatives of $f$ are discontinuous is also a singular
point in the
geometry of the worldsheet. This is precisely what happens
in a dcc.  Indeed, the following simple reasoning
shows the presence of solutions of $\tilde w=0$  in
\NonChiralSection\  which are not in $\tcF^\pm$.

Consider the space
$\Xi= Map( \Sw, \ST ) \times Met( \Sw )$ where we have
added singular geometries to form a ``boundary.''
\foot{
 We have sketched how dcc's arise in the bundle
approach to Hurwitz space when we consider the full
space of maps and metrics.
It would be very interesting to construct a careful compactification
of  the
space of
Maps $\times$ Metrics in such a way that
it is manifest why
 degenerated coupled covers should be counted
with the degeneracy  $m(e)$.
}
The situation is illustrated schematically in
Fig. 16.
\vskipabit
\ifig\fhhpiv{$\tcF^+$ and $ \tcF^-$ in $Map \times Met$}
{\epsfxsize3.5in\epsfbox{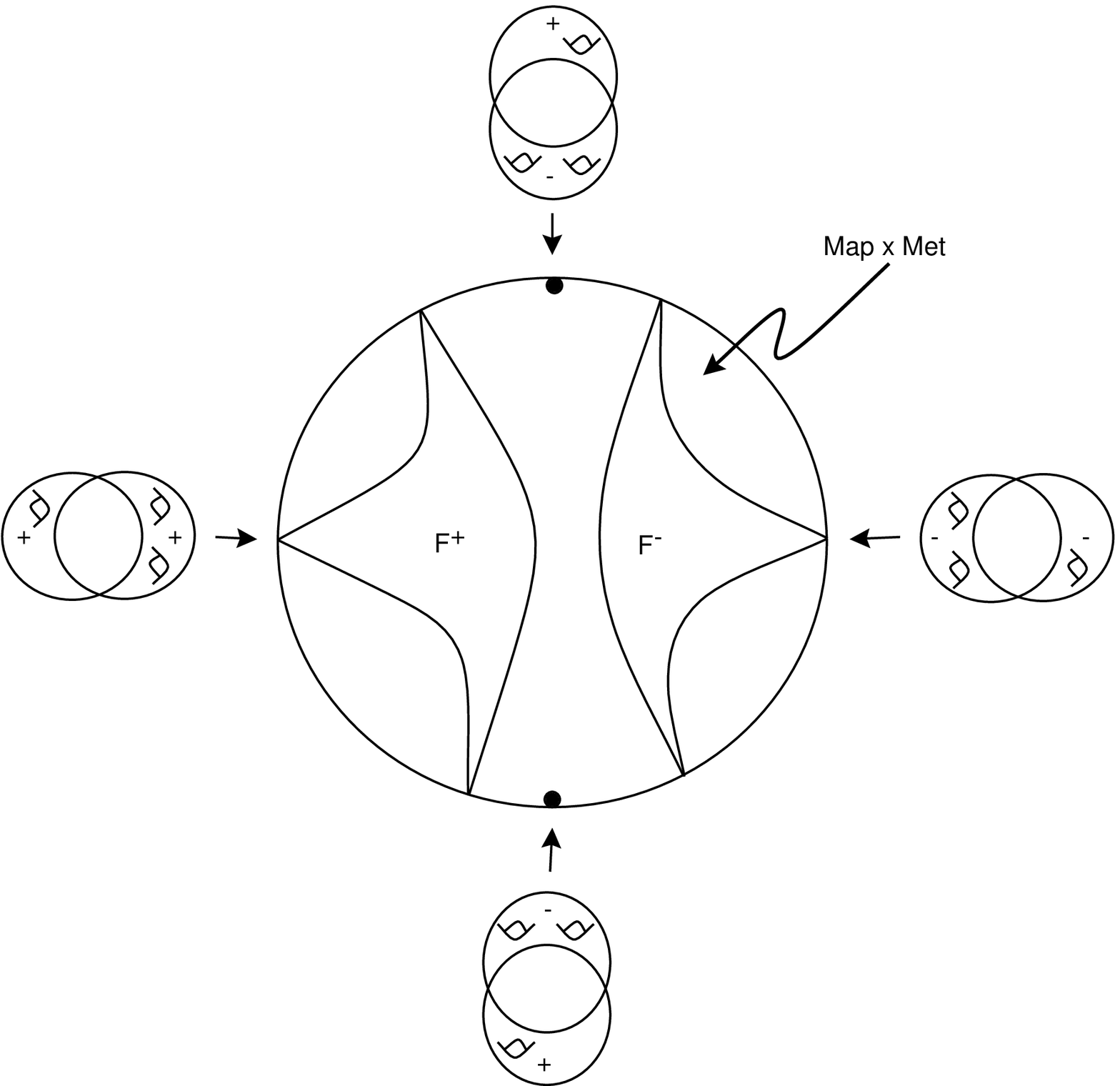}}

The Hurwitz spaces $\tcF^\pm$ lie in the interior
of $\Xi$, but extend out to the boundary because,
as we have already seen in section 4.5, type 3 collisions between
branch points can give rise to a singular worldsheets.
We saw there that we can obtain
ramification points  of equal index
on each of the components which are joined by
a tube. Such degenerations of holomorphic covers
are labeled by $(++)$ on the LHS of Figure 16.

Now, the theory of section 11.1 is invariant only under
the group of orientations preserving diffeomorphisms
$Diff^+(\Sw)$, so configurations related by
diffeomorphisms of type $(\pm, \mp)$, which are
orientation preserving on one component but orientation
reversing on the other are considered gauge-inequivalent.
The $4$  points we have indicated on the boundary  of  $\Xi$
which lie on the outer
circle are related to one another by such diffeomorphisms.
The configurations indicated at the top and bottom of
Fig. 16 are additional, singular configurations which
can contribute to the localisation of the path integral.
The way in which we handle these surface contributions
corresponds to a choice of contact terms, similar to the
contact terms that accounted for area polynomials
in section 8.

One choice of contact terms simply declares that
the dcc's do not contribute: we cut out all singular field
configurations and
define the integral by a limiting procedure. From the results of
section 7 we can immediately conclude that
with this choice of contact terms the  partition
function of the topological string theory becomes:
\eqn\result{
exp\Biggl\{\sum_{h\geq 0}\biggl({1\over N}\biggr)^{2h-2} Z_{\rm
string}(\Sw\to
\ST)\Biggr\}=Z^+(A=0,N)Z^-(A=0,N)
.}
That is, we produce the ``chiral'' \ymt\ theory
obtained by replacing:
$$\Omega_{n^+,n^-} \longrightarrow
\Omega_{n^+,0} \Omega_{0,n^-}$$
in the full \ymt\ sum.  (Note this is a product of
chiral theories in the sense that ``chiral'' is usually
employed.)
It is not at all clear that this is a sensible
(e.g., BRST invariant) choice of contact terms
from the point of view of the topological string theory.
 A  contact term analysis similar to that used
in section 8  is  needed to explain why dcc's contribute whereas
degenerations of $(++)$ or $(--)$ type do not contribute.

\newsec{Conclusions}

\subsec{General Remarks}

In this paper we have used the
results of \GrTa\ to make some progress towards
a  formulation of $YM_2$  as a topological string theory.
We have seen that the $1/N$  expansion of  \ymt\  may always be
formulated in  terms of quantities associated to branched covers,
provided we admit sufficiently singular geometries.
In sections 6-9  we formulated a string theory which
reproduces chiral \ymt\ when proper account is taken
of singular geometries. In sections 10,11 we have
partially extended the results to nonchiral \ymt.

Reproducing all  \ymt\ in complete detail, involves highly
complicated
considerations of the contributions  of multiple contact terms.
In some cases, for example, for the area polynomials
associated with simple Hurwitz space a heuristic
analysis of contact terms  allows us to {\it derive} the singular
contributions to the string path integral,
as in section 8.
Amplitudes we have {\it not} explicitly calculated
from the string theory picture are :

\noindent
1.)Area polynomials associated with non-simple coverings, and the
contributions
of handles and tubes (from expanding  \tubhan).

\noindent
2.) Modifications of area polynomials from including dcc's , and
expanding $\exp(-n^+n^-A/N^2) $ in \fullgt.

\noindent
3.) Nonchiral Wilson loop amplitudes.

There is an interesting analogy between the
topological string formulation of $YM_2$ and the topological approach
to 2D
gravity.
The exact \ymt\ answers like \pfun\wilavi\
play a role analogous to the results of
the double-scaled matrix models.
As in 2D gravity,
these  nontrivial and exact answers
show
the importance of singular geometries to any
topological formulation. We are lucky to have
these answers, since they guide us through the
dense thicket of singular
boundary contributions which are, presumably,
ubiquitous in all theories of gravity.
\foot{An interesting related example where the
exact answers have {\it not} been  previously available
 is curve-counting in
Calabi-Yau three-folds \bcov.} Contact terms
are at once the Achilles heel and the rock of
salvation of topological gravity. They make the
truly interesting theories nearly impossible to
analyze,
yet  provide a mechanism whereby the theory
can be nontrivial enough to be worthy of attention.

\subsec{Open problems, future directions}

The present work suggests several possible
generalizations  and further directions for research.

One important generalization is the string-theoretic
version of \ymt\ based on the
other series of classical compact
gauge groups $Sp(N)$, $O(N)$.
The $1/N$ expansion of these theories has been
worked out in \ref\nars{S.G.~ Naculich, H.A.~ Riggs, and
H.G.~ Schnitzer, ``2D Yang Mills theories are string theories,''
Mod. Phys. Lett. {\bf A8} (1993) 2223. }\ref\skr{S. Ramgoolam,
``Comment on two dimensional $O(N)$ and $Sp(N)$ Yang Mills
theories as string theories,'' hep-th/9307085, to appear in Nuc. Phys.
B. }.
For $O(N)$ and $Sp(N)$ Yang Mills
one has to deal with new subtleties associated
with string theories on
non-orientable
surfaces.  A given
branched cover (possibly with double points)
 in these
theories appears only once. There is no analog
of the two sectors in the $SU(N)$ theory.
In a bundle description of
the corresponding Hurwitz space analogous to section 4.3,
one would replace $Diff^+(\Sigma_W)$ with $Diff(\Sigma_W)$ .

Douglas \dougrev\ has observed that for $\ST$ of
genus one there is a ``near'' $A\to 1/A$ duality of
\ymt\ since the amplitudes are expressed in terms
of Eisenstein series in $q=e^{-A}$. As noted by
many physicists, target space duality is quite natural
for a \top\ string theory based on a topological
conformal field theory (TCFT). In order to make
the answer truly modular covariant it is necessary
\dougrev\ to modify  \ymt\ in seemingly
unmotivated ways. One may search for an
explanation of this in the string formulation.
Further, when $\ST$ does {\it not} have genus one
we are coupling topological gravity to a topological
$\sigma$-model which is not a TCFT.

As is well known, an important challenge to
the string approach is the derivation of
quantities such as the meson spectrum.
It is nontrivial to
rederive the standard
results of the  't Hooft model.

Finally, the construction of
the anomaly-cancelling co-model $S_c$, can be
applied to a wide class of topological field theories.
One may wonder if the resulting theories are of
any interest. For example, the analogous
construction with Donaldson theory would compute
the Euler character of the moduli space of instantons.
What is the physical interpretation of this theory?

\subsec{What About QCD4?}

Aside from the intrinsic beauty of the subject, one of the main
reasons we
are interested in a string action for \ymt\ is the hope that this
action might contain essential features of a hypothetical string
action
for $YM_4$. In this respect our construction is disappointing since
it relies
on topological field theory. Our construction {\it does} have a
natural
generalization to 4 dimensions: Choose an almost
complex structure on the four-dimensional target and form
the appropriate topological sigma model. Then follow the
construction of the co-model to cancel
the anomalous $R$-symmetry and guarantee that
we get the Euler density for the moduli space of
curves in the target.
Following the general ideas outlined above one will find
that the \top\ string theory localizes onto the family of
\holo\ maps, ${\cal F} ( \Sigma_h, X )$. A formula for
the  dimension
of this moduli space  is  given in eq. $(A.16)$ below.

Of course, any given topological theory admits several
different formulations, and it might be that these different
formulations admit different generalizations to four dimensions.
Perhaps one of these will teach us something about
$YM_4$. Some indication that this might be possible is given by the
construction of  \Kostov which also has a natural
generalization to four dimensions.

\bigskip
\centerline{\bf Acknowledgements}

We would like to thank V.I. Arnold,
M. Bershadsky, P. Bouwknegt, S. Cecotti,
 R. Dijkgraaf, M. Douglas, E. Getzler, D. Gross,
M. Guest, J. Harris, J. Horne,
V. Kazakov, M. Khovanov, I. Kostov, B. Lian, V. Mathai, J. McCarthy
M. Newman, H. Ooguri, R. Rudd, J. Segert, W. Taylor, A. Wilkins
and G. Zuckerman for discussions and correspondence.
We would like to thank R. Dijkgraaf, R. Plesser, and W. Taylor
for useful comments on the manuscript.
This work is supported by DOE grants DE-AC02-76ER03075,
 DE-FG02-92ER25121, DE-FG05-900ER40559,
and by a Presidential Young Investigator Award.
GM is grateful to the Rutgers Dept. of Physics for
hospitality while this paper was being finished.

\appendix{A}{The Deformation Theory Approach to Hurwitz Space}
\par
In this appendix we derive a formula for the dimension of the space
of families of holomorphic maps.
Our treatment here is valid also for the case of maps into higher
dimensional target spaces.
\par
The deformation theory of holomorphic maps is a subject developed by
Horikawa\ref\Horikawa{E.~ Horikawa, ``On deformations of holomorphic
maps, I, II, III", J. Math. Soc. Japan, {\bf 25} (1973) 647; ibid {\bf 26}
(1974) 372; Math. Ann. {\bf 222} (1976) 275.}, Mijayima\ref\Mijayima{M.~
Mijayima, ``On the existence of Kuranishi Family for deformations of
holomorphic maps", Science Rep. Kagoshima Univ., {\bf 27} (1978) 43.}
and Namba\ref\Namba{M.~ Namba, ``Families of Meromorphic Functions
on Compact Riemann Surfaces", Lecture Notes in Mathematics, Number
767, Springer Verlag, New York, 1979.}.
Let $U$ be a fixed compact complex manifold.
A family of holomorphic maps into $U$ is, by definition a family
$( X, \pi, S ) = \{ V_s \}_{s \in S}$ of compact complex manifolds,
together with a holomorphic map ${\cal F} : X \to  U$.
We denote it by $( X, \pi, S, {\cal F} )$.
Set
$$
f_s = {\cal F} \vert_{V_s} : V_s \longrightarrow U.
$$
Sometimes this family of maps is denoted by
$\{ V_s, f_s \}_{s \in S}$.
\par
Next we define an infinitesimal deformation of a family
$( X, \pi, S, {\cal F} ) = \{ V_s, f_s \}_{s \in S}$ at $o \in S$.
Let $f = f_o$.
The data which characterize ${\cal F}$ locally are:
\item{(a)} An open covering of $V$, ${\cal V} = \{ V_m \}$  with local
coordinates $z^\alpha_{\scriptscriptstyle ( m )}$ and
transition functions $g_{mn} : V_n \times S \to V_m$ which vary with
$s \in S$,
\item{(b)} An open covering of $U$, ${\cal U} = \{ U_m \}$ with local
coordinates
$w^i_{\scriptscriptstyle ( m )}$ and {\it fixed}\foot{There is a
natural extension of the deformation theory of holomorphic maps
developed by Namba\refs{\Namba}, wherein the complex structures of
both $V$ and $U$ are varied.
This, however, would only become relevant if we studied
${\rm YM}_2$ coupled to 2d (spacetime) gravity.}
transition functions $h_{mn} : U_n  \to U_m$,
\item{(c)} A family of holomorphic maps
$f_{\scriptscriptstyle ( m )}\colon V_m \times S \to U_m$
which vary with $s \in S$.
\par\noindent
The $\{ f_m \}$, $\{ g_{mn} \}$, and $\{ h_{mn} \}$ must satisfy the
compatibility condition
\eqn\compcond{
h_{mn} \circ f_n = f_m \circ g_{mn},}
which ensures commutativity of the following:
$$
\matrix{
U_n \cap U_m & \maprightu{h_{mn}} & U_m \cap U_n\cr
   \mapup{f_n}    &                                  &
\mapup{f_m} \cr
V_n \cap V_m  & \maprightu{g_{mn}} & V_m \cap V_n  \cr}
$$
Differentiating \compcond\ with respect to $s$ and contracting with
${\partial\over{\partial w^i_{\scriptscriptstyle ( m )}}}$, we find
\eqn\infdef{
\sum_\alpha \partial_s f_{\scriptscriptstyle ( m )}^i
{\partial\over{\partial w^i_{\scriptscriptstyle ( m )}}}
- \sum_{\alpha,\beta} \partial_s f^i_{\scriptscriptstyle ( n )}
{\partial h^j_{mn} \over{\partial w^i_{\scriptscriptstyle (n )}}}
\circ f_n { \partial\over{\partial w^j_{\scriptscriptstyle ( m )}}}
{}~=~ - \sum_{\alpha\beta} \partial_s g_{mn}^\beta
{\partial {f_{\scriptscriptstyle ( m )}^\alpha} \over {\partial
z_{\scriptscriptstyle ( m )}^\beta}}
\circ g_{mn} {\partial \over {\partial w_{\scriptscriptstyle ( m )}^i}}.}
Now define the \u{C}ech $1$-cocycle, $\theta$, valued in the sheaf of
germs of holomorphic sections of $TV$ to be:
\eqn\thedef{
\theta ~=~ \{ \theta_{mn} \} ~\in~ Z^1 ( {\cal V}, \Theta_V ),
\qquad\qquad
\theta_{mn} ~=~ \partial_s g_{mn}.}
Further define a \u{C}ech $0$-cochain, $\eta$, valued in the inverse
image sheaf\foot{Recall that if $f: V \to U$ is a continuous map of
topological spaces, then the {\it inverse image sheaf},
$f^\ast \Theta_U$, of $\Theta_U$ by the map $f$ is defined as
\eqn\defiis{
f^\ast \Theta_U ~=~ {\cal O} ( f^\ast TU ),}
the sheaf of germs of holomorphic sections of the pullback $f^\ast TU$ of
the holomorphic tangent bundle over $f$.
\par
$f_\ast : \Theta_V \to \Theta_U$ is the {\it push-forward} map.},
$f^\ast \Theta_U$, by
\eqn\etadef{
\eta ~=~ \{ \eta_i \} ~\in~ C^0 ( {\cal V}, f^\ast \Theta_U ),
\qquad\qquad \eta_m ~=~ \partial_s f_m.}
Then \infdef\ expresses the fact that
\eqn\infdefp{
( \delta \eta )_{mn} ~=~ -f_\ast \theta_{mn}.}
\par
The deformation theory of holomorphic maps has a very concise formulation
in terms
of a {\it characteristic map} from a tangent  vector
${\partial\over{\partial s}} \in T_o S$ to
a certain cohomology class.
This is the analogue of the Kodaira-Spencer map which arises in the
study of deformations of the complex structures of complex manifolds.
In order to present this description of the tangent space to ${\cal F}$,
we need to introduce a few more notions:
First we introduce the following complex of sheaves\refs{\Namba}:
\eqn\shcomp{
{\cal L}^\ast :\qquad
0 ~\maprightu{f_\ast}~ {\cal L}^0 = \Theta_V ~\maprightu{f_\ast}~
{\cal L}^1 = f^\ast \Theta_U ~\maprightu{f_\ast}~ 0,}
where $f_\ast$ is the sheaf map which tautologically satisfies
$( f_\ast )^2 = 0$.
Associated to this complex of sheaves\ref\GriHar{ P.Griffiths and
J.Harris,``Principles of Algebraic Geometry,'' p. 445, J.Wiley and
Sons, 1978. } are the cohomology sheaves
${\cal H}^q = {\cal H}^q ( {\cal L} )$.
Setting ${\cal L}^q ( {\cal U} ) = H^0 ( {\cal V}, {\cal L}^q )$, the
presheaf
$$
V \longmapsto {{\ker~ \{ f_\ast: {\cal L}^q ( V )
\to {\cal L}^{q+1} (V ) \}}\over{f_\ast {\cal L}^{q-1} ( V )},}
$$
gives rise to a sheaf ${\cal H}^q$ whose stalk is
$$
{\cal H}^q_x ~=~
\lim_{V \ni x} {{\ker~ \{ f_\ast: {\cal L}^q ( V ) \to {\cal L}^{q+1} ( V ) \}}
\over{f_\ast {\cal L}^{q-1} ( V )}.}
$$
A section $\eta$ of ${\cal H}^q$ over $V$ is given by a
covering $\{ V_m \}$ of $V$ and $\eta_m \in {\cal L}^q ( V_m )$
such that
\eqn\hsect{
f_\ast \eta_m ~=~ 0}
A section is zero in the case when
\eqn\zsect{
\eta_m ~=~ f_\ast \theta_m,\qquad\qquad
\theta_m \in {\cal L}^{q-1} ( V_m ),}
after perhaps refining the cover.
\par
Let  $C^p ( {\cal V}, {\cal L}^q )$ be \u{C}ech cochains valued in
${\cal L}^q$.
Then we have the two operators:
\eqn\dcoh{\eqalign{
\delta: C^p ( {\cal V}, {\cal L}^q )
{}~\longrightarrow&~ C^{p+1} ( {\cal V}, {\cal L}^q),\cr
f_\ast : C^p ( {\cal V}, {\cal L}^q )
{}~\longrightarrow&~ C^p ( {\cal V}, {\cal L}^{q+1}),\cr}}
which satisfy $( f_\ast )^2 = \delta^2 =  \{ f_\ast, \delta \} = 0$,
so we have a double complex
$\{ C^{p,q} = C^p ( {\cal V}, {\cal L}^q ), f_\ast, \delta \}$.
The associated single complex $( C^\ast, D )$ is defined by
\eqn\singcomp{
C^n ~=~ \bigoplus_{p+q=n} C^{p,q},\qquad\qquad
D ~=~ f_\ast + \delta.}
We define the {\it hypercohomology} as follows:
\eqn\hycohom{
\IH^\ast ( V, {\cal L}^\ast )
{}~=~ \lim_{\cal V} H^\ast ( C^\ast ( {\cal V} ), D).}
The pair $( \eta, \theta ) \in \IH^1 ( {\cal V}, {\cal L}^\ast )$ is called an
{\it infinitesimal deformation of the family} $\{ V_s, f_s \}_{s \in S}$
{\it at} $s \in S$ in the direction ${\partial\over{\partial s}} \in T_o S$.

One denotes $\alpha_o ( {\partial\over{\partial s}} )\equiv ( \eta,
\theta )$.
$\alpha_o$ is a linear map
\eqn\chmap{
\alpha_o : T_o S ~\longrightarrow~ \IH^1 ( {\cal U}, {\cal L}^\ast ).}
called the {\it characteristic map}.
\par\noindent
{\bf Definition}: The family $\{ V_s, f_s \}_{s \in S}$
is said to
be
{\it effectively parametrized} at $o \in S$ if $\alpha_0$ is
injective.
\par\noindent
{\bf Definition}: A {\it morphism} of $( X^\prime, \pi^\prime,
S^\prime, {\cal F}^\prime )$ to $( X, \pi, S, {\cal F})$ is by definition
a morphism $( h, \widetilde h )$ which makes the following diagram
commutative
\medskip
$$
\matrix{
     X^\prime   &                  & \maprightu{\tilde h} &
&
     X          \cr
                         & \mapse{\cal F^\prime} &
&
\mapsw{\cal F}  &
                       \cr
 \mapdown{\pi^\prime}  &                  &           U
&
& \mapdown{\pi}  \cr
                         &                  &
&
               &                        \cr
S^\prime         &                  & \maprightu{h} &
&
      S           \cr}
$$
\medskip\noindent
{\bf Definition}: A family $\{ V_s, f_s \}_{s \in S}$ is said to be
{\it complete at} $o \in S$ if for every family
$\{ V_{s^\prime}, f_{s^\prime} \}_{s^\prime \in S^\prime}$
with a point $o^\prime \in S^\prime$ and a biholomorphic map
$i : V_{o^\prime}^\prime \to V_o$ which makes the following
diagram commutative
\medskip
$$
\matrix{
V^\prime_{o^\prime} &                  & \maprightu{i} &
&
V_o \cr
                                      & \mapse{f^\prime_{o^\prime}} &
& \mapsw{f_o}  &          \cr
                                      &                  &
U
     &                    &          \cr
}
$$
\medskip\noindent
there is an open neighborhood $U^\prime$ of $o^\prime \in S^\prime$
and a morphism
$( h, {\widetilde h} )$ of $\{ V_{s^\prime},
f_{s^\prime} \}_{s^\prime \in S^\prime}$
such that
\item{(i)} $h ( o^\prime )= o$,
\item{(ii)} ${\widetilde h}_{o^\prime} = i : V_0^\prime \to V_0$.
\par\noindent
If a family is complete at $o \in S$, then it contains all small
deformations of $f$.
$\{ V_s, f_s \}_{s \in S}$ is {\it complete}, if it is complete at
every point of $S$.
\par
If a family  $\{ V_s, f_s \}_{s \in S}$ is complete and effectively
parametrized at $o \in S$, then it is said to be {\it versal}.
In this case, it is the smallest among complete families.
One important property of ${\cal F}$ which is of obvious interest is its
dimension.
For this purpose the following two theorems due to
Horikawa\refs{\Horikawa} and Namba\refs{\Namba} are useful:
\medskip\noindent
{\bf Theorem}: [Horikawa] For a \def\IH{\relax{\rm I\kern-.18em H}}
holomorphic map $f: V \to U$, if
\item{(A)} $H^1 ( {\cal V}, \Theta_V )
\to H^1 ( {\cal V}, f^\ast \Theta_U )$ is surjective,
\item{(B)} $H^2 ( {\cal V}, \Theta_V )
\to H^2 ( {\cal V}, f^\ast \Theta_U )$ is injective,
\par\noindent
there exists a complete family $\{ V_s, f_s \}_{s \in S}$
of holomoprhic maps into $U$ with a point $o \in S$, such that
\item{(1)} $V_o = V$,
\item{(2)} $f_o = f$,
\item{(3)} it is effectively parametrized at $o$,
\item{(4)} $o$ is a non-singular point of
$S$ and ${\rm dim}_o~ S
={\rm dim}~ \IH^1 ( {\cal V} , {\cal L}^\ast ).$
\medskip\noindent
{\bf Theorem}: [Namba] The following sequence is exact:
\eqn\les{\matrix{
0 & \longrightarrow
& \IH^0 ( {\cal V}, {\cal L} ) & \longrightarrow
& H^0 ( {\cal V}, \Theta_V ) & \longrightarrow
& H^0 ( {\cal V}, f^\ast \Theta_U ) & \longrightarrow\cr
  & \longrightarrow
& \IH^1 ( {\cal V}, {\cal L} ) & \longrightarrow
& H^1 ( {\cal V}, \Theta_V ) & \longrightarrow
& H^1 ( {\cal V}, f^\ast \Theta_U ) & \longrightarrow\cr
  & \longrightarrow
& \IH^2 ( {\cal V}, {\cal L} ) & \longrightarrow
& H^2 ( {\cal V}, \Theta_V ) & \longrightarrow
& H^2 ( {\cal V}, f^\ast \Theta_U ) & \longrightarrow\cr}}
For $h \ge 2$, $\IH^0 ( {\cal V}, {\cal L}^\ast ) = 0$, we find for
{\it versal} families ${\cal F} (  \Sigma_h, \Sigma_G, J)$ that
\eqn\dimF{\eqalign{
{\rm dim}~ T {\cal F} (  \Sigma_h, \Sigma_G, J)
{}~=&~  {\rm dim}~ \IH^1 ( {\cal V}, {\cal L}^\ast )\cr
{}~=&~  - {\rm dim}~ H^0 ( {\cal V}, \Theta_V )
+ {\rm dim}~ H^0 ( {\cal V}, f^\ast \Theta_U )\cr
&\qquad+ {\rm dim}~ H^1 ( {\cal V},  \Theta_V )
- {\rm dim}~ H^1 ( {\cal V}, f^\ast \Theta_U )
{}~=~ B\cr}}
where $B$ is precisely the branching number!

As another application, one can use \dimF\ to determine
the dimension of the moduli space of holomorphic maps,
$f\colon \Sigma_W \to X$, from a Riemann surface, $\Sigma_W$, into
a higher dimensional target space, $X$.
If $\Sigma_W$ has genus $h$, then it is easy to establish
that
\eqn\Fdim{
{\rm dim}~ {\cal F} ( \Sigma_h, X )
{}~=~ 3 (h - 1 ) + {\rm dim}~ H^0 ( \Sigma_h, f^\ast \Theta_X )
        -  {\rm dim}~ H^0 ( \Sigma_h, K_h \otimes f^\ast K_X ).}

\appendix{B}{Derivation of the Variation of Gromov's Equation}

We shall remain general and consider the target space to be an arbitrary
complex manifold, $X$.
It is important to note that the Gromov equation is non-linear in $f$.
We can make this clear by explicitly indicating that $J$ is evaluated at
$f ( \sigma )$:
\eqn\Gromov{
d f ( \sigma ) + J [ f ( \sigma )] df ( \sigma ) \epsilon ( \sigma ) = 0}
Now consider a one parameter family of holomorphic maps
$$\eqalign{
F\colon \Sigma \times I \to& X\cr
F ( \sigma; t ) \mapsto& f_t ( \sigma )\cr}
$$
with $f_0 ( \sigma ) = f ( \sigma )$.
This family must also satisfy the Gromov equation
$$
d f_t ( \sigma ) + J [ f_t  ( \sigma )] df_t ( \sigma ) \epsilon ( \sigma ) =
0\qquad
\forall \sigma \in \Sigma~ {\rm and}~ \forall t \in I
$$
Now take the derivative with respect to $t$ and evaluate at $t=0$.
We suppress worldsheet indices where they are obvious.
$$\left [
d \dot f^\mu_t ( \sigma )
+ \partial_\kappa J^\mu_{~ \nu} [ f_t ( \sigma )] \dot f^\kappa_t ( \sigma )
   d f^\nu_t ( \sigma ) \epsilon ( \sigma )
+ J^\mu_{~ \nu} [ f_t ( \sigma )] d \dot f^\nu_t ( \sigma ) \epsilon ( \sigma )
\right ]_{t=0} = 0
$$
Now consider the covariant derivative of $J$
$$
\nabla_\kappa J^\mu_{~ \nu}
= \partial_\kappa J^\mu_{~ \nu}
+ \Gamma^\mu_{~ \kappa\lambda} J^\lambda_{~ \nu}
 - \Gamma^\lambda_{~ \kappa\nu} J^\mu_{~ \lambda}
$$
Then we may write (setting $\delta f^\mu = \dot f^\mu_t \vert_{t=0}$)
$$
\partial_\kappa J^\mu_{~ \nu} \delta f^\kappa d f^\nu \epsilon
= - \Gamma^\mu_{~ \kappa\lambda} J^\lambda_{~ \nu} \delta f^\kappa d f^\nu
\epsilon
+  \Gamma^\lambda_{~ \kappa\nu} J^\mu_{~ \lambda} \delta f^\kappa d f^\nu
\epsilon
+ \nabla_\kappa J^\mu_{~ \nu} \delta f^\kappa d f^\nu \epsilon
$$
Now since $f_t ( \sigma )$ is, by fiat, a family of holomorphic maps,
$d f_t^\mu \epsilon = J^\mu_{~ \nu} d f_t^\nu$, for all $t$, so that
$$
\partial_\kappa J^\mu_{~ \nu} \delta f^\kappa d f^\nu \epsilon
= \Gamma^\lambda_{~ \kappa\nu} J^\mu_{~ \lambda} \delta f^\kappa d f^\nu
\epsilon
-  \Gamma^\mu_{~ \kappa\lambda} \delta f^\kappa d f^\lambda
+ \nabla_\kappa J^\mu_{~ \nu} \delta f^\kappa d f^\nu \epsilon
$$
In the case that $X$ is a complex manifold, $\nabla_\kappa J^\mu{}_\nu
= 0$ and we deduce the following equation for the tangent space:
\eqn\TanSpEq{
D ( \delta f ) + J~ D ( \delta f )~ \epsilon[ h ] + J~ d f~ k [ \delta h ]
{}~=~ 0}
where $D$ is the pulled-back connection
$(D_\alpha \delta f )^\mu = \partial_\alpha \delta f^\mu
+ \Gamma^\mu_{\kappa\lambda} \partial_\alpha f^\kappa
\delta f^\lambda$.

\listrefs
\bye